\documentclass[bibyear]{aa}                   
\usepackage{graphicx}                
\usepackage{txfonts}                 
\usepackage{rotating}

\newcommand{\Halpha}{H$\alpha$}
\newcommand{\Hbeta}{H$\beta$}

\newcommand{\kms}{km\,s$^{-1}$}
\newcommand{\ms}{m\,s$^{-1}$}

\begin{document}

\title{VPNEP: Detailed characterization of TESS targets around the Northern~Ecliptic~Pole\thanks{Based on data acquired with the Potsdam Echelle Polarimetric and Spectroscopic Instrument (PEPSI) using the Vatican Advanced Technology Telescope (VATT; i.e. the Alice P. Lennon Telescope and the Thomas J. Bannan Astrophysics Facility) in Arizona and the STELLA Echelle Spectrograph (SES) using the Stellar Activity (STELLA) robotic facility in Tenerife.}}

\subtitle{I. Survey design, pilot analysis, and initial data release\thanks{The data and the table with the observing logs are only available in electronic form at the CDS via anonymous ftp to cdsarc.cds.unistra.fr (130.79.128.5) or via https://cdsarc.cds.unistra.fr/cgi-bin/qcat?J/A+A/}}

\author{K. G. Strassmeier\inst{1,2}, M. Weber\inst{1}, D. Gruner\inst{1,2}, I. Ilyin\inst{1}, M. Steffen\inst{1}, M. Baratella\inst{1}, S. J\"arvinen\inst{1}, T. Granzer\inst{1},\\ S. A. Barnes\inst{1}, T. A. Carroll\inst{1}, M. Mallonn\inst{1}, D. Sablowski\inst{1}, P. Gabor\inst{3}, D. Brown\inst{3}, C. Corbally\inst{3}, \and M. Franz\inst{3}}

\institute{
    Leibniz-Institute for Astrophysics Potsdam (AIP), An der Sternwarte 16, D-14482 Potsdam, Germany; \\ \email{kstrassmeier@aip.de},
    \and
    Institute for Physics and Astronomy, University of Potsdam, D-14476 Potsdam, Germany;
    \and
    Vatican Observatory Research Group, Steward Observatory, 933 N. Cherry Avenue, Tucson, U.S.A.}

\date{Received ... ; accepted ...}

\abstract{We embarked on a high-resolution optical spectroscopic survey of bright Transiting Exoplanet Survey Satellite (TESS) stars around the Northern Ecliptic Pole (NEP), dubbed the Vatican-Potsdam-NEP  (VPNEP) survey.}{Our NEP coverage comprises $\approx$770 square degrees with 1067 stars, of which 352 are bona fide dwarf stars and 715 are giant stars, all cooler than spectral type F0 and brighter than $V$=8\fm5. Our aim is to characterize these stars for the benefit of future studies in the community.}{We analyzed the spectra via comparisons with synthetic spectra. Particular line profiles were analyzed by means of eigenprofiles, equivalent widths, and relative emission-line fluxes (when applicable).}{Two $R$=200\,000 spectra were obtained for each of the dwarf stars with the Vatican Advanced Technology Telescope (VATT) and the Potsdam Echelle Polarimetric and Spectroscopic Instrument (PEPSI), with typically three $R$=55\,000 spectra obtained for the giant stars with STELLA and the STELLA Echelle Spectrograph (SES). Combined with $V$-band magnitudes, \emph{Gaia} eDR3 parallaxes, and isochrones from the Padova and Trieste Stellar Evolutionary Code, the spectra can be used to obtain radial velocities, effective temperatures, gravities, rotational and turbulence broadenings, stellar masses and ages, and abundances for 27 chemical elements, as well as isotope ratios for lithium and carbon, line bisector spans, convective blue-shifts (when feasible), and levels of magnetic activity from \Halpha , \Hbeta,\ and the \ion{Ca}{ii} infrared triplet. In this initial paper, we discuss our analysis tools and biases,  presenting our first results from a pilot sub-sample of 54 stars (27 bona-fide dwarf stars observed with VATT+PEPSI and 27 bona-fide giant stars observed with STELLA+SES) and making all reduced spectra available to the community. We carried out a follow-up error analysis was carried out, including systematic biases and standard deviations based on a joint target sample for both facilities, as well as a comparison with external data sources. }{}


\authorrunning{K. G. Strassmeier et al.}

\titlerunning{VPNEP: characterizing TESS targets around the NEP}

\maketitle

\section{Introduction}

The main aim of this survey is to provide precise spectroscopic parameters for potential planet-host stars from the NASA Transiting Exoplanet Survey Satellite (TESS; Ricker et al. \cite{tess}) mission and the future ESA PLAnetary Transits and Oscillations of stars (PLATO; Rauer et al.~\cite{plato}) mission. The two ecliptic poles are the best-sampled fields for TESS and each covers a core area on the sky of approximately 24\degr$\times$24\degr . Because its sampling over time decreases beyond a radius of $\approx$16\degr\ from the pole, we limited our survey to a radius of 16\degr\ from the Northern~Ecliptic~Pole (NEP). The coordinates of the center of the NEP are right-ascension 18h00m00s and declination +66\degr33\arcmin38.55\arcsec\ for equinox 2000.0.

Stellar host characteristics must be known in order to describe and understand the planets discovered (e.g., Udry \& Santos \cite{udr:san}). Several high-resolution surveys have been initiated worldwide, also partly for other purposes. For example, ESA's \emph{Gaia} mission initiated a large spectroscopic survey of FGK-type stars at ESO with low-to-moderate resolution (Gilmore et al. \cite{gil}) but also with HARPS at high resolution (Delgado Mena et al. \cite{delgado}), while NASA's \emph{Kepler}, K2, and TESS missions have spurred follow-up programs at several U.S., European, and Australian observatories (e.g., Brewer et al. \cite{brewer}, Furlan et al. \cite{furlan}, Sharma et al. \cite{sharma}, and Tautvaisiene et al. \cite{taut}). These analyses have focused on global stellar parameters such as effective temperature, gravity, and metallicity, as well as on detailed chemical abundances (e.g., Valenti \& Fischer \cite{val:fis}, Heiter et al. \cite{heit}, Delgado Mena et al. \cite{delgado21}).

Included among the more elusive stellar parameters are convective blue shift (Bauer et al. \cite{bauer}), stellar rotation (Barnes~\cite{syd03}, Lanzafame et al. \cite{lanza}), and stellar activity (Strassmeier et al. \cite{vie-kpno}). In addition, certain chemical abundances and isotopic ratios can be related to the evolutionary status of the stellar system once a particular measurement precision has been achieved (as per Tucci Maia et al. \cite{tucci}, Soto \& Jenkins \cite{sot:jen}, and Adibekyan et al. \cite{adi21}). Prior spectroscopic surveys of exoplanet-host stars have revealed that Jupiter-like giants are preferentially found around metal-rich stars (Fischer \& Valenti \cite{fis:val},  Udry \& Santos \cite{udr:san}), while smaller planets are apparently equally likely in all kinds of stars (Sousa et al. \cite{sousa}; Buchhave et al. \cite{buch12}, \cite{buch14}; Schuler et al. \cite{sch15}). Additionally, studies by Adibekyan et al. (\cite{adi15}) and Santos et al. (\cite{santos17}) have hinted that metal-poor planet hosts tend to have a higher [Mg/Si] ratio than metal-poor stars without planets. They also predicted a higher Fe-core mass fraction for rocky planets. We note that for [Mg/Si] ratios in solar-type stars  more details are given in Su\'arez-Andr\'es et al. (\cite{sua}). Several authors, for example, Mel\'endez et al. (\cite{mel09}, \cite{mel14}) and Adibekyan et al. (\cite{adi16}), have shown that Sun-like stars possibly show an increase in elemental abundance as a function of condensation temperature, but see Mack et al. (\cite{mack}) for a counter example. While the Sun itself appears to be depleted in elements with high condensation temperatures, it is the refractory elements that are likely to be the main constituents of terrestrial planets. We may hypothesize that the abundance of the host star's refractory elements hints, in particular via the C/O and C/Fe ratios (cf. Delgado Mena et al. \cite{delgado21}), toward the existence and age of rocky planets.

Our survey in this paper is based on high-resolution spectra and conducted with two different telescope-spectrograph combinations. Dwarf stars are observed with the VATT and PEPSI from Arizona, while giants stars are observed with STELLA and SES from Tenerife. The PEPSI dwarf-star spectra resemble the spectra of the PEPSI \emph{Gaia} benchmark stars presented earlier by Strassmeier et al. (\cite{gaia}) based on the sample of Blanco-Cuaresma et al. (\cite{blanco}). We refer to these higher-quality data, mostly taken with the Large Binocular Telescope (LBT), for a detailed wavelength-by-wavelength comparison with the present VATT-PEPSI data. The giant-star spectra from STELLA and its SES \'echelle spectrograph resemble the data analyzed, for example, in Strassmeier et al. (\cite{hde}), with additional details given therein. An independent survey done in parallel with spectra with a comparable resolution to SES was presented recently by Tautvaisiene et al. (\cite{taut}) and included 302 stars brighter and cooler than F5 near the NEP. These data were used to extract the absolute stellar parameters and chemical abundances for 277 of them. Additional abundances for neutron-capture elements for 82 of these stars were presented just recently by Tautvaisiene et al. (\cite{taut21}). Ayres (\cite{ayres}) also recently published HST/COS far-ultraviolet spectra for 49 dwarf targets near the ecliptic poles and derived individual ultraviolet emission lines as well as basal fluxes for these targets.

The present paper is organized as follows. Section~\ref{S2} describes the design of the survey, instrumentation, observations, and data reduction. Section~\ref{S3} describes the analysis tools, along with their strengths and limitations. Section~\ref{S4} shows the first data from the pilot observations of 54 of the 1067 targets, presenting initial results for these targets and enabling a first quality assessment of the data products. Section~\ref{S5} offers a description of the data release, including some notes on individual stars of the pilot sample. Section~\ref{S6} provides a summary and an outlook. Detailed numerical results are given for the pilot stars in form of tables and figures in the appendix. All the spectra for all 1067 stars are to be made public and can be retrieved via CDS.

\begin{figure}
\includegraphics[angle=0,width=8cm,clip]{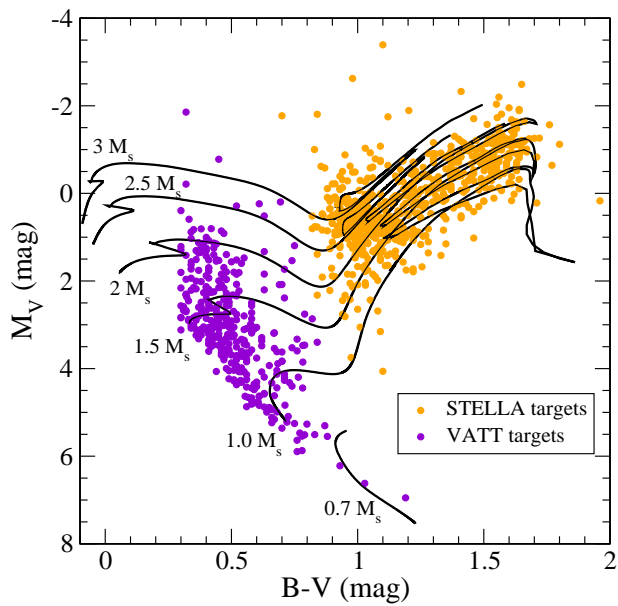}
\caption{Color-magnitude diagram of the NEP stars in this survey. The evolutionary tracks are the Basti tracks from Pietrinferni et al.~\cite{basti} and are intended only for the purposes of orientation. The arbitrary sample cut-off was set at $B-V$ of 0\fm3. The most luminous targets in the respective two samples are $\psi^{02}$\,Dra (F2III, in the VATT sample) and $\beta$\,Dra (G2Ib-IIa, in the STELLA sample). }
 \label{F1}
\end{figure}

\section{Survey design, observations, and data reduction}\label{S2}

\subsection{Sample selection, timeline, and deliverables}

\begin{figure*}
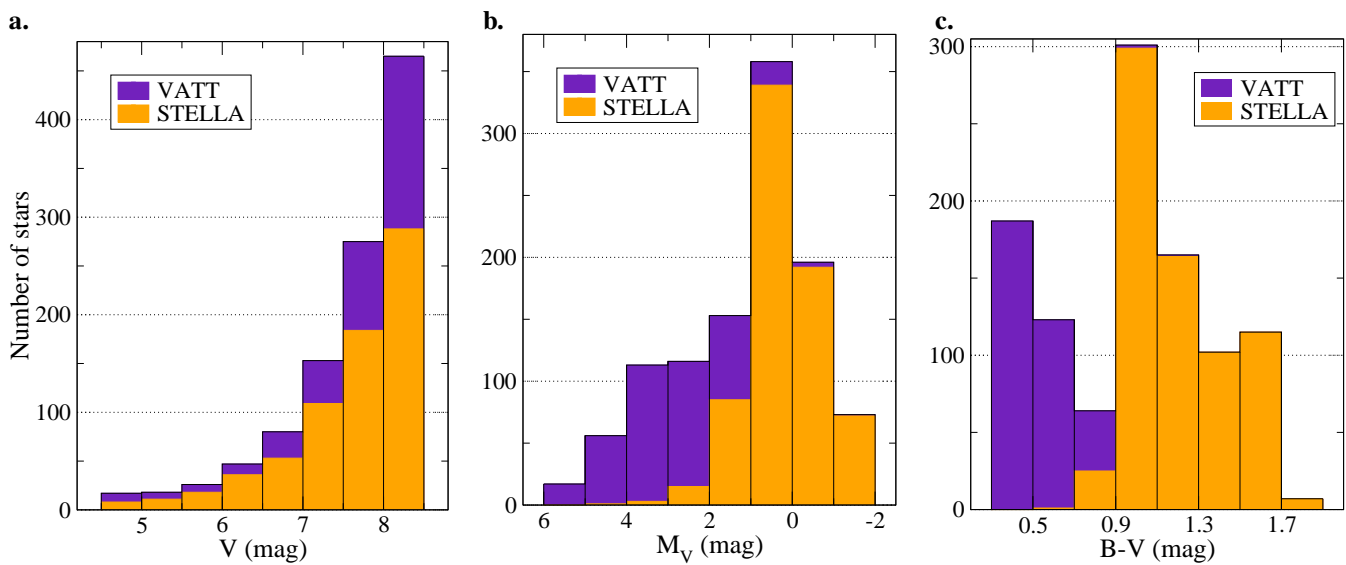

{\bf a.} \hspace{58mm} {\bf b.} \hspace{55mm} {\bf c.} \\
\includegraphics[angle=0,width=58mm,clip]{F2a.eps}\hspace{3mm}
\includegraphics[angle=0,width=55mm,clip]{F2b.eps}\hspace{3mm}
\includegraphics[angle=0,width=55mm,clip]{F2c.eps}
\caption{Stellar parameters of the target sample. The three panels show the number of stars observed as a function of selection parameter. Panel \emph{a} shows the apparent visual brightness, panel \emph{b} shows the absolute visual brightness based on the \emph{Gaia} eDR3 parallax, and  panel \emph{c} shows the $B-V$ color from the Tycho catalog. Total sample of the full survey is 1067 stars cooler than spectral type F0. }
 \label{F2}
\end{figure*}

We adopted a circular field around the NEP with a radius of 16\degr\ from the pole. Its $\approx$770 square degrees contain a total of 1093 stars brighter than 8\fm5 in Johnson $V$. The limit of 8\fm5 was chosen solely due to observational constraints dictated by the telescope diameter, the high spectral resolution and the desired signal-to-noise ratio (S/N) of at least 100. Of these 1093 stars, 1067 are cooler than spectral type F0, that we defined by using a value of $B-V$ of +0\fm3. Of these, 352 stars can be classified as main-sequence stars or close to it based on their \emph{Gaia} eDR-3 parallaxes and 715 as evolved stars on the red giant branch (RGB), in the clump region, or perhaps on the asymptotic giant branch (AGB). Figure~\ref{F1} is a color-magnitude diagram (CMD) of the entire sample and Fig.~\ref{F2} displays histograms of the sample distribution in apparent brightness, $B-V$ color and absolute magnitude. Because the giant stars are less likely to be detectable as planet-hosting systems compared with the dwarf stars, and because the dwarf-star sample already consumes all of our telescope time on the VATT+PEPSI, we deferred the giant-star sample to another telescope-spectrograph combination: STELLA+SES. Both samples require about 50 nights per year for three years around the culmination of the NEP ($\approx$June). Targets were observed independently but approximately contemporaneously around the same epochs in May through July.

The PEPSI data in this paper provide $R$=$\lambda/\Delta\lambda$=200\,000 in three wavelength windows, totaling a coverage of 3480\,\AA,\ while SES provides $R$=55\,000 but full optical wavelength coverage of 4800\,\AA . The PEPSI spectrograph is physically located at the Large Binocular Telescope Observatory (LBTO) on Mt.~Graham and is fed by an underground fiber from the nearby VATT. In order to identify RV variations, for example, due to unknown binaries, we carried out at least two (often up to five) visits of each target to obtain independent radial velocities, but distributing the observing load to both telescopes. For the PEPSI set-up, we decided to maintain the reddest cross disperser (CD) in the blue arm (CD\,III) for both visits and only change from CD\,V to CD\,VI for the second visit for increased wavelength coverage. We note that no such change is needed for STELLA+SES because of its fixed spectral format.

A ten-night pilot survey was first conducted with the VATT and PEPSI during May 18-27, 2017. The same instrumental set-up as for the main survey was employed. A total of 27 target stars were observed: 21 of these in CD\,III and V, and again in CD\,III and VI; the missing 6 targets were re-observed at a later point in time. A similar pilot survey of a randomly selected sample of 27 giant stars was conducted with STELLA and SES during March 2018. Both samples together constitute the pilot targets in this paper. The individual pilot stars are identified in Tables~\ref{table-1} and \ref{table-2} along with $B$ and $V$ magnitudes, a CDS-Simbad based spectral type, and the distance, $d$, from the \emph{Gaia} eDR3 parallax (Gaia collaboration \cite{eDR3}) whenever available, otherwise from \emph{Hipparcos} (van Leeuwen \cite{hip}). There are several targets in common with the survey by Tautvaisiene et al. (\cite{taut}), with 6 dwarfs and 8 giants and by Ayres (\cite{ayres}), with 4 dwarfs, which we will use to make our comparison, presented in the results (Sect.~\ref{S4}).

\begin{table}[!t]
\caption{Pilot sample observed with VATT+PEPSI.}\label{table-1}
\begin{flushleft}
\begin{tabular}{lllll}
\hline\hline
\noalign{\smallskip}
Star  & $B$ & $V$                 & Sp. type & $d$  \\
      & \multicolumn{2}{c}{(mag)} &          & (pc) \\
\noalign{\smallskip}\hline\noalign{\smallskip}
\object{BY Dra}     & 9.23 & 8.04 &K4Ve+K7.5Ve&  16.56$\pm$0.01\\
\object{HD 135119}  & 7.42 & 7.12 &F2         & 114.07$\pm$0.12\\
\object{HD 135143}  & 8.43 & 7.83 &G0         &  48.319$\pm$0.038\\
\object{HD 138916}  & 8.61 & 8.27 &F0         & 109.04$\pm$0.18\\
\object{HD 139797}  & 9.33 & 8.54 &G0         & 143.5$\pm$11 \\
\object{HD 140341}  & 8.43 & 8.03 &F5         & 207.1$\pm$0.8\\
\object{HD 142006}  & 8.26 & 7.76 &F8         &  59.30$\pm$0.61\\
\object{HD 142089}  & 8.41 & 7.92 &F5         &  94.1$\pm$5.4 \\
\object{HD 144061}  & 7.90 & 7.28 &G2V        &  29.602$\pm$0.014\\
\object{HD 145710}  & 8.88 & 8.47 &F0V        & 186.2$\pm$0.25\\
\object{HD 150826}  & 8.71 & 8.24 &F6IV       & 109.38$\pm$0.30\\
\object{HD 155859}  & 7.94 & 7.38 &G0         &  52.163$\pm$0.042\\
\object{HD 160052}  & 8.44 & 8.04 &F2         & 161.7$\pm$1.8\\
\object{HD 160076}  & 8.79 & 8.47 &F5         & 138.80$\pm$0.44\\
\object{HD 160605}  & 9.05 & 8.02 &K2         & 157.03$\pm$0.50\\
\object{HD 161897}  & 8.33 & 7.61 &K0         &  29.980$\pm$0.017\\
\object{HD 162524}  & 8.99 & 8.48 &F8         &  83.18$\pm$0.07\\
\object{HD 165700}  & 8.21 & 7.80 &F8         &  88.11$\pm$0.16\\
\object{HD 175225}  & 6.35 & 5.51 &G9IVa      &  26.534$\pm$0.048\\
\object{HD 176841}  & 8.30 & 7.62 &G5         &  43.315$\pm$0.028\\
\object{HD 180005}  & 8.54 & 8.21 &F2         & 137.58$\pm$0.65\\
\object{HD 180712}  & 8.57 & 7.99 &F8         &  45.075$\pm$0.030\\
\object{HD 192438}  & 7.88 & 7.51 &F2         & 168.61$\pm$0.53\\
\object{HD 199019}  & 8.98 & 8.21 &G5         &  35.224$\pm$0.020\\
\object{VX UMi}     & 6.48 & 6.18 &F2IVnn     &  46.545$\pm$0.045\\
\object{$\psi^{01}$ Dra A}& 5.00 & 4.56 &F5IV-V &  22.75$\pm$0.44\\
\object{$\psi^{01}$ Dra B}& 6.32 & 5.78 &F8V    &  22.702$\pm$0.027\\
\noalign{\smallskip}
\hline
\end{tabular}
\end{flushleft}
\end{table}

\begin{table}[!t]
\caption{Pilot sample observed with STELLA+SES.}\label{table-2}
\begin{flushleft}
\begin{tabular}{lllll}
\hline\hline
\noalign{\smallskip}
Star  & $B$ & $V$                 & Sp. type & $d$  \\
      & \multicolumn{2}{c}{(mag)} &          & (pc) \\
\noalign{\smallskip}\hline\noalign{\smallskip}
\object{11 UMi}     & 6.38 & 5.01 &K4III      & 126.2$\pm$1.4\\
\object{HD 136919}  & 7.71 & 6.68 &K0         & 101.81$\pm$0.16\\
\object{HD 138020}  & 9.15 & 7.76 &K0         & 498.9$\pm$3.9\\
\object{20 UMi}     & 7.64 & 6.34 &K2IV       & 213.20$\pm$0.70\\
\object{HD 138116}  & 9.16 & 8.10 &M0-1       & 224.65$\pm$0.85\\
\object{HD 143641}  & 8.26 & 7.23 &K0         & 122.97$\pm$0.26\\
\object{HD 142961}  & 7.66 & 6.72 &G5         & 234.5$\pm$0.90\\
\object{HD 150275}  & 7.34 & 6.34 &K1III      & 110.90$\pm$0.29\\
\object{HD 151698}  & 9.13 & 8.06 &K0         & 275.80$\pm$0.11\\
\object{HD 155153}  & 7.63 & 6.62 &G5         & 235.9$\pm$1.0\\
\object{HR 6069}    & 7.21 & 6.21 &G8III      & 136.10$\pm$0.40\\
\object{HR 6180}    & 7.64 & 6.30 &K2III      & 143.9$\pm$9.5\\
\object{HD 145310}  & 8.32 & 7.08 &K0         & 210.42$\pm$0.69\\
\object{HD 144903}  & 8.57 & 7.39 &K0         & 337.9$\pm$1.8\\
\object{HD 148374}  & 6.63 & 5.67 &G8III      & 157.8$\pm$5.1 \\
\object{HD 150142}  & 8.52 & 7.12 &K0         & 384.3$\pm$2.6\\
\object{HD 147764}  & 8.59 & 7.54 &G5         & 233.64$\pm$0.83\\
\object{$\upsilon$ Dra} & 5.96 & 4.81 &K0IIICN0.5 & 110.2$\pm$2.8\\
\object{HD 148978}  & 9.14 & 8.09 &K0         & 293.1$\pm$1.3\\
\object{HD 149843}  & 8.75 & 7.71 &K0         & 263.7$\pm$1.0\\
\object{HD 137292}  & 8.78 & 7.62 &K2         & 221.55$\pm$0.88\\
\object{HD 138852}  & 6.72 & 5.74 &K0III-IV   & 102.30$\pm$0.43\\
\object{HD 145742}  & 8.60 & 7.55 &K0         & 238.26$\pm$0.81\\
\object{HR 5844}    & 7.01 & 5.63 &M0III      & 234.9$\pm$6.1\\
\object{HD 138301}  & 8.73 & 7.70 &K0         & 240.87$\pm$0.83\\
\object{$\varepsilon$ Dra A} & 4.73 & 3.84 &G7IIIb& 46.90$\pm$0.28\\
\object{HD 194298}  & 7.25 & 5.69 &K5III      & 272.7$\pm$4.0\\
\noalign{\smallskip}
\hline
\end{tabular}
\end{flushleft}
\end{table}

In addition to the giant star sample, we steadily moved those stars from the VATT sample to STELLA that turned out to be hitherto unknown binaries or that are known binaries, but without an appreciably well-defined orbit. Furthermore, a total of $\approx$200 main-sequence targets from the VATT+PEPSI sample were also observed with STELLA+SES  to identify and define systematic differences, for example, due to spectral resolution. At the time of writing, six stars in the full sample are already known to have planetary candidates (11~UMi, HD\,150010, HD\,150706, HD\,156279, HD\,163607, and $\psi^{01}$\,Dra\,B). It is worth mentioning that the full dwarf star sample from the VATT contains a number of G-type Hertzsprung-gap giants and even a few F-type giants, for example, $\psi^{02}$\,Dra (F2III; Soubiran et al.~\cite{pastel}), due to the simple $B-V$ selection criterium. Similarly, the giant-star sample on STELLA contains a few bright giants, for example, $\beta$\,Dra (G2Ib-IIa; Keenan \& McNeil~\cite{kee:mcn}). Because these are just a few (and interesting) targets, we kept them in the respective observing programs and will make their data available, along with the other survey targets.

The survey deliverables are the same for both subsamples. The only difference is that the lower spectral resolution of STELLA+SES does not allow for a meaningful convective velocity determination (see L\"ohner-B\"ottcher et al. \cite{lars} for a detailed discussion). To summarize, our survey deliverables are as follows.
Radial velocities:\ one radial velocity (RV) per spectrum, good to up to 5--30\,\ms\ for the VATT sample and good to 30--300\,\ms\ for the STELLA sample, depending on the target.
Global stellar parameters:\ effective temperature ($T_{\rm eff}$), logarithmic gravity ($\log g$), metallicity ($[$M/H$]$), micro turbulence ($\xi_t$) and macro turbulence ($\zeta_t$), and projected rotational broadening ($v\sin i$). From this and the \emph{Gaia} eDR3 parallaxes (Gaia collaboration \cite{eDR3}), we infer the luminosity, radius, mass, and age via a comparison with evolutionary tracks and isochrones using the most recent models from PARSEC (Padova and Trieste Stellar Evolutionary Code; Bressan et al. \cite{bress}).
Chemical abundances:\ abundances with respect to the Sun and hydrogen of as many elements as possible, most importantly, for the $\alpha$ elements and CNO. Whenever possible nonlocal thermal equilibrium (NLTE) and/or three-dimensional (3D) corrections were applied. Isotope ratios:\ for lithium, carbon, and helium, if feasible.
Surface granulation:\ from an average line bisector (BIS) we derived its maximum velocity span as a measure for convective granulation. Convective blue shift:\ from a systematic wavelength shift of weak lines with respect to strong lines we obtain a measure for convective blue shift. Finally, magnetic activity:\ absolute line-core fluxes for \Halpha, \Hbeta, the Ca\,{\sc ii} infrared-triplet (IRT) lines, and He\,{\sc i} 5876\,\AA\ (if present).

\subsection{Instrumental set-up for VATT+PEPSI observations}

Very high-resolution spectra with $R$ of 200\,000 are being obtained with PEPSI (Strassmeier et al. \cite{pepsi}) connected to the 1.8m VATT through a 450\,m long fiber (Sablowski et al. \cite{vatt}). Due to the fiber losses in the blue, we focus consequently only on the wavelengths of its red CDs, which provide enough throughput even after 450\,m of glass. PEPSI is a fiber-fed white-pupil \'echelle spectrograph with two arms (blue and red) covering the wavelength range 3830--9120\,\AA\ with six different CDs. The observations in this paper were performed using only three of the six cross-dispersers, covering the wavelength ranges 4800--5440\,\AA\ (CD~III), 6280--7410\,\AA\ (CD~V), and 7410--9120\,\AA\ (CD~VI). The high spectral resolution is made possible with a nine-slice image slicer and a 200-$\mu$m fiber (called 130L mode) with a projected sky aperture of 3\arcsec . Its resolution element is sampled with two pixels and corresponds to 1.5\,\kms\ (30\,m\AA ) at 6000\,\AA . Two 10.3k$\times$10.3k STA1600LN CCDs with 9-$\mu$m pixels record 51 \'echelle orders in the three wavelength settings. The average dispersion is 10~m\AA/pixel. Every target exposure is made with a simultaneous recording of the light from a Fabry-Perot etalon provided through the sky fibers.

\begin{figure*}
\includegraphics[angle=0,width=175mm,clip]{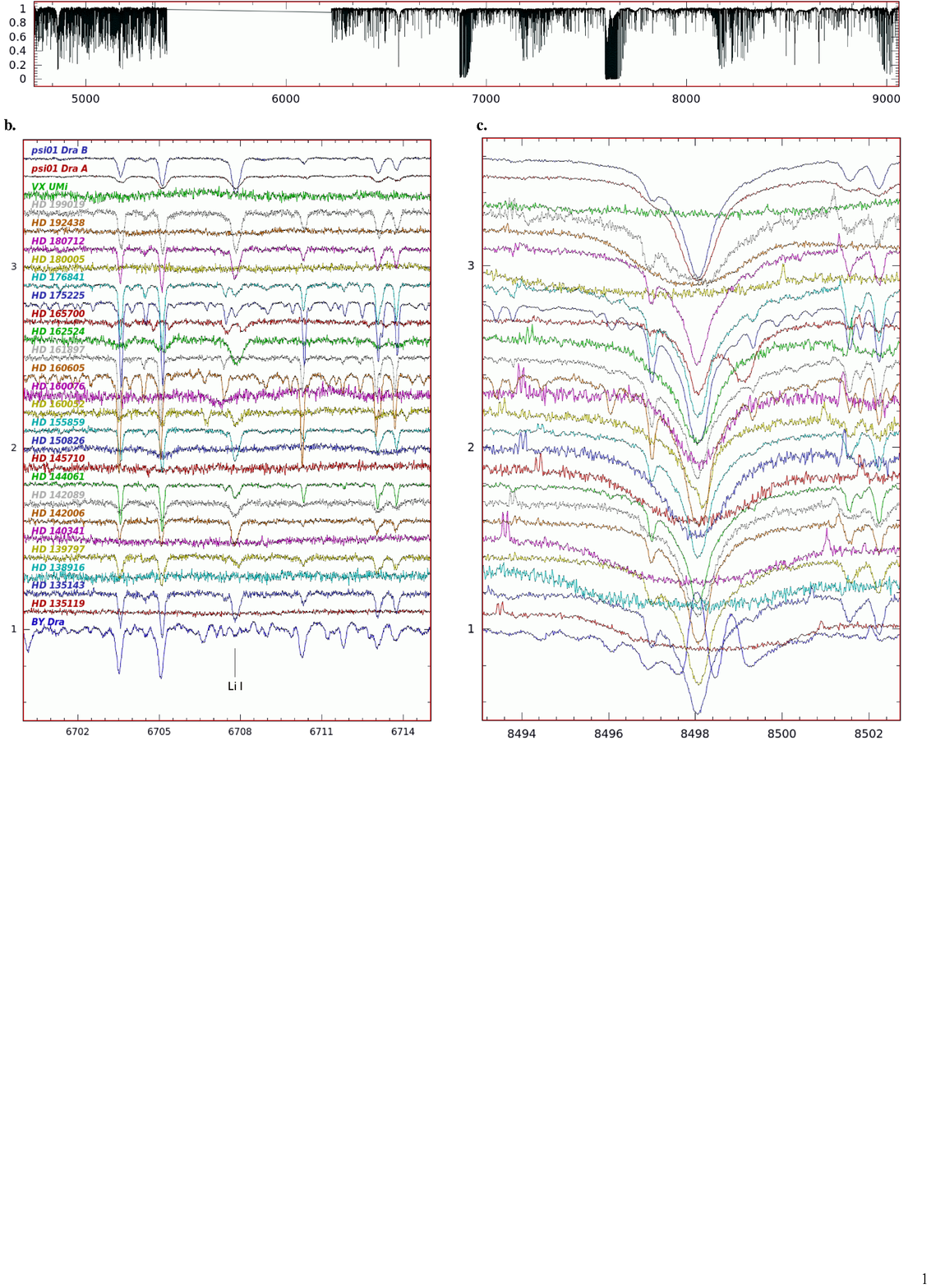}
\caption{Pilot-survey spectra for the dwarf-star sample observed with VATT and PEPSI at $R\approx$200\,000. The respective x-axes are wavelength in Angstroem. Panel \emph{a} shows the available wavelength coverage with cross dispersers III (4800--5440\,\AA), V (6280--7410\,\AA), and VI (7410--9120\,\AA). Panel \emph{b} shows sample stars for a wavelength window around the Li\,{\sc i} 6707.8-\AA\ line. Panel \emph{c} shows the same stars but for a wavelength window around the core of the chromospheric Ca\,{\sc ii} IRT $\lambda$8498-\AA\ line. We note that all stars with lines shifted with respect to the laboratory wavelength appear to be spectroscopic binaries. We also note the large line broadening for some of the targets. }
 \label{F3}
\end{figure*}

Observations with the VATT commenced on May 26, 2018 and ended on June 30, 2022. We adopted integration times of between 10--90\,min. The S/N ranges from 80--870 per pixel in CD\,V and VI to 30--370 in CD\,III depending on target and on weather conditions. A sample spectrum showing the full wavelength range is shown in Fig.~\ref{F3}a. Two selected wavelength windows of interest are plotted for all 27 VATT targets of the pilot survey in Figs.~\ref{F3}b and \ref{F3}c.

The PEPSI data reduction was initially described in Strassmeier et al. (\cite{Sun}) but will be laid out in more detail in a forthcoming paper by Ilyin (2023, in prep.). The VATT PEPSI data in the present paper do not differ principally from the LBT PEPSI data other than S/N. We employed the Spectroscopic Data System for PEPSI (SDS4PEPSI) based upon the 4A software (Ilyin \cite{4A}) developed for the SOFIN spectrograph at the Nordic Optical Telescope. It relies on adaptive selection of parameters by using statistical inference and robust estimators. The standard reduction steps include bias overscan detection and subtraction, scattered light surface extraction from the inter-order space and subtraction, definition of \'echelle orders, optimal extraction of spectral orders, wavelength calibration, and a self-consistent continuum fit to the full 2D image of extracted orders. Wavelengths in this paper are given for air and radial velocities are reduced to the barycentric motion of the solar system.

\subsection{Instrumental set-up for STELLA+SES observations}

Giant stars are observed with AIP's robotic 1.2\,m STELLA-II telescope on Tenerife in the Canary islands (Strassmeier et al.~\cite{stella}, Weber et al. \cite{spie12}). Its fiber-fed echelle spectrograph SES with an e2v 2k$\times$2k CCD detector was used. All spectra cover the full optical wavelength range from 3900--8800\,\AA\ at a resolving power of $R$=55\,000 with a sampling of three pixels per resolution element. The spectral resolution is made possible with a two-slice image slicer and a 50-$\mu$m (octagonal) fiber with a projected sky aperture of 3.7\arcsec . At a wavelength of 6000\,\AA\ this corresponds to an effective resolution of 5.5\,\kms\ or 110\,m\AA . The average dispersion is 45~m\AA/pixel. Further details on the performance of the system can be found in previous applications to giant stars, for example in Weber \& Strassmeier (\cite{capella}) or Strassmeier et al. (\cite{hde}).

\begin{figure*}
\includegraphics[angle=0,width=175mm,clip]{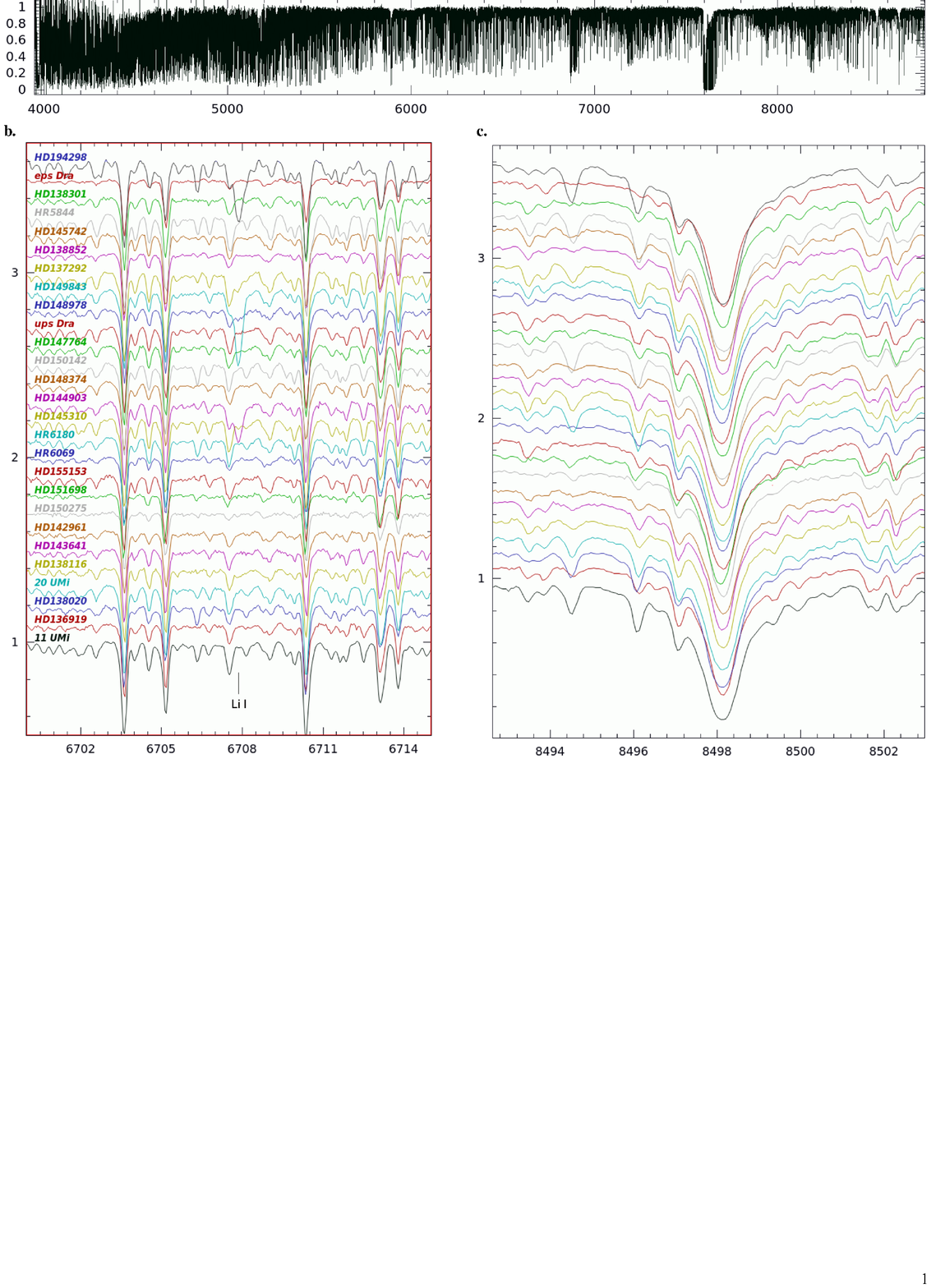}
\caption{Pilot-survey spectra for the giant-star sample observed with STELLA and SES at $R$=55\,000. Otherwise, the details are the same as in Fig.~\ref{F3}.}
 \label{F4}
\end{figure*}

Observations with STELLA commenced in March 2018 and lasted until November 2022. We adopted integration times of 10--15\,min for stars brighter than 6\fm0, 20\,min for 6\fm0--6\fm5, 30\,min for 6\fm5--7\fm0, 40\,min for 7\fm0--7\fm5, 60\,min for 7\fm5--8\fm0, and 90\,min for stars fainter than this. The achieved S/N ranges between 210--450 per pixel depending on target and on weather conditions. As for the VATT+PEPSI observations, a minimum of at least two visits per target were scheduled. A full STELLA+SES example spectrum is shown in Fig.~\ref{F4}a, while the same two selected wavelength windows as for the VATT+PEPSI sample are shown in Fig.~\ref{F4}b and \ref{F4}c.

The SES spectra are automatically reduced using the IRAF-based STELLA data-reduction pipeline SESDR (Weber et al. \cite{spie08}, \cite{spie16}). All images were corrected for bad pixels and cosmic-ray impacts. Bias levels were removed by subtracting the average overscan from each image followed by the subtraction of the mean of the (already overscan subtracted) master bias frame. The target spectra were flattened by dividing by a nightly master flat which has been normalized to unity. The nightly master flat itself is constructed from around 50 individual flats observed during dusk and dawn. Following the removal of scattered light, the 1D spectra were extracted using an optimal-extraction algorithm. The blaze function was then removed from the target spectra, followed by a wavelength calibration using consecutively recorded Th-Ar spectra. Finally, the extracted spectral orders were continuum normalized by dividing with a flux-normalized synthetic spectrum of comparable spectral classification as the target.

\section{Analysis tools and techniques}\label{S3}

\subsection{Spectroscopic Data System (SDS)}

Data preparation for the analysis was done with the same software package that was used for the data reduction. This generic software platform is written in C++ under Linux and is called the Spectroscopic Data System (SDS; Ilyin~\cite{4A}). It is specifically designed to handle the PEPSI data calibration flow and image-specific content and is then referred to as SDS4PEPSI. However, the same SDS tools were used for data obtained with different \'echelle spectrographs, that is, SES in our case (SDS4STELLA), as well as to perform data analysis such as the determination of line bisectors in wavelength space. The numerical toolkit and graphical interface was designed based on the parameters of Ilyin (\cite{4A}), some applications to PEPSI spectra are initially described in Strassmeier et al. (\cite{Sun}).

SDS exploits the many advantages of object-oriented and template-class programming. An extensive Numerical Template Library (NTL) is specifically tailored to the needs of the data analysis tools. Basic approximation tools include smoothing, circular, and polar splines with adaptive regularization parameter to minimize the curvature or the second derivative of the residuals, orthogonal polynomial series, and Chebyshev polynomials in one, two, and three dimensions. Non-linear least-squares approaches include multi-profile fitting, multi-harmonic fitting to periodic signal (also circular splines), and an orbital solution to radial velocities. In addition, NTL implements a special class of numeric iterators that allow for its use as a vector language (like in Java or IDL). An extensive matrix template class incorporates a variety of different types of 2D matrices and a 3D cube matrix necessary for solving least-squares fitting problems with a number of matrix decomposition template classes.

\subsection{Parameters from SES (ParSES)}

Our numerical tool, namely, Parameters from SES (ParSES; Allende Prieto \cite{all04}, Jovanovic et al.~\cite{parses}) is implemented in the STELLA data analysis pipeline as a suite of Fortran~programs. It utilizes a grid of pre-computed template spectra built with Turbospectrum (Plez~\cite{turbo}).  The line-list used by Turbospectrum is the most up-to-date VALD version (VALD-3; Ryabchikova et al. \cite{vald3}). The best-fit search is based on the minimum distance method, with a non-linear simplex optimization described, among others, in Allende Prieto et al.~(\cite{all}). The program is applied to SES as well as to PEPSI spectra. The free parameters per spectrum are $T_{\rm eff}$, $\log g$, $[$M/H$]$, $\xi_t$, and $v\sin i$.

We note that ParSES cannot handle macroturbulence ($\zeta$) implicitly. Instead, it uses a predefined table of macroturbulence velocities as a function of effective temperature and gravity. We implemented the relation given by Gray (\cite{gray}), except for main-sequence targets with $\log g\ge 4$ which we took from Valenti \& Fischer (\cite{val:fis}). Therefore, the ParSES fit to the observed line broadening is the geometric sum of the macroturbulence broadening and the rotational broadening, $\sqrt{\zeta^2 + (v\sin i)^2}$.
Because double-lined binary spectra (SB2) cannot be run automatically through the ParSES routines, we treated  the two SB2 targets in the pilot sample (BY\,Dra and HD\,165700) just as single stars and, thus, we extracted only approximate values for the respective primary components.

\subsection{Singular Value Decomposition of line profiles ($i$SVD)}\label{S_svd}

The multi-line Stokes-profile reconstruction code $i$SVD is part of the $i$MAP software package (Carroll et al.~\cite{carr12}). We applied it only in its non-polarimetric Stokes-I mode. The code uses a preset candidate line list (currently with data from VALD) to extract and reconstruct an average line profile. The method relies on a Singular-Value-Decomposition (SVD) algorithm and a bootstrap-permutation test to determine the dimension (rank) of the signal subspace. The eigenprofiles of the signal subspace are then cross-correlated with the candidate lines to reconstruct a mean Stokes-I profile. The actual number of lines varies from star to star and depends on the richness of the spectrum and the initial S/N. Typically, a few thousand spectral lines are averaged for a K star and many hundreds for a G star. The S/N values for the resulting averaged line profiles typically then amount to several thousands.

\subsection{q2/MOOG and Turbospectrum}

For determining the chemical abundances, we measured the equivalent widths (EWs) with the ARES v2 (Sousa et al. \cite{ares2}\footnote{http://www.astro.up.pt/$^\sim$sousasag/ares}) package. Additionally, the SPECTRE analysis package (Fitzpatrick \& Sneden \cite{fitz}) and the task {\tt splot} of IRAF\footnote{http://iraf.noao.edu/} were used to check the EWs. Then we used the qoyllur-quipu (q2) code (Ramirez et al. \cite{q2}), which is a free python package\footnote{https://github.com/astroChasqui/q2}  that allows for the use the 2019 version of MOOG (Sneden \cite{moog}) in silent mode to derive automatically the chemical abundances of different elements at once and/or the stellar parameters. The code allows the user to employ both the {\tt abfind} and {\tt blends}  drivers of MOOG to take into account for hyper-fine-structure (HFS) details. The abundances of the elements with many spectral lines are always determined from the measurements of their EWs on a line-by-line basis.

For selected wavelength regions, we employed Turbospectrum (Plez \cite{turbo}) and its fitting program TurboMPfit for a direct line synthesis fit. The synthesis approach with TurboMPfit compares an observed spectrum with a library of pre-computed 1D LTE synthetic spectra. It is employed for elements where practically only a single wavelength range is available. For example for the lithium 6708\,\AA\ region, we compute synthetic spectra for different abundances A(Li), isotope ratios $^6$Li/$^7$Li, and iron abundances $[$Fe/H$]$. We then identified the optimal fit by an automatic Levenberg-Marquardt minimization procedure included in TurboMPfit, which was designed specifically for the present purpose: providing the multi-dimensional library of synthetic spectra computed with Turbospectrum as input for MPFIT (Markwardt \cite{mpfit}), together with a list of fitting parameters that are to be optimized to find the minimum $\chi^2$.

\subsection{A Python tOol Fits Isochrones to Stars (APOFIS)}\label{apofis_section}

For the determination of stellar ages and masses, we developed \emph{A Python tOol Fits Isochrones to Stars} (APOFIS) which finds the point closest to an observation in a set of stellar isochrones (or evolutionary tracks) based on a quadratic minimization between observational and isochrone parameters. The selection of input parameters is only limited by the parameter coverage of the isochrone and the sample star. For each pre-calculated point along an isochrone, APOFIS calculates the (error weighted) deviation over the included parameters for the fit. The smallest deviation and its associated point along an isochrone are adopted as the best fit.

The sampling rate of the isochrones has a strong impact on the best match found. Therefore, APOFIS first oversamples each individual isochrone with a linear interpolation between consecutive points along their mass dimension. It can handle multiple isochrones at once. However, there is no interpolation between different isochrones, that is, along the age dimension. Therefore, sets of isochrones with a dense age sampling are preferred. The details of the oversampling are dependent on the chosen isochrones and are described in the application in Sect.~\ref{mass_age}. It can handle evolutionary tracks as well, which reverses the above mentioned problem with the sampling.

To estimate the fitting error that results from the uncertainty in the input stellar parameters, APOFIS creates random samples of new data points based on the error of the input parameter. Thereby, the parameter errors are assumed to be Gaussian and the input error is set as $3\sigma$ error. The fit is carried out for each point which returns a distribution of solutions. This distribution is subjected to a $K$-means clustering algorithm (Lloyd \cite{lloyd}),  with the number of applied clusters determined from a silhouette analysis, to identify patterns in the data that allows the separation into subgroups. This is relevant since stellar isochrones may intersect and the same parameters may correspond to vastly different types of stars, for example, higher mass pre-main sequence and lower mass Hertzsprung-gap stars. The clustering allows to (automatically) separate those solutions based on the information in the isochrones. The highest populated cluster is selected for the solution and the mean and the standard deviation of the distribution are adopted as fitted parameter and error, respectively. Hereby, APOFIS accounts for the general asymmetry of the error by calculating a separate standard deviation for above and below the fitted parameter.

\begin{figure*}
\includegraphics[angle=0,width=\linewidth,clip]{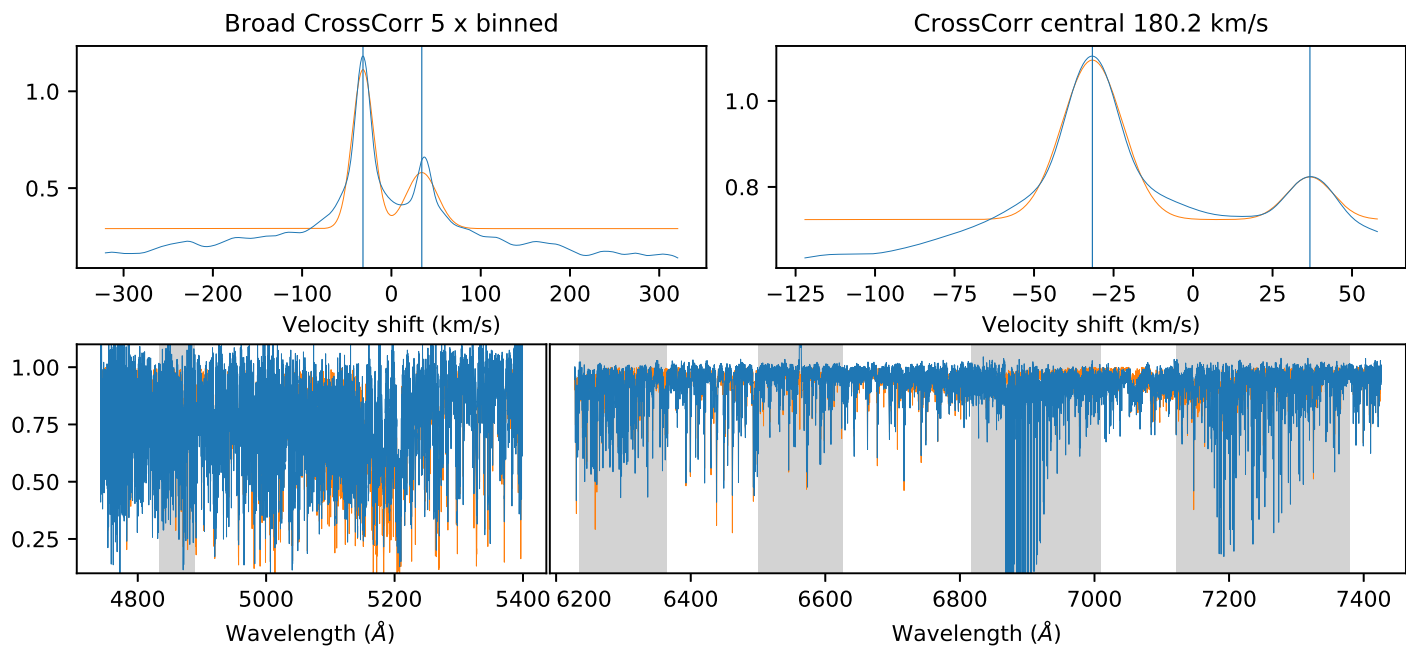}
\caption{Radial velocities from PEPSI spectra. Shown is the CCF (top panels) and the wavelength coverage for one target for one visit (bottom panels). The target is the double-lined spectroscopic binary BY\,Dra. Panels show the full CCF with pixels binned by a factor five (top left, blue line) and the unbinned central 180\,\kms\ of the CCF (top right, blue line), both along with its Gaussian fit (orange line). The bottom panels are overplots of the observed (blue) and the synthetic spectrum (orange) for CD\,III (left) and CD\,V (right). The wavelength zones excluded from the CCF are shaded in grey.}
 \label{F_RV}
\end{figure*}

\section{Analysis of the pilot sample and first results}\label{S4}

\subsection{Radial velocity}

The PEPSI long-term wavelength calibration with a Th-Ar lamp is precise to 10--30\,\ms\ (see Strassmeier et al.~\cite{Sun}). We achieve higher precision with a simultaneous exposure of the fringes from the Fabry-Perot etalon calibrator. This set-up was used for the survey targets in this paper. The resulting long-term radial-velocity rms of PEPSI is then 5-6\,\ms , although the applicable time range is on the order of months due to instrument maintenance. A precision of $\approx$1\,\ms\ can be achieved throughout an individual night. When PEPSI was used for an eight-hour series of solar spectra, a 47~cm\,s$^{-1}$ peak-to-peak amplitude due to the five-minute oscillation was detected at a 5$\sigma$ level (Strassmeier et al. \cite{Sun}). For SES, the long-term precision in terms of radial velocity is much better known and is between 30 and 150\,\ms\ (rms), depending on the target's line broadening and spectral properties (Weber et al.~\cite{spie12}). Its value is well determined over the years and is based on 16 radial-velocity standard stars that are being observed since the inauguration of the telescope in 2006 (Strassmeier et al. \cite{orbits}).

Relative radial velocities in this paper were obtained from a fit to the cross correlation function (CCF) with synthetic spectra from Turbospectrum, itself based on MARCS atmospheres (Gustafsson et al. \cite{gus}) and the VALD-3 (Ryabchikova et al. \cite{vald3}) line list. We first re-sampled both spectra to velocity space and then define the peak of a Gaussian fit to the CCF as the velocity shift with respect to the template. We did not use the entire spectral range is used for the CCF. Regions strongly affected by telluric lines as well as regions including Fraunhofer lines such as \Halpha\ and \Hbeta\ were excluded from the fit. The actual wavelength ranges used are 4880--5387\,\AA\ for the blue-arm CD\,III in VATT+PEPSI, 6392--6540\,\AA , and 6586--6857\,\AA;\ and 6969--7162\,\AA\ for the red-arm CD\,V. The red wavelength end of CD\,VI is severely contaminated by water vapor and O$_2$ bands and we use only the range 8389--8906\,\AA\ for its CCF. We also excluded the regions of the \ion{Ca}{ii} IRT lines because of their sensitivity to chromospheric activity. For STELLA+SES, all of these wavelength ranges are employed for the CCF simultaneously and result in a single velocity per epoch. For VATT+PEPSI, we treat velocities from the blue and the red arm as if they were independent. The first VATT visit usually was with CD\,III and CD\,V, while the second visit with CD\,III and CD\,VI. Therefore, the VATT+PEPSI data  (in principle) provide four radial velocities for two epochs. We then averaged the equal-epoch velocities with equal weight and give their standard deviation. The SB2 spectra were basically treated  as single-star spectra, but just that the template is applied twice to fit the CCF peaks from both components. Figure~\ref{F_RV} shows an example of a RV determination for a SB2 spectrum. It specifies the wavelength ranges for the CCF, the wavelength coverage used, and shows the resulting CCF with a fit for the primary star. For all spectra, we use the {\tt pdferror} setting in FERRE (Allende Prieto et al. \cite{ferre}) to derive the uncertainty estimates. Standard deviations of the parameter values are derived from the covariance matrix, which, in turn, is derived by randomly sampling the likelihood function. We note that once an initial RV had been determined, the ParSES routine is called for a fit for the astrophysical parameters including $v\sin i$ and metallicity. If these differ from the initially adopted solar template spectrum the RV CCF fit is redone with a proper and broadened template spectrum.

Absolute velocities are determined by applying a zero-point shift to the measured relative velocities. The zero point for STELLA+SES was determined previously in Strassmeier et al. (\cite{orbits}) relative to CORAVEL and was linked to the IAU standard star list (see Weber et al.~\cite{spie12}). PEPSI is stabilized and consequently would require a much more accurate zero point correction at the level of a few \ms,\ which is not available. An RV offset of 74\,\ms\ between PEPSI solar spectra and the laser-comb based IAG FTS solar atlas (Reiners et al. \cite{iagfts}) had been noticed previously (Strassmeier et al. \cite{lunar}). However, we refrain from making a correction in the present paper. Due to telescope-time constraints on VATT+PEPSI, we could not re-observe the same RV standard stars that were used for the STELLA+SES zero point. Instead, we used stars that are commonly observed with SES and PEPSI to determine a relative offset between the two telescope-spectrograph combinations. The (preliminary) mean RV difference $\Delta RV$ in \kms\ for SES-minus-PEPSI and its standard deviation is:
\begin{equation}
\Delta RV = RV_{\rm SES} - RV_{\rm PEPSI} = -0.395\pm 0.209 \ {\rm km\,s^{-1}} .
\end{equation}
We note that no RV shifts are applied for the present-paper pilot sample but that the individual RVs for the full survey sample, with a joint zero point will be refined in a forthcoming paper. All stellar RVs in this paper are reduced to the barycentric motion of the solar system using the JPL ephemeris. Table~\ref{table-A1} lists the RVs for the spectra of the pilot stars.

\begin{figure*}[ht!]
{\bf a.} \hspace{50mm} {\bf b.} \hspace{40mm} {\bf c.} \hspace{40mm} {\bf d.}\\
\includegraphics[angle=0,width=\linewidth,clip]{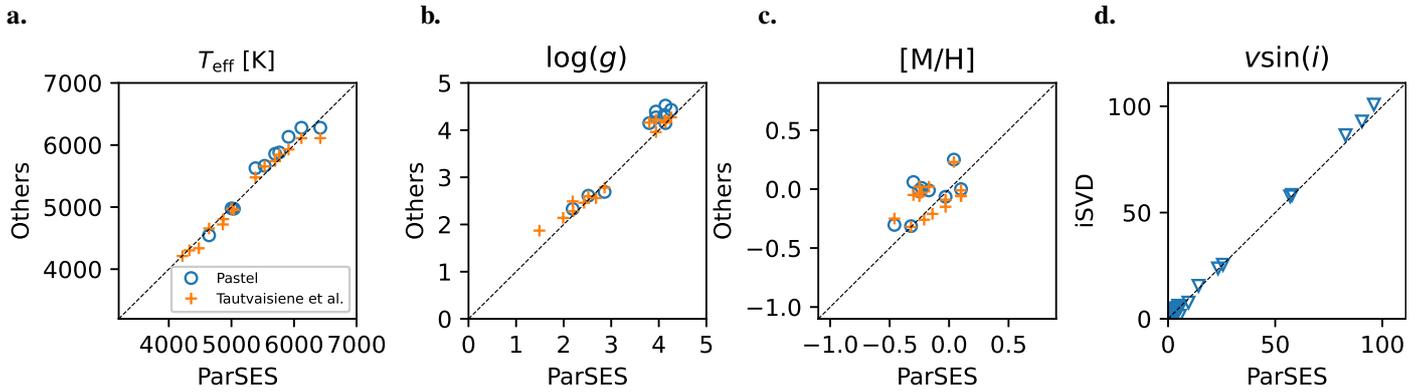}
 \caption{Biases of the ParSES output for the pilot sample with respect to PASTEL (Soubiran et al. \cite{pastel}; circles) and Tautvaisiene et al. (\cite{taut}), shown as pluses, both collectively labeled ``others''. \emph{Panel a}:   Effective temperature in Kelvin. \emph{Panel b}:  Logarithmic gravity in cm\,s$^{-2}$. \emph{Panel c:} Global logarithmic metallicity relative to solar. \emph{Panel d:} Comparison between $v\sin i$ results (inverse triangles) from the two independent methods ParSES and $i$SVD. The dashed line in each panel is just a 45-degree line to guide the eye. }
 \label{F_parses}
\end{figure*}

\begin{figure*}
{\bf a.} \hspace{50mm} {\bf b.} \hspace{40mm} {\bf c.} \hspace{40mm} {\bf d.}\\
\includegraphics[angle=0,width=\linewidth,clip]{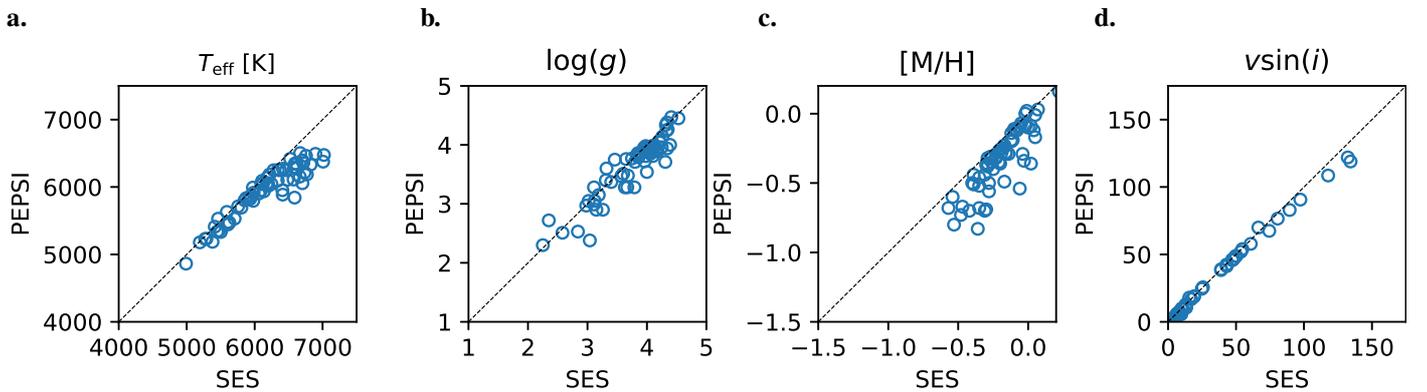}
 \caption{Comparison of the ParSES parameters for a mixed stellar subsample observed with STELLA+SES and VATT+PEPSI. \emph{Panel a.} Effective temperature in Kelvin. \emph{Panel b.} Logarithmic gravity in cm\,s$^{-2}$. \emph{Panel c.} Global metallicity relative to solar. \emph{Panel d.} Rotational line broadening $v\sin i$ in \kms . Dashed line is again just a 45-degree line to guide the eye. }
 \label{F_sespepsi}
\end{figure*}

\subsection{Temperature, gravity, and metallicity}\label{S_tgm}

The main astrophysical parameters were obtained from a comparison with synthetic spectra. The comparison is done with the multi-parameter minimization program ParSES for a masked wavelength range. Like for the RVs, the selection mask identifies and removes wavelength regions from the minimization at the echelle order edges, of strong telluric contamination, and of spectral resonance lines with broad wings like \Halpha , \Hbeta , or the Na\,D lines. As a basis, we adopted the \emph{Gaia}-ESO ``clean'' line list (Jofr\'e et al.~\cite{jof1}, Heiter et al.~\cite{hei21}) with various mask widths around the line cores between $\pm$0.05 to $\pm$0.25\,\AA . The number of free parameters in ParSES is five ($T_{\rm eff}$, $\log g$, $[$M/H$]$, $v\sin i$, $\xi_t$).
Table~\ref{table-A1} lists the results for the pilot stars. We note that in this table a line entry without a JD date (or RV) comes from the combined spectrum built from the individual epochs. These are the values that should be used in case an average is desired.

The synthetic spectra are pre-computed and tabulated with metallicities between --2.5\,dex and +0.5\,dex of that of the Sun in steps of 0.5\,dex, logarithmic gravities between 1.5 and 5 in steps of 0.5, and temperatures between 3500\,K and 8000\,K in steps of 250\,K for a wavelength range of 3800--9200\,\AA . This grid is then used to compare with the selected wavelength regions of our spectra. The model atmospheres are again those from MARCS (Gustafsson et al. \cite{gus}). Plane-parallel atmospheres are used for the dwarfs, and spherical atmospheres are used for the giants. The latter are available for a grid from 2800\,K to 7000\,K, $\log g$ between 0.0 and 3.0, along with a micro turbulence between 2.0 and 5.0\,\kms,\ but otherwise using the same metallicities as the plane-parallel atmospheres. Therefore, we restricted the application of spherical models to $\log g \geq 1.5$ for $7000\leq T_\mathrm{eff} < 6000$\,K, and to $\log g \geq 0.5$ for $6000\leq T_\mathrm{eff} < 2800$\,K.

Internal errors are given from the median rms of the fits to the various selected wavelength regions. These uncertainties are small:\ typically less than 50\,K for $T_{\rm eff}$, 0.1\,dex for $\log g$, and 0.05\,dex for $[$M/H$]$. External errors are estimated from comparisons with benchmark stars (Jofr\'e et al. \cite{jof2}, list V2.1) and the Sun (see Strassmeier et al.~\cite{gaia}). For the STELLA+SES spectra, the external errors for $T_{\rm eff}$ are typically 70~K, for $\log g$ typically 0.2~dex, and for $[$M/H$]$ typically 0.1\,dex. For the VATT+PEPSI spectra in this paper, the external errors are comparable, that is, 70~K for $T_{\rm eff}$, 0.15\,dex for $\log g$, and 0.1\,dex for $[$M/H$]$.

For a methodical systematics check, we compared the stellar parameters from the ParSES synthesis with the parameters from the EW method for the case when only the iron lines from the updated \emph{Gaia}-ESO ``clean'' line list (Heiter et al. \cite{hei21}) are used. We selected only lines with reliable atomic data, in particular, setting ${\rm gf\_flag}=Y$. We also selected lines with ${\rm syn\_flag}=Y/U$, which are not blended either in the Sun nor in Arcturus (Y) or which might be blended in one of the two (U). All the lines were carefully checked on the Sun and we discarded those which delivered anomalous abundances, that is, more than 0.3--0.4~dex off the scale of Asplund et al. (\cite{asp21}). We employed ARES v2 for EW measurements together with q2 to compute the stellar parameters in the standard way. Figure~\ref{F_C1} shows the differences synthesis-minus-EW for the STELLA+SES subsample (giants) and Fig.~\ref{F_C2} for the VATT+PEPSI subsample (mostly dwarfs). While we were able to analyze all the giants with the EW method, this was not the case for the dwarfs. We analyzed only those with $v\sin i<20$\,\kms\ and with S/N$>$50 in order to avoid problems with heavy line blending and continuum setting. We found an excellent agreement between the two methods for the STELLA+SES subsample with systematic deviations always less than or approximately 1$\sigma$ ($T_{\rm eff}$ +18$\pm$63\,K, $\log g$ $-0.05$$\pm$0.13\,dex, $\xi_t$ --0.11$\pm$0.12\,\kms, [Fe/H] +0.07$\pm$0.06\,dex). The VATT+PEPSI subsample shows larger deviations, between 1$\sigma$ and 3$\sigma$ ($T_{\rm eff}$: --176$\pm$99\,K, $\log g$: --0.35$\pm$0.28\,dex, $\xi_t$: +0.12$\pm$0.12\,\kms, and [Fe/H]: --0.28$\pm$0.10\,dex). In order to re-check the impact of the line list, we created a separate line list consisting of 42 Fe\,{\sc i} and 11 Fe\,{\sc ii} lines plus 25 Ti\,{\sc i} and 10 Ti\,{\sc ii} lines (Ti lines also from Heiter et al. \cite{hei21}). Its aim was to increase the number of spectral lines for the analysis given the limited wavelength range of the VATT+PEPSI spectra and to verify that ionization equilibrium had indeed been reached for these two abundant metals. We found that this was the case for practically all stars, both with the EW method and with ParSES. There are a few stars in the VATT+PEPSI sample for which equilibrium was not achieved but these were due to their large $v\sin i$ and/or low S/N which limited the number of ionized lines reliably measured by ARES. Further tests on the adopted line lists and methods were made with the (full) PEPSI spectrum of the Sun and four \emph{Gaia} benchmark stars with parameters similar to our stellar sample (18\,Sco, $\tau$\,Cet, $\varepsilon$\,Eri and $\beta$\,Vir). There, the Fe+Ti line list causes only a slight overestimation of $\log g$ for the Sun and the four benchmark dwarfs with respect to their literature values of always less than 0.1\,dex, compared to the --0.34\,dex difference between synthetic-fit abundances and EW abundances for the VATT+PEPSI subsample with the iron-only line list. However, both analysis (synthesis and EW method) are robust on their own, in particular when the same line list is used. The remaining systematics can be explained with the limited wavelength range of the VATT+PEPSI spectra.

A cross check for instrumentally induced systematics is done with a set of $\approx$70 targets common to VATT+PEPSI and STELLA+SES. At this point, we recall the different spectral resolutions and wavelength coverage of the two subsamples (see Fig.~\ref{F3}a viz. Fig.~\ref{F4}a). Figure~\ref{F_sespepsi} shows the comparison for $T_{\rm eff}$, $\log g$, [M/H], and $v\sin i$, based on the ParSES and Jofr\'e et al.~(\cite{jof1}) line list. The overall rms in $T_{\rm eff}$, $\log g$, [M/H], and $v\sin i$ are 70\,K, 0.14\,dex, 0.05\,dex, and $\approx$1\,\kms\ (for $v\sin i\leq 50$\,\kms), respectively.

The following systematics appeared:\ 1) a general offset in $T_{\rm eff}$ of 70\,K for $T_{\rm eff}\geq 6400$\,K with an increasing deviation of up to $\approx$300\,K for targets with $v\sin i\geq 50$\,\kms; 2) a general offset of 0.10\,dex in $\log g$; 3) an increasing deviation of on average 0.08\,dex for [M/H]$\leq$--0.4\,dex; and 4) an increasing deviation in $v\sin i$ of up to 10\,\kms\ for $v\sin i\geq 100$\,\kms. The differences are always in the sense that SES values are larger, while the PEPSI values are smaller. However, for the vast majority of targets, the differences remain always in the range of 1$\sigma$, or sometimes even less, but with the higher the $v\sin i $ value, the higher the difference. Together with the comparison with the \emph{Gaia} benchmark stars, we conclude that the SES spectra usually overestimate $T_{\rm eff}$, [M/H] and $v\sin i$ with respect to PEPSI, likely due to the lower spectral resolution, while PEPSI underestimates $\log g$, likely because of the lower number of available gravity-sensitive lines due to the smaller wavelength coverage and the overall lower S/N of the dwarf stars with respect to the giant stars. For the further analysis in this paper, we adopted a $\log g$ correction of +0.10\,dex for the ParSES VATT+PEPSI gravities and $-70$\,K for the STELLA+SES temperature, but we will revisit this issue with the full sample in a forthcoming paper.

\begin{figure}[ht!]
\includegraphics[width=5.5cm,clip]{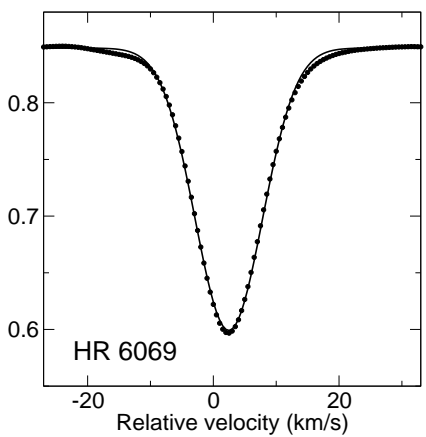}
 \caption{$i$SVD fit for HR\,6069 (G8\,III). The dots comprise the observed mean line profile reconstructed from 1242 individual spectral lines. The line is the best fit with $v\sin i$=3.0$\pm$0.4\,\kms , $\zeta_{\rm RT}$=5.4$\pm$0.5\,\kms , and a bisector span of +56$\pm$31\,\kms. Error bars are from a bootstrap re-sampling procedure and represent a fair external error, but are smaller than the plotting symbol and cannot be seen by eye.}
 \label{F_vsini}
\end{figure}

For an external systematics check, we compared the (uncorrected) ParSES parameters with the corresponding values in Tautvaisiene et al. (\cite{taut}), Ayres (\cite{ayres}), and the PASTEL catalog (Soubiran et al. \cite{pastel}). Figure~\ref{F_parses}a-c summarizes the comparison. Our pilot sample has 14 stars (6 dwarfs, 8 giants) in common with Tautvaisiene et al. (\cite{taut}) and 4 (3 dwarfs, 1 giant) with Ayres (\cite{ayres}); but with one of the four targets in Ayres (\cite{ayres}) being the SB2 binary HD\,165700. Thirty-one of our pilot targets are also in the PASTEL catalog (10 dwarfs, 8 giants, and 13 without a gravity or  metallicity determination). No info is listed for $v\sin i$ in either of the above literature sources. The biases (average difference and rms) for the three atmospheric parameters from our sample are for $T_{\rm eff}$ 44$\pm$59\,K with respect to Tautvaisiene et al. (\cite{taut}), 143$\pm$49\,K to Ayres (\cite{ayres}), and 85$\pm$113\,K to PASTEL (excluding stars with $v\sin i\geq 25$\,\kms), respectively. The biases for $\log g$ are 0.18$\pm$0.18\,dex with respect to Tautvaisiene et al. (\cite{taut}) and 0.21$\pm$0.16\,dex to PASTEL. Accordingly, the biases for [M/H] are 0.07$\pm$0.10\,dex with respect to Tautvaisiene et al. (\cite{taut}), and 0.13$\pm$0.14\,dex to PASTEL. Ayres (\cite{ayres}) did not derive $\log g$ and [M/H] on their own, but basically adopted PASTEL values. These biases are comparable to our estimated instrumental systematics, which are themselves comparable to our expected external errors per measurement.

Table~\ref{T5} at the end of the paper is a summary of all the expected average uncertainties from this survey. Naturally, they can deviate significantly from star to star. We note that external errors in the form of random errors will be revisited again once the full data set from the survey has been analyzed.

\subsection{Rotation and micro- and macroturbulence}\label{Srot}

Our final value for the projected rotational line broadening, $v\sin i$, was measured from the SVD-averaged line profiles in conjunction with atmospheric turbulence broadening (Table~\ref{table-A2}). The SVD-averaging has the advantage of returning a relatively blend-free spectral line in velocity space with  superior S/N. Spectral lines are selected on the basis of residual intensity $>$0.2 and Land\'e factor $>$1.2. A total of 1242 lines were then available from the fixed-format STELLA+SES spectra and 193 from the VATT+PEPSI spectra. For the very rapid rotators, these selection criteria were not applied and then resulted in 843 lines for the PEPSI spectra. Micro turbulence, $\xi_t$, is a free parameter for the spectrum synthesis and is being initially solved for implicitly during the ParSES fit for $T_{\rm eff}$, $\log g$, and $[$M/H$]$ and then kept fixed for the final fit. It is also used as a fixed parameter for the SVD-averaging. At this point, we note that the microturbulence grid of the MARCS plane-parallel models is available just for values of 1\,\kms\ and 2\,\kms , while for the spherical model grid, it is available for 2\,\kms\ and 5\,\kms .

Macro turbulence, $\zeta_{\rm RT}$, is assumed to be of a radial-tangential nature, with equally strong spatial components (Gray~\cite{gray}). It contributes to the line broadening differently than the rotational Doppler broadening and can be de-convolved from the line profiles. In this respect, we followed the analysis of Valenti \& Fischer (\cite{val:fis}) and similar works in the literature. We emphasize that for slowly rotating inactive (cool) stars the rotational broadening and the macro-turbulence broadening are of comparable strength. Both values, $v\sin i$ and $\zeta_{\rm RT}$, are listed along with their bootstrap errors in Table~\ref{table-A2}. Figure~\ref{F_vsini} is an example of the application to a giant star from the STELLA+SES sample.

We emphasize that the SVD $v\sin i$ values supersede the initial $v\sin i$ values from ParSES in Table~\ref{table-A1} because they are more accurate. In general, the two values agree to within an expected range depending on its overall size and the impact of macro turbulence. Figure~\ref{F_parses}d shows a comparison of the $v\sin i$ values from Table~\ref{table-A1} (ParSES) and \ref{table-A2} ($i$SVD).  We note that their respective errors are not directly comparable. The ParSES $v\sin i$ errors are internal errors from the fit with synthetic spectra (for several wavelength chunks containing different number of lines). It is part of a larger error matrix driven by the error of and cross talk with the other free parameters. The SVD-averaged $v\sin i$ error is more what we define an external error and was based on a bootstrap procedure with 1000 re-samplings per spectral line. It represents the best estimate of the error for our $v\sin i$ measurements. Its average for the pilot stars from the SES spectra is 0.48$\pm$0.05(rms)\,\kms  and from the PEPSI spectra, it is 1.1$\pm$0.2\,\kms . The latter value is larger because the PEPSI pilot sample contains 36\%\ rapidly rotating F-type stars. We note that one target showed a dramatic difference between the ParSES and the SVD $v\sin i$ value (HD\,194298: ParSES 1.04\,\kms\ with an internal error of $\pm$0.09; SVD 3.96$\pm$0.50\,\kms ). It turned out that this target is also the target with the largest bisector span in our sample ($-180\pm28$\,\ms; see Sect.~\ref{sBIS})  and also turned out to be a supergiant with $\log g$=0.65.

\subsection{Granulation and convective blue shift}\label{sBIS}

Observations of line bisectors have shown the presence of asymmetries which result from the up- and down-flows of convective granulation (Dravins et al. \cite{drav}; Gray \cite{gray}). By studying the asymmetries in different types of stars, for lines of different excitation potential and ionization stages (i.e., formed at different levels in the atmosphere), it is possible to quantify granulation in stars. Additionally, if RV variations are due to a planet, we would see no change in shape or orientation of the bisector in consecutive spectra. Another challenging task for our understanding and correction of planet-related Doppler shifts is the effect of a suppressed convective blue-shift in active surface regions (e.g., Bauer et al.~\cite{bauer}). It adds to the RV ``noise'' of a star and, together with magnetic activity in form of starspots or limits -- it may even prevents detections of Earth-sized planets. Only recently have  there been first attempts to measure convective blue-shifts in other stars more systematically (Meunier et al. \cite{meu1}, \cite{meu2}).

Our survey deliverable is a bisector velocity span. This parameter is just a snapshot in time because we did not resolve rotational or pulsation modulation. Bisector velocity spans in this survey are derived from averaged line profiles only. Profile averaging is done with our $i$SVD method described in Sect.~\ref{S_svd}. We actually use the same average line profiles as for the $v\sin i$ and macroturbulence measurement in Sect.~\ref{Srot}. The velocity spans are determined following Queloz et al. (\cite{que:hen}). Two points of the bisector at the top 1/3 and the bottom 1/3 of the line profile were selected and their respective radial velocity was measured ($RV_{\rm top}$ and $RV_{\rm bottom}$). The difference between the two is then the bisector velocity span $BIS$, given in \ms \ is:
\begin{equation}
BIS = RV_{\rm top} - RV_{\rm bottom}  \ . \label{eq-bis}
\end{equation}
The bisector spans for the pilot targets are listed in Table~\ref{table-A2}. Its error is the standard deviation from a bootstrap procedure with 1000 re-samplings per spectral line.

\begin{figure*}[ht!]
 \includegraphics[width=\linewidth]{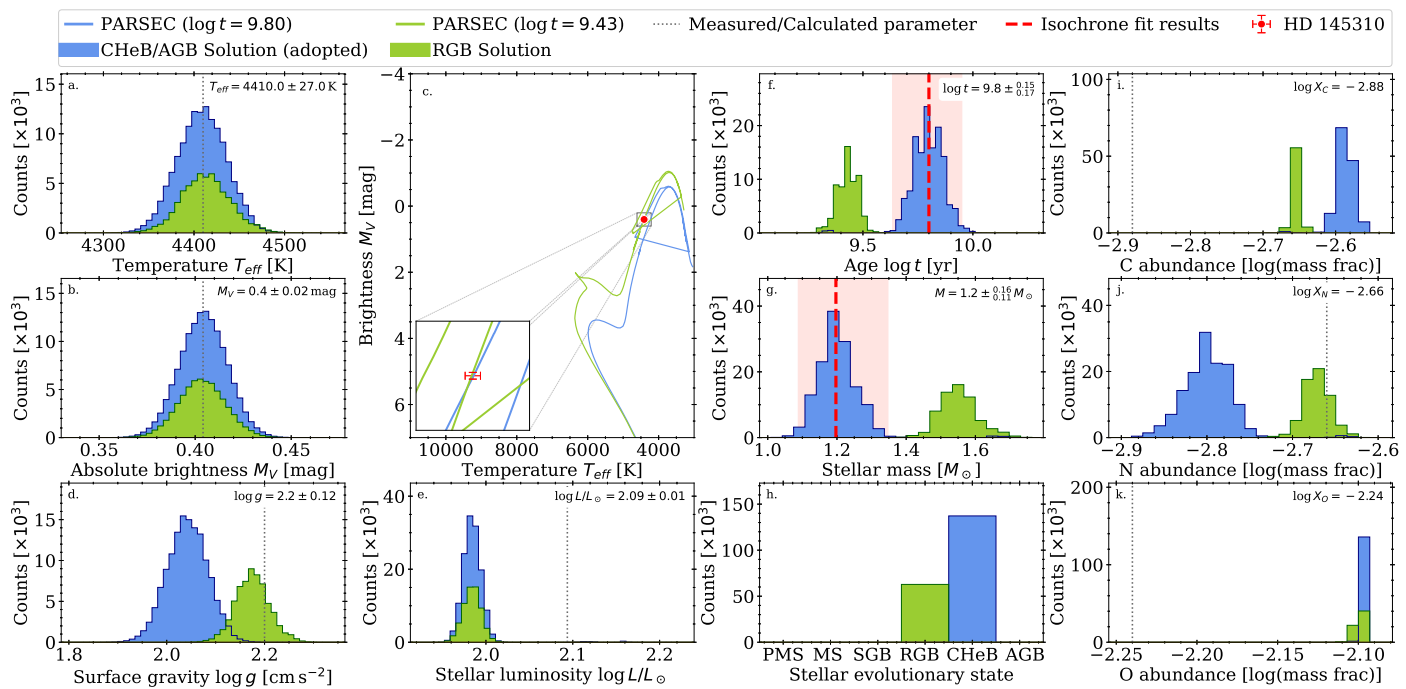}
 \caption{Example result from APOFIS for HD\,145310. APOFIS finds two possible solutions, indicated in green and blue. One is a red giant branch (RGB) star and the other is a core-He burning
(CHeB) star. The adopted solution (CHeB) is in blue.
Panels $a$ and $b$ show the probability distribution of the 200\,000 sample points used as input parameter for $T_{\rm eff}$ and $M_V$, respectively. The H-R diagram in panel $c$ shows the isochrone fits for the two solutions. Their colors match the corresponding distributions in the other panels. Panels $d-k$ show the resulting distributions from the fit for selected isochrone parameters ($d$ gravity; $e$ luminosity; $f$ age; $g$ mass; $h$ evolutionary stage according to Table~\ref{isofit_results_flag_notation}; $i$ C abundance; $j$ N abundance; and $k$ O abundance). Dashed gray lines in the figures indicate the measured values. Dashed red line in panels $f$ and $g$ show the final mass and age from the isochrone fit, with the shaded areas indicating its error range.}
 \label{isofit_example}
\end{figure*}

\subsection{Mass and age}\label{mass_age}

Stellar masses and ages are obtained from a comparison of the target position in the Hertzsprung-Russell (H-R) diagram with theoretical isochrones using APOFIS. The fit is carried out in two dimensions: temperature, $T_{\rm eff}$, and absolute brightness, $M_V$. The value for $T_{\rm eff}$ is taken from ParSES; $M_V$ is calculated from distance and extinction-corrected $V$-band photometry. We used $V$-band photometry from CDS/Simbad, supplementing missing values with measurements from the \emph{Guide Star Catalog 2.4.2} (Lasker et al. \cite{GSC}) and the \emph{U.S. Naval Observatory Robotic Astrometric Telescope Catalog} (Zacharias et al. \cite{URAT}). We combined the results from ParSES with astrometric and photometric measurements to obtain additional stellar parameters. Stellar parallaxes were taken from \emph{Gaia} eDR3 whenever available. Two of the stars of our pilot sample are not covered by eDR3 and for those, we used \emph{Hipparcos} parallaxes instead. We assumed interstellar extinction according to the dust maps of Green et al. (\cite{green}); however, we emphasize that most of the stars in the present sample are nearer than 100\,pc and, thus, the extinction is almost negligible.

We used the stellar evolution models from PARSEC\footnote{{stev.oapd.inaf.it/cgi-bin/cmd}} (Bressan et al. \cite{bress}; Tang et al. \cite{tang}) and the newest generation of isochrones based on PARSEC (Marigo et al. \cite{mari}; Pastorelli et al. \cite{past}). The PARSEC evolutionary models cover a large range of masses and metallicities (Chen et al. \cite{chen14}), with special attention given to late-type dwarfs (Chen et al. \cite{chen15}). For each sample star, we used a set of 353 isochrones covering $6.6 \leq \log t \leq 10.12$ with $\Delta \log t=0.01$\,dex and the corresponding metallicity from ParSES. We first oversampled the isochrones by a factor of 50 and used 200\,000 points for the error estimate. The PARSEC isochrones provide a rough estimate on the stellar evolutionary state noted as the "label"\ for each point in an isochrone. We find that this categorization provides a good constraint for the $K$-means clustering to disentangle solutions. The clustering is based on the parameters age, mass, and the  above-mentioned label.

The result of the fitting process is displayed in Fig.~\ref{isofit_example} for the example target HD\,145310. We show the two different possible solutions, along with their probability distributions in green and blue color. The most highly populated solution is selected as the best fit, while (for the time being) a pre-MS solution is omitted if another solution is possible. We emphasize that the errors quoted for mass and age are the fit errors and thus are internal errors. As panels $j,k,l$ in Fig.~\ref{isofit_example} illustrate, better constrained CNO abundances will eventually allow a more physically motivated selection of the solution in the final data product. Just for comparison, we have included a calculation of the stellar luminosity $L_\star$ based on $M_V$ and a temperature-dependent bolometric correction (Flower \cite{flow}).

We recall that an isochrone fit fails if the star is an SB2 spectroscopic binary with light contributions from more than one component. It requires a prior disentangling of the stellar components, which is currently not included in this pilot study. Therefore, we omitted stars with combined spectra from the APOFIS fit. The SB1 binaries are treated as effectively single stars and for those, the results are to be taken with proper caution. Information on stellar multiplicity is either obtained from the literature or from our multi-epoch RVs, for example, stars are marked as binaries when they are listed as such in the Tycho-2 catalog (H\o g et al. \cite{tycho}). In the pilot sample, there was only one such star ($\psi^{01}$~Dra\,A) because its spectrum was found recently to be contaminated by a close companion of mass $\approx$0.7\,M$_\odot$ (dubbed the C component; Gullikson et al. \cite{gulli}). However, at the epoch of our observations, we do not see the C component in the RV cross correlation, although it was likely included in all of the mentioned photometry. We further omitted stars for which there is a large inconsistency between the ParSES results and photometry or the isochrones. We also omitted stars where very fast rotation ($v\sin i\geq 50$\,\kms ) causes large parameter uncertainties in the ParSES results.
The stellar ages and masses from APOFIS are listed in Table~\ref{table-A3}. We also list the stars without an APOFIS fit with a corresponding note, indicating the reason for it. The notations are explained in Table~\ref{isofit_results_flag_notation}.

\begin{table}[ht!]
\caption{Notations for the flags in Fig.~\ref{isofit_example} and Table~\ref{table-A3}.} \label{isofit_results_flag_notation}
\begin{flushleft}
\begin{tabular}{ll}
\hline\hline
\noalign{\smallskip}
Flag  & Description \\
\noalign{\smallskip}\hline\noalign{\smallskip}
SB    & Spectroscopic binary star, \\
      & (1,2,3 single-, double-, or triple-lined), \\
par?  & inconsistent parameter coverage, \\
v50   & star with $v\sin i>50$\,km\,s$^{-1}$, \\
\noalign{\smallskip}\hline\noalign{\smallskip}
PMS   & pre main sequence star,\\
MS    & main sequence star, \\
SGB   & subgiant branch or Hertzsprung gap star,\\
RGB   & red giant branch star, \\
CHeB  & core He-burning (for low mass stars), \\
AGB   & asymptotic giant branch star. \\
\noalign{\smallskip}
\hline
\end{tabular}
\end{flushleft}
\end{table}

We compared our derived ages with the corresponding targets in Tautvaisiene et al. (\cite{taut}). From the 14 stars in common, 7 have an age estimate provided here. We basically find that our results agree within the errors given. However, we arrive at slightly older ages (0.2--0.25~dex) for the dwarf stars. This spread may actually indicate the true (external) error for the ages in Table~\ref{table-A3}. Both surveys utilize the PARSEC stellar models and the deviation of the dwarf-star ages likely originates from different stellar parameters. The culprits are primarily the stellar metallicities. Tautvaisiene et al.'s (\cite{taut}) metallicities are generally higher than ours for the dwarfs in question, which leads to older ages for MS/PMS stars and younger ages for SGB stars. Age uncertainties are of similar magnitude in both studies though. Tautvaisiene et al. (\cite{taut}) do not provide an estimate for mass or luminosity. Thus, we cannot investigate the impact of the different photometry chosen here and there. We note again that the final errors in our survey will be revisited once the full data set is analyzed.

\subsection{Chemical abundances of 27 different elements}

We determine the abundances of the following elements: Li, C, N, O, \ion{Na}{i}, \ion{Mg}{i}, \ion{Al}{i}, \ion{Si}{i}, \ion{S}{i}, \ion{K}{i}, \ion{Ca}{i}, \ion{Sc}{i} and \ion{Sc}{ii}, \ion{Ti}{i} and \ion{Ti}{ii}, \ion{V}{i}, \ion{Cr}{i}, \ion{Mn}{i}, \ion{Fe}{i} and \ion{Fe}{ii},  \ion{Co}{i},  \ion{Ni}{i},  \ion{Cu}{i},  \ion{Zn}{i},  \ion{Y}{ii}, \ion{Zr}{i} and \ion{Zr}{ii}, \ion{Ba}{ii}, \ion{La}{ii}, \ion{Ce}{ii}, and \ion{Eu}{ii}. The abundances of Li, C, N, O were derived through spectral synthesis. Abundances of the remaining elements are obtained with the EW of carefully chosen spectral lines. Additionally, the carbon $^{12}$C/$^{13}$C and the lithium $^{6}$Li/$^{7}$Li isotopic ratios are measured. These are all elements that are important in many astrophysical fields. In particular, they are crucial for understanding the evolutionary status of stars (e.g., Chanam\'e et al. \cite{2005ApJ...631..540C}), constraining stellar nucleosynthesis theories, and for studying the link between properties and frequency of exoplanetary systems and host stars (e.g., Adibekyan et al. \cite{adi12}, Santos et al. \cite{2013santos}, Maldonado et al. \cite{2019maldonado}, Adibekyan et al. \cite{adi21}, Biazzo et al. \cite{2022biazzo} and references therein).

\begin{table}[t!]
\caption{Solar abundances from STELLA+SES and VATT+PEPSI Sun-as-a-star spectra.} \label{Tsun}
\begin{flushleft}
\begin{tabular}{llll}
\hline\hline
\noalign{\smallskip}
Element &       STELLA+SES      &       VATT+PEPSI      &       Asplund      \\
        &               &               &   et al. (2021) \\
\noalign{\smallskip}\hline\noalign{\smallskip}
C (atomic)      &       8.42$\pm$0.01   &       8.45$\pm$0.02   &       8.46$\pm$0.04 \\
CH      &       8.31$\pm$0.05   & \dots & \dots \\
N (CN)  &       7.93$\pm$0.05   &       \dots & 7.83$\pm$0.07   \\
O (NIR-NLTE)    &       8.63$\pm$0.01   &       8.65$\pm$0.01   &       8.69$\pm$0.04 \\
O (6300\AA )    &       8.64$\pm$0.10   &       \dots & \dots \\
Na      &       6.26$\pm$0.05   &       6.13$\pm$0.06   &       6.22$\pm$0.03 \\
Mg      &       7.47$\pm$0.05   &       7.58$\pm$0.05   &       7.55$\pm$0.03 \\
Al      &       6.41$\pm$0.01   &       6.39$\pm$0.03   &       6.43$\pm$0.03 \\
Si      &       7.47$\pm$0.02   &       7.55$\pm$0.02   &       7.51$\pm$0.03 \\
S       &       7.12$\pm$0.01   &       7.19$\pm$0.01   &       7.12$\pm$0.03 \\
K       &       5.16$\pm$0.02   &       5.11$\pm$0.06   &       5.07$\pm$0.03 \\
Ca      &       6.27$\pm$0.03   &       6.29$\pm$0.03   &       6.30$\pm$0.03 \\
Sc$^a$  &       3.11$\pm$0.07   &       3.08$\pm$0.06   &       3.14$\pm$0.04 \\
Ti      &       4.88$\pm$0.04   &       4.94$\pm$0.01   &       4.97$\pm$0.05 \\
V$^a$   &       3.90$\pm$0.02   &       3.94$\pm$0.04   &       3.90$\pm$0.08 \\
Cr      &       5.50$\pm$0.02   &       5.64$\pm$0.03   &       5.62$\pm$0.04 \\
Mn$^a$  &       5.29$\pm$0.03   &       5.56$\pm$0.08   &       5.42$\pm$0.06 \\
Fe      &       7.40$\pm$0.02   &       7.46$\pm$0.01   &       7.46$\pm$0.04 \\
Co$^a$  &       4.85$\pm$0.02   &       4.95$\pm$0.04   &       4.94$\pm$0.05 \\
Ni      &       6.21$\pm$0.03   &       6.22$\pm$0.03   &       6.20$\pm$0.04 \\
Cu$^a$  &       4.04$\pm$0.10   &       4.14$\pm$0.05   &       4.18$\pm$0.05 \\
Zn      &       4.49$\pm$0.04   &       4.56$\pm$0.06   &       4.56$\pm$0.05 \\
Y       &       2.14$\pm$0.06   &       2.29$\pm$0.08   &       2.21$\pm$0.05 \\
Zr      &       2.66$\pm$0.01$^b$ &     2.62$\pm$0.03   &       2.59$\pm$0.04 \\
Ba      &       2.15$\pm$0.05   &       2.08$\pm$0.01   &       2.27$\pm$0.05 \\
La$^a$  &       1.45$\pm$0.01   &       1.22$\pm$0.09   &       1.11$\pm$0.04 \\
Ce      &       1.65$\pm$0.04   &       1.57$\pm$0.07   &       1.58$\pm$0.04 \\
Eu$^a$  &       \dots           &       0.62$\pm$0.01   &       0.52$\pm$0.04 \\
\noalign{\smallskip}
\hline
\end{tabular}
\tablefoot{Mean values and standard deviations between neutral and ionized atoms are given for Sc, Ti, Fe, and Zr. $^a$HFS included in the analysis. $^b$Abundance is from \ion{Zr}{ii} only.}
\end{flushleft}
\end{table}

\subsubsection{C, N, and O abundances}

The abundances of C, N, and O are of highest priority both to constrain stellar evolutionary status and for planetary formation models via the C/O ratios, see Brewer et al. (\cite{2017brewer}) and Delgado-Mena et al. (\cite{delgado21}). Therefore, accurately determining  their abundances is of primary importance. We used  spectral synthesis with different indicators of C, N, and O abundances depending on the evolutionary status of the star. The general methodology is that we first determine the C abundance. Once it has been fixed, we derived the N abundance. Finally, fixing C and N, we derived the O abundance.

Regarding C, it has been demonstrated that high-excitation atomic lines of C (and other elements) show increasing abundances for decreasing $T_{\mathrm{eff}}$ both in giants and dwarf stars (Schuler et al. \cite{sch15}, Baratella et al. \cite{bara}, Maldonado et al. \cite{mal}, Delgado-Mena et al. \cite{delgado21}, and Biazzo et al. \cite{2022biazzo}). This is also the case for our pilot sample stars, which show this peculiar pattern since they cover a large range of different temperatures. In particular, the analysis of the giant sample produced a $\approx$1\,dex increase from 5100\,K to 3900\,K with an rms of 0.4\,dex. In Baratella et al. (\cite{bara}) it has been shown also that the use of CH bands is a solid approach to derive reliable estimates of C abundances. For these reasons, we used the $^{12}$CH and $^{13}$CH molecular features around 4320\,\AA, adopting molecular parameters from Masseron et al. (\cite{mass}) for the giant stars. Because the PEPSI data do not cover this wavelength range, we derived C from the C$_2$ Swan line system at 5135\,\AA\ (molecular data from Ram et al. \cite{ram} for the dwarfs).

Regarding N, there are no useful lines in the optical and the high-excitation lines at $\lambda$7468.3, 8216.3, and 8683.4\,\AA\ (all of them within CD\,VI in the PEPSI set-up and in the SES spectra) are extremely weak in dwarf stars and they are undetectable for stars with $T_{\mathrm{eff}}<5200$\,K. We therefore used the CN molecular bands at 4215\,\AA\ for the giants and at 7980-8005\,\AA\ for the dwarfs. The line lists for the $^{12}$C$^{14}$N, $^{13}$C$^{14}$N and $^{12}$C$^{15}$N isotopes are taken from Sneden et al. (\cite{2014sneden}).

Finally, O abundances are derived from the [\ion{O}{i}] forbidden line at 6300\,\AA, which is not affected by NLTE deviations (Caffau et al. \cite{2008caffau}). We also investigated the NIR triplet lines at 7774\,\AA, which is covered by both instruments. The code q2 allows us to obtain NLTE-corrected abundances for the triplet, using the grid of NLTE corrections from Ramirez et al. (\cite{ram:all}). However, our results for the giant stars of the pilot survey showed a positive trend with decreasing temperatures with an increase of 0.4\,dex from 5000 to 4400\,K. Most likely, this is related to the NLTE effects that strongly affect this triplet. On the contrary, the dwarf stars sample showed a trend that is less significant, but given the large range of temperatures, we discarded the NIR O triplet as a diagnostic for the time being -- however, we will revisit the issue in a forthcoming paper. Nevertheless, a separate table in the appendix (Table~\ref{table-A6}) lists the NLTE O-triplet abundances for comparison. We show [X/H] versus effective temperature for our pilot stars in the appendix in Fig.~\ref{F_C3}. For both giants and dwarfs, we then synthesized the forbidden line at 6300\,\AA, taking into account the known blend with the Ni line (Allende Prieto et al. \cite{all:lam}). The line lists and atomic data come from Caffau et al. (\cite{2008caffau}) and Johansson et al. (\cite{2003johansson}).

\begin{figure*}
{\bf a.} \\
\includegraphics[angle=0,width=14cm]{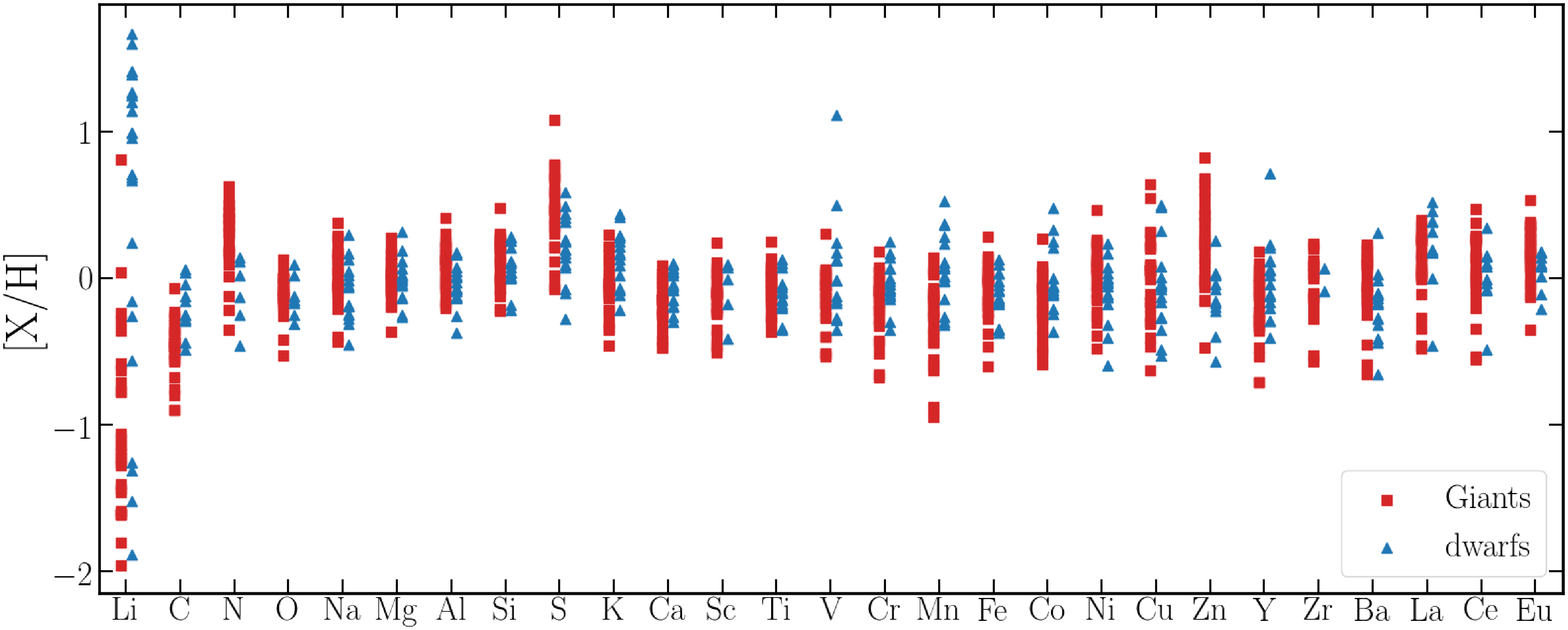}

{\bf b.} \\
\includegraphics[angle=0,width=14cm]{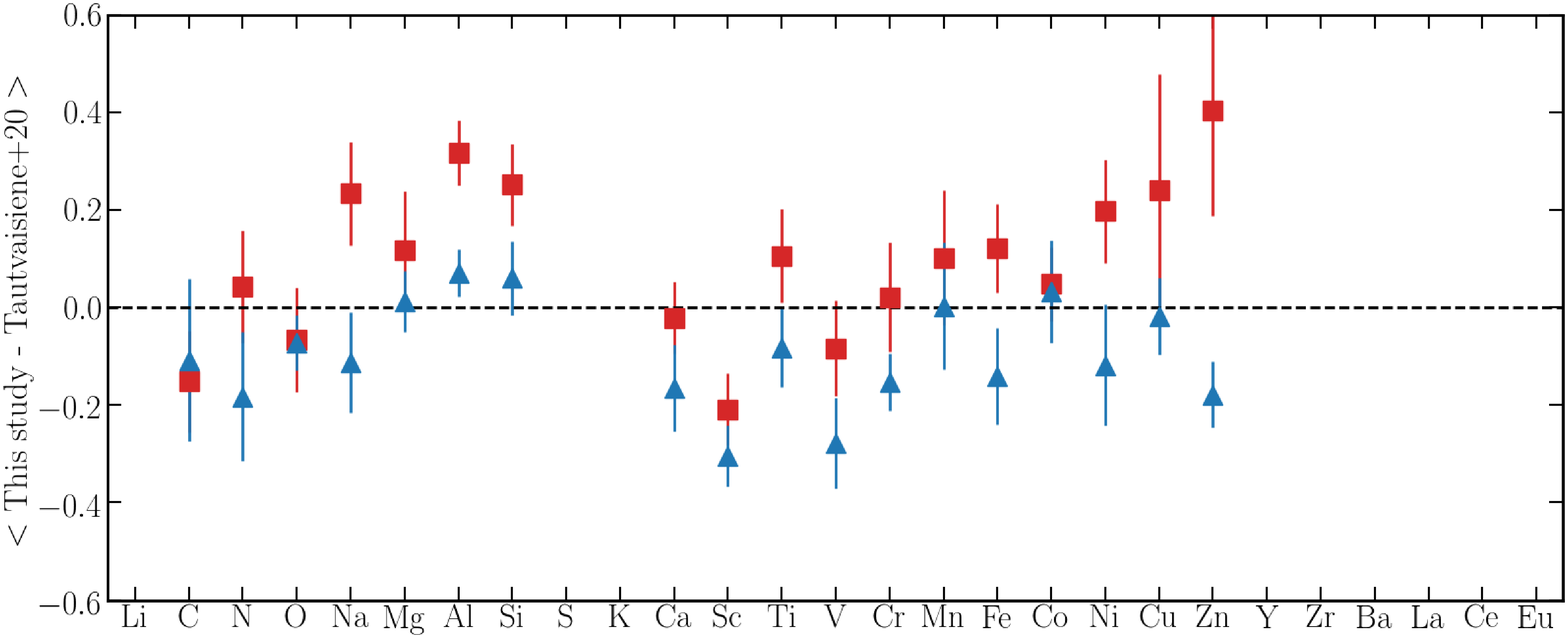}
\caption{Elemental abundances. Panel \emph{a}: Individual abundances for the pilot stars in this survey (squares denote giants, triangles denote dwarfs). Panel \emph{b}: Comparison with stars and elements in common with the survey of Tautvaisiene et al. (\cite{taut}).}
 \label{Fabund}
\end{figure*}

\subsubsection{Light, $\alpha$, iron-peak, and neutron-capture elements}

As mentioned earlier in this paper, the abundances of the different elements were derived through their EWs measured with ARES, and then used in q2. First, a proper line list of non-blended and discernible lines from 4600--9000\,\AA\ was compiled using the VALD-3 database (Ryabchikova et al. \cite{vald3}). This initial line list was then cross-matched with the most recent \textit{Gaia}-ESO line list from Heiter et al. (\cite{hei21}) in order to select only lines with good quality atomic data in the same way as described in Sec.~\ref{S_tgm}. For Sc, V, Mn, Co, Cu, La, and Eu, we considered the HFS details from the same list. We also searched the NIST database (Kramida et al. \cite{2019kramida}) for atomic data on those lines not covered by the \textit{Gaia}-ESO list. After this, we split them into two separate lists for analyzing the dwarf stars (mostly VATT+PEPSI spectra with smaller wavelength range but higher resolution) and the giant stars (mostly STELLA+SES spectra with larger wavelength ranges, but lower resolution).

First, we analyzed a deep PEPSI ($R$=250\,000) and a deep SES ($R$=55\,000) spectrum of the Sun-as-a-star in the same way as the target stars, meaning that the same wavelength ranges as for the survey stars are considered. In particular for the PEPSI solar spectrum, we selected only lines from the 4800--5400\,\AA\ plus the 6280--9120\,\AA\ range. The adopted solar parameters are $T_{\mathrm{eff}}$=5777\,K, $\log g$=4.44\,dex, $\xi_{t}$=1.0\,\kms\ and [M/H]=0.0. In addition to this, for the giants line list, the spectrum of the evolved star EPIC211413402 of M67 (an open cluster with solar age and chemical composition; Liu et al. \cite{liu:m67}) was scrutinized to check the reliability of the selected spectral lines. The combined line list is given in the Appendix in Table~\ref{table-A4}, where we list the ion (column~1), the wavelength (column~2), the excitation potential $\chi$ (column~3), and the $\log gf$ (column~4), together with the EW measured from the PEPSI (column~5) and the SES (column~6) solar spectra. The spectral lines in column~5 and column~6 are thus the actual ones used for the survey giants and the dwarfs, respectively.

Our PEPSI and SES solar abundances in Table~\ref{Tsun} agree very well within the errors with each other and with Asplund et al. (\cite{asp21}). The median of the individual element standard deviation is 0.04\,dex, except for O (6300\,\AA ) and Cu from STELLA+SES, where it is $\approx$0.10\,dex. This confirms the consistency of the data and the basic EW measurement technique, despite the different instrumental setups. The low standard deviation also suggest that the quality of our abundances is mostly S/N-limited. Griffith et al. (\cite{griff}) have used a sample of 89 metal-poor subgiants observed with PEPSI, and analyzed for chemical elements very similar as in the present paper, concluding that there is a photon-noise standard deviation of $<0.05$\,dex in [X/H] for an average S/N of $\approx$200, which is comparable to the median S/N of the spectra in this paper.

A total of 284 and 295 spectral lines from 27 elements were considered for the PEPSI and SES spectra, respectively. However, not all these lines were always measured, either due to a low-quality spectrum (low S/N or large $v\sin i$) or due to the stellar physical conditions (e.g., the Eu line becomes weaker as $T_{\mathrm{eff}}$ increases, with an EW of 36\,m\AA\ in EPIC211413402 and 5\,m\AA\ in the Sun).

Stellar abundances are given relative to the usual $\log N$(H)=12.00 scale for hydrogen and also either relative to the Sun, [X/H], or in terms of absolute abundance, A(X):
\begin{equation}
[X/H] = \log_{10}(N_X)_\star - \log_{10}(N_X)_\odot
\end{equation}
with
\begin{equation}
A(X) = \log_{10}(N_X) = \log_{10}(N_X/N_H) + 12 \ .
\end{equation}

\begin{figure}
\includegraphics[angle=0,width=87mm]{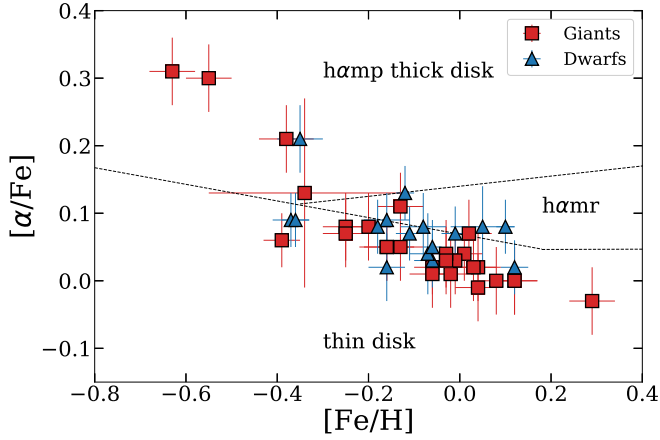}
\caption{Relative $\alpha$-element abundance [$\alpha$/Fe] versus iron abundance [Fe/H] for the pilot sample. Symbols are the observations as identified in the insert. The dashed lines mark the separations of the stellar populations from the APOGEE survey (Lagarde et al. \cite{lagarde}) for thin disk stars and thick disk high-$\alpha$ metal-poor (h$\alpha$mp) and high-$\alpha$ metal-rich (h$\alpha$mr) stars. }
 \label{Falpha}
\end{figure}

To test the reliability and to estimate the systematics of the EW measurements obtained, we compared the values measured with four different packages: ARES v2, iSpec (Blanco-Cuaresma et al. \cite{ispec}), SPECTRE (updated version, Gregersen \cite{dsgspectre}), and {\tt splot} from IRAF. We found insignificant differences for the selected lines; indeed, the ARES approach is the fastest and easiest to use. Therefore, we will use ARES for the whole survey. Rotational line broadening in excess of $\approx$20\,\kms\ renders the measurements of EWs so uncertain, or sometimes impossible to measure, which we refrain from giving abundances for altogether nine pilot targets.

The LTE abundances are computed with the q2 code, employing both the \texttt{abfind} and \texttt{blends} MOOG drivers in an automatic way for those elements with or without HFS. The MARCS (Gustaffson et al. \cite{gus}) model atmospheres were selected for both samples. In this way, the spherical and the plane-parallel geometries are considered for the giant and dwarf stars. We adopted the stellar parameters ($T_{\mathrm{eff}}$, $\log g$, [M/H], $v \sin i$, $\xi_{t}$) obtained from ParSES (Table~\ref{table-A1}) for the model atmospheres. The final error for the abundances in Table~\ref{table-A5} is the quadratic sum of both EW scatter and uncertainties of the atmospheric parameters.

The distribution of the pilot sample of all the elements is shown in Fig.~\ref{Fabund}a, while in Fig.~\ref{Fabund}b we show the mean difference between Tautvaisiene et al. (\cite{taut}) and our abundances. As it can be seen, some elements differ significantly between the two studies because of differences in technique and line lists adopted. The trends between individual [X/H] ratios, computed as [X/H]=$\log (X)_{\star}-\log (X)_{\odot}$, with the solar scale from  Asplund et al. (\cite{asp21}) and the ParSES $T_{\mathrm{eff}}$, are shown in Fig.~\ref{F_C3}a and \ref{F_C3}b for the dwarf and giant stars, respectively. It is clear from these plots that there are some significant trends with temperature, most notably for \ion{K}{i}. These are most likely due to the line list through the presence of undetected blends in some lines, NLTE effects, or due to intrinsic limitations of the analysis of metal-rich or young stars (e.g., Baratella et al. \cite{2020baratella_2}, \cite{2021baratella}, Carrera et al. \cite{2022carrera}). For example, severe NLTE effects for the \ion{K}{i} 7699\,\AA\ transition were found by Takeda et al. (\cite{takeda}), and recently refined by Reggiani et al. (\cite{reggi}) who also provided a grid of 1-D corrections. We applied the Reggiani et al. (\cite{reggi}) corrections to our LTE K abundances and present the NLTE K abundances for our pilot sample in the separate Table~\ref{table-A6}. The differences LTE versus NLTE for the Sun and solar-type stars are large and amount to 0.3--0.4\,dex in the sense that LTE abundances larger than the NLTE abundances. Other element abundances are also affected by NLTE deviations, but we will investigate these discrepancies in a forthcoming paper when all the survey targets have been analyzed.

Finally, we computed the $\alpha$-element abundance of each star. As shown by Adibekyan et al. (\cite{adi12}), metal-poor host stars with small planets are enhanced in $\alpha$-elements (Ne, Mg, Si, S, Ar, Ca, and Ti) with respect to metal-poor stars without known planets. To determine the $\alpha$ abundance, we averaged [Mg/H], [Si/H], [Ca/H], and [Ti/H] to obtain a relative value of [$\alpha$/Fe] with respect to the Sun via
\begin{equation}
[\alpha /{\rm{Fe}}]=\log _{10}{\left({\frac {N_{\alpha}}{N_{{\rm{Fe}}}}}\right)_\star} - \ \log _{10}{\left({\frac {N_{\alpha }}{N_{{\rm{Fe}}}}}\right)_{\odot}}. \label{eq-alpha}
\end{equation}
For the solar term, we used the solar abundance from Asplund et al. (\cite{asp21}). The [$\alpha$/Fe] values and its respective uncertainties are reported in the last column of Table~\ref{table-A5}. Figure~\ref{Falpha} plots the distribution of our pilot targets in the [$\alpha$/Fe] versus [Fe/H] plane. For comparison, we also show the thin disk and the thick-disk $\alpha$-depleted metal-rich and metal-poor domains from Lagarde et al. (\cite{lagarde}).

\begin{figure}
\includegraphics[angle=0,width=87mm]{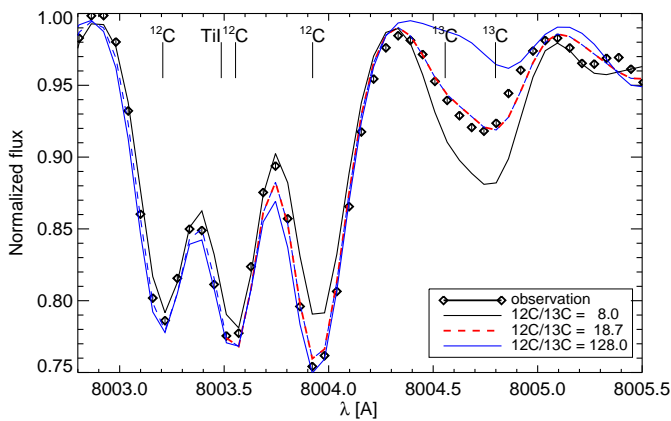}
\caption{TurboMPfit for the 8000-\AA\ CN region of HR\,5844. Diamond symbols are the observed spectrum and the lines are fits with three different $^{12}$C/$^{13}$C isotope ratios, using the CN line list of Plez. Fixed parameters were ($T_{\rm eff}$, $\log g$, $\xi_t$, $v\sin i$) = (4296, 1.54, 2.00, 4.53) in the usual units. The best average from eight fits is $^{12}$C/$^{13}$C=17.8 using a relative nitrogen abundance $\Delta$A(N) of 0.50$\pm$0.05 and a relative iron abundance of --0.25$\pm$0.03 (the best fit shown in the plot is with a slightly different isotope ratio). A total of 41 pixels were fitted to a $\chi^2$ of 575. Average S/N per pixel in this wavelength range is 400. }
 \label{F1213C}
\end{figure}

\begin{figure}
\includegraphics[angle=0,width=87mm]{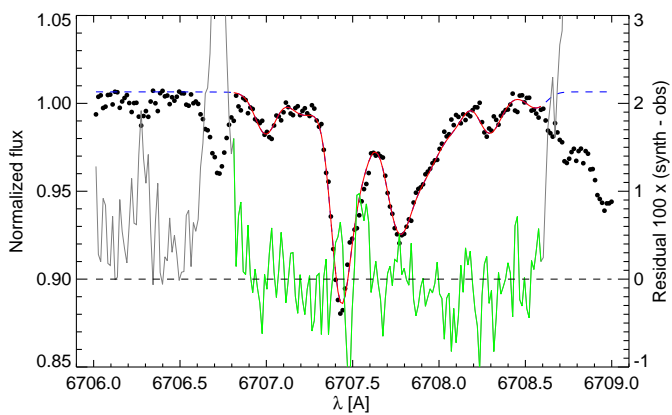}
\caption{TurboMPfit for the lithium region of HD\,175225. Fixed parameters were ($T_{\rm eff}$, $\log g$, $[$M/H$]$, $\xi_t$, $v\sin i$) = (5180, 3.52, 0.0, 1.0, 2.0) in the usual units. The best (average) fit was achieved with A(Li) of 1.22$\pm$0.03, an isotope ratio of 0.00$\pm$0.00, and a relative iron abundance of 0.20$\pm$0.04. A total of 131 pixels were fitted to a $\chi^2$ of 236. Average S/N per pixel in this wavelength range is 311. The residuals in the plot are enhanced by a factor five for better visibility.}
 \label{Fli}
\end{figure}

\subsubsection{Carbon isotope ratio}

The carbon $^{12}$C/$^{13}$C isotope ratio is measured whenever the line broadening allows to do so. The Turbospectrum package is employed under the assumption of LTE. We synthesize the CN line regions around 8003.5\,\AA\ (due to $^{12}$C$^{14}$N) and 8004.7\,\AA\ (due to $^{13}$C$^{14}$N) and fit them simultaneously to the data. Based on a four-dimensional grid of MARCS model atmospheres, characterized by $T_{\rm eff}$, $\log g$, [Fe/H], and microturbulence $\xi_t$, a 6D grid of synthetic spectra covering the wavelength range 8001--8006\,\AA\ was calculated with Turbospectrum (V\,19.1.2). The two additional parameters are the $^{12}$C/$^{13}$C isotopic ratio and the correction of the nitrogen abundance, $\Delta$A(N), relative to the scaled solar value ($\approx$90 and A(N)=7.83; Asplund et al. \cite{asp}). The strength of the CN lines is controlled by both A(C) and A(N). While the carbon abundance A(C) scales with [Fe/H], the parameter $\Delta$A(N) is used to adjust the strength of the CN lines relative to that of the atomic metal lines. Figure~\ref{F1213C} is a representative fit from the giant-star sample.

First tests were run with the line list of Carlberg et al. (\cite{carl}), both for the CN and the atomic lines. We note that Carlberg et al. reduced the transition probabilities $\log gf$ of some important Fe\,{\sc i} and Ti\,{\sc i} lines by 0.2 to 0.6\,dex relative to the ones given by the current VALD-3 data base to achieve reasonable fits at the appropriate [Fe/H]. We ran further tests with another CN line list from B.~Plez (priv. comm.) based on a paper by Hedrosa et al. (\cite{hed}), with the atomic lines unchanged relative to Carlberg et al. (\cite{carl}). We find that this line list gives mostly (but not always) better fits to the observed SES spectra.

A reference fit was also obtained for the extremely high S/N PEPSI spectrum of the Sun in order to verify the method and line list applied. The original spectrum had S/N$\approx$9250 per pixel but was heavily contaminated by telluric lines, making a determination of $^{12}$C/$^{13}$C impossible. We thus first cleaned the spectrum by applying the telluric-fitting program \emph{TelFit} (Gullikson et al. \cite{telfit}), which removed the telluric lines but also reduced the S/N of the corrected spectrum to $\approx$1500 per pixel. Analyzing this solar spectrum in the same way as the stellar spectra, we find the formal result $^{12}$C/$^{13}$C=110$\pm$30, which is consistent with the accepted solar value (based on different spectral lines). We note that the solar CN lines are much weaker than in most of the stellar targets and the fits are relatively poor.

For a given target, the stellar parameters $T_{\rm eff}$, $\log g$, $\xi_t$, and $v\sin i$ remain fixed to the values in Table~\ref{table-A1} from the ParSES pipeline. The best fit to the observed spectrum is determined with the IDL procedure MPFIT (Markwardt \cite{mpfit}). The fit with the minimum $\chi^2$ is found by iteration, adjusting the five (or six) fitting parameters [Fe/H], $\Delta$A(N), $^{12}$C/$^{13}$C, $v_{\rm b}$, $\Delta\lambda$, and optionally $F_{\rm c}$. Here $v_{\rm b}$ is the full width half maximum (FWHM) in velocity space of a Gaussian line broadening kernel (representing instrumental broadening plus macroturbulence), $\Delta\lambda$ is a global wavelength shift applied to the synthetic spectrum for compensating possible radial velocity shifts, and $F_{\rm c}$ is a scaling factor effecting a re-normalization of the local continuum. We point out that [Fe/H] is of low quality due to the somewhat arbitrary assignment of the $\log gf$ values, and A(C) and A(N) cannot be determined unambiguously from the CN lines alone. We therefore refrain from listing all six free parameters.

Eight different fits are obtained for each target. Two for a fitting range over which $\chi^2$ is evaluated of 8001.6--8005.9\,\AA\ and 8003.5--8005.9\,\AA, with\ the latter being similar to what was used by Sablowski et al. (\cite{sab:capella}). For each of these two runs, the continuum was either fixed or free, and the line list is either that taken from Carlberg et al. (\cite{carl}) or from Plez (priv. comm.). The final $^{12}$C/$^{13}$C results given in Table~\ref{table-A7} are simply the unweighted mean and standard deviation of the eight individual fitting runs. These should represent reasonable best estimates and internal uncertainties.

Our immediate result from the above analysis of the (pilot) evolved stars is that all $^{12}$C/$^{13}$C isotopic ratios appear to be significantly lower than solar ($\approx$90) with an average of $\approx$18 (range between 10--30). The pilot dwarf stars are comparably difficult if not impossible to measure, mostly because the CN lines are dramatically weaker than in the cooler giants. Additionally, the generally larger rotational line broadening combined with lower S/N due to being mostly fainter and observed at a higher spectral resolution, adds to this. The sum of these constrains led us to only two successful measures in the (pilot) dwarf star sample for which, however, one turned out to be an evolved star (HD\,160605). The spectra of these two stars were also dehydrated as in the case of the solar spectrum and then analyzed in the same way again. However, the corrections for the telluric absorption turned out to be on the order of or less than 10\%\ for the ratio (in contrast to the cleaned solar spectrum where the corrections made a difference). The cleaned spectra showed that the $^{12}$C/$^{13}$C isotopic ratio appears significantly lower for HD\,160605 (40$\pm$20) than for HD\,175225 (99$\pm$39). While the latter is consistent with the solar value, the former may indicate that some dredge-up had already occurred in HD\,160605, which appears plausible since this star's low gravity ($\log g = 3.2$) indicates that it had clearly evolved off the main sequence.

\begin{figure*}
{\bf a.} \hspace{58mm} {\bf b.} \hspace{55mm} {\bf c.} \\
\includegraphics[angle=0,width=55mm,clip]{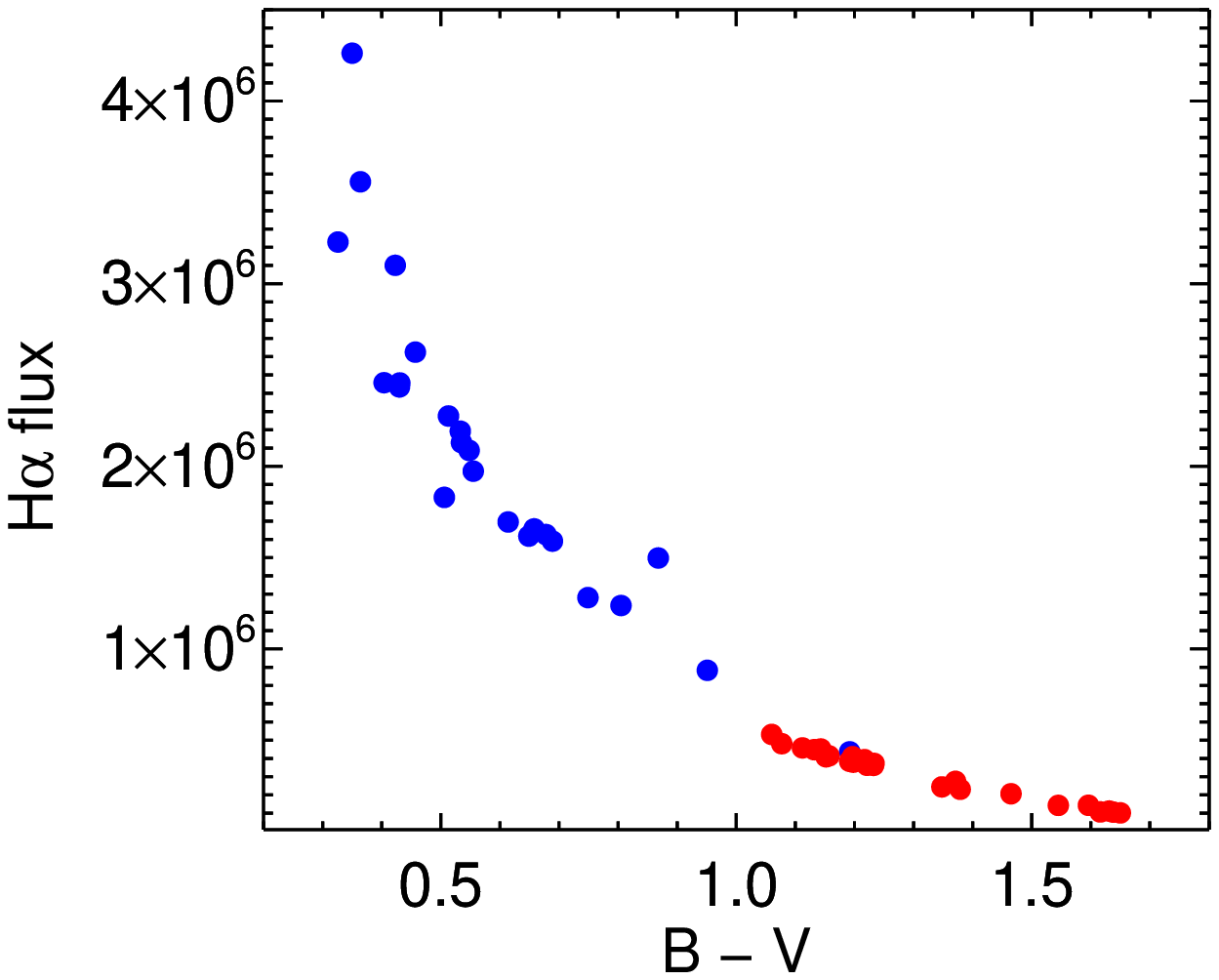} \hspace{3mm}
\includegraphics[angle=0,width=55mm,clip]{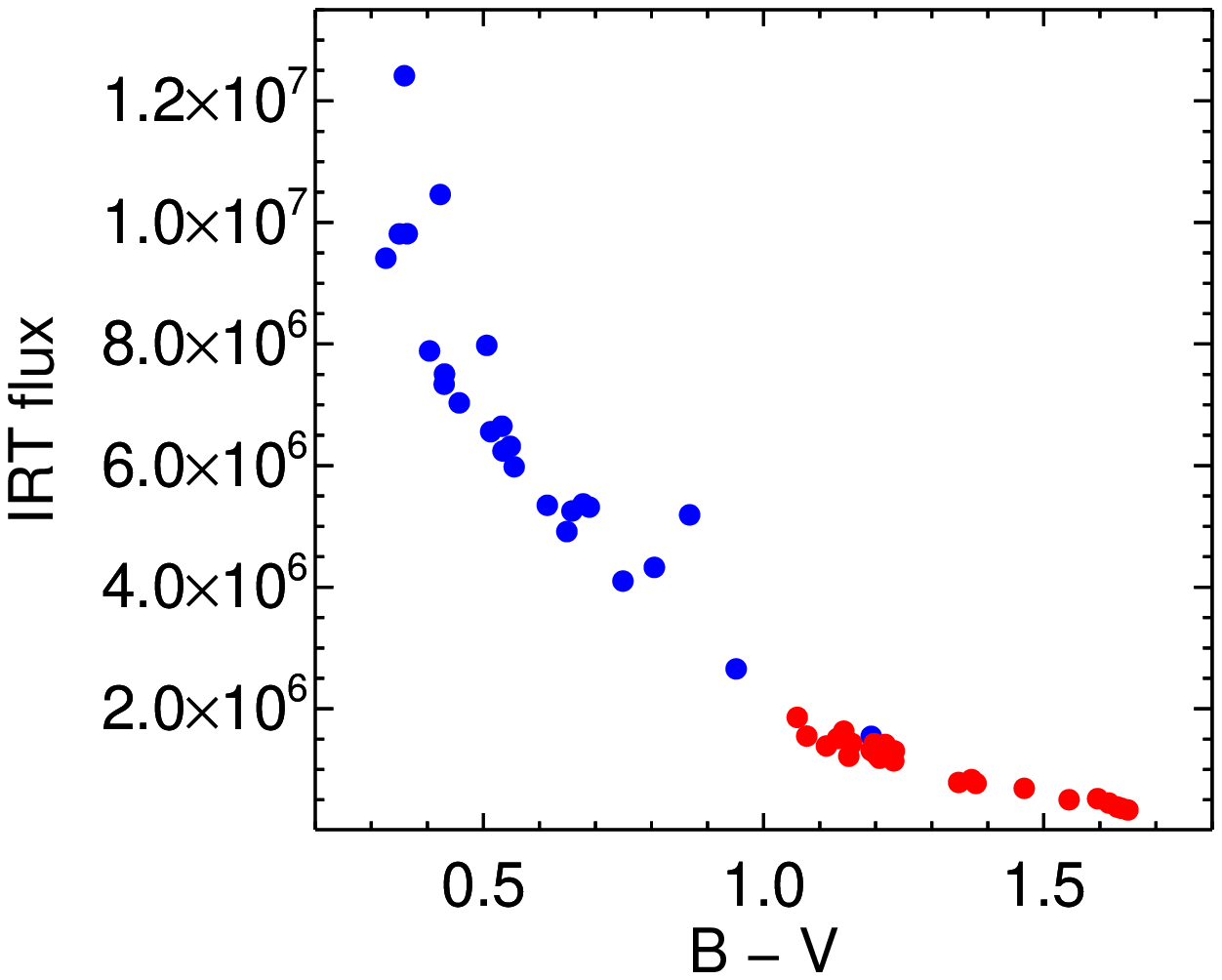} \hspace{3mm}
\includegraphics[angle=0,width=55mm,clip]{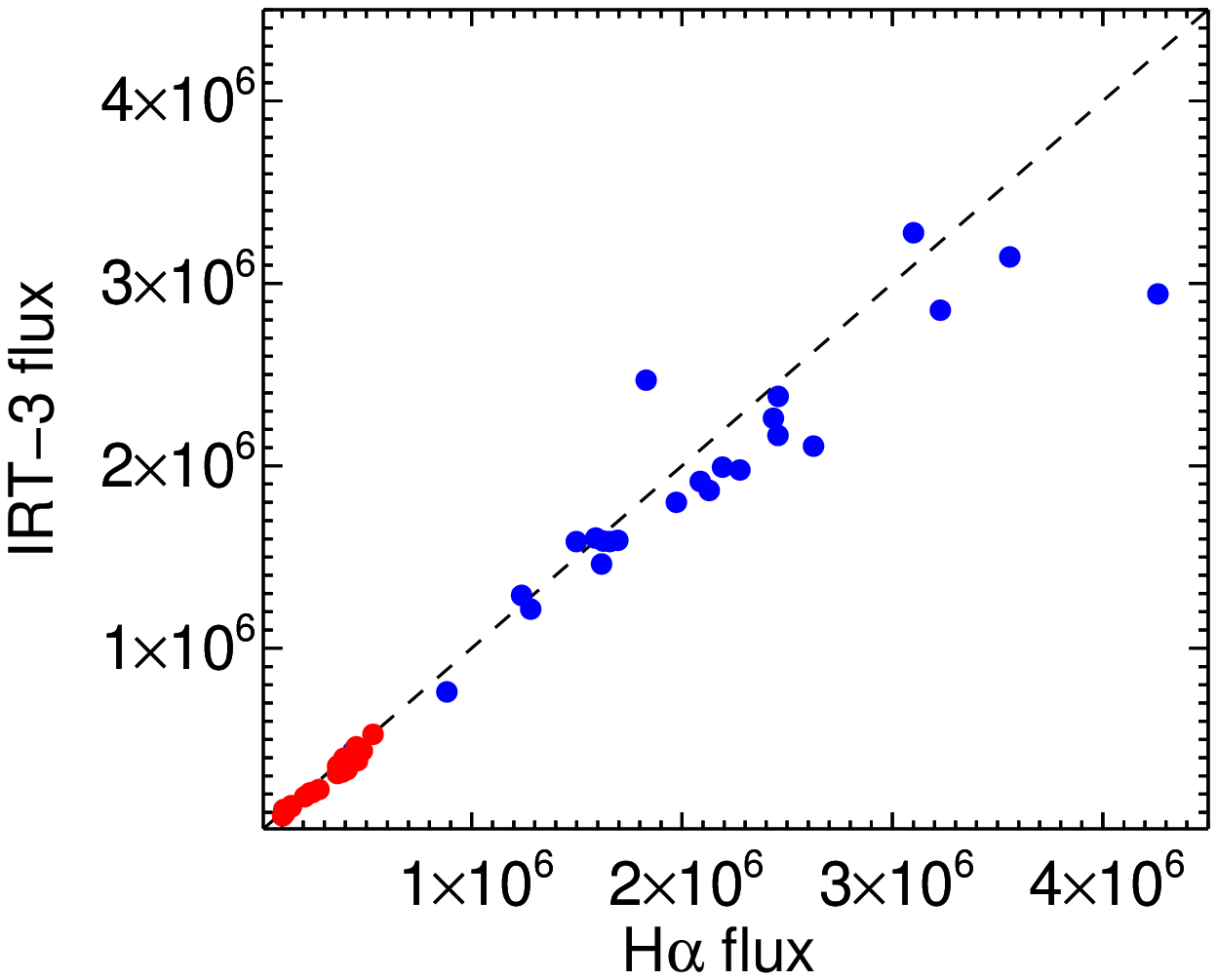}
\caption{Absolute emission line surface fluxes for the pilot sample. Panel \emph{a}: Balmer \Halpha\ line-core flux vs. $B-V$. Panel \emph{b}: Summed \ion{Ca}{ii} IRT line-core flux vs. $B-V$. Panel \emph{c}: Flux-flux relation for Balmer \Halpha\ vs. \ion{Ca}{ii} IRT-3. Dashed line is the 1:1 relation. Red dots are from STELLA+SES spectra, blue dots from VATT+PEPSI spectra.}
 \label{Fact}
\end{figure*}

\subsubsection{Lithium}

Another important element is lithium. Figures~\ref{F3}b and \ref{F4}b show the wavelength region around the lithium doublet at 6708\,\AA\ for the pilot stars. In the Sun, the photospheric lithium has been largely depleted as a result of thermonuclear reactions occurring during stellar evolution. Its total abundance is more than hundred times smaller than the meteoritic value, with the more fragile $^6$Li isotope being completely destroyed. An unambiguous detection of this fragile isotope in an atmosphere of a cool star would possibly be an indication of an external pollution process by planetary matter (Israelian et al. \cite{israel01}, \cite{israel03}; Mott et al. \cite{mott}). We provide the total lithium abundance A(Li) as defined in Eq.~\ref{eq-ALi} and the $^6$Li/$^7$Li isotopic ratio whenever possible:
\begin{equation}
A({\rm{Li}}) = \log_{10} { \left( { \frac{N_{\rm{Li}}}{N_{\rm{H}}}} \right) } + 12 \ . \label{eq-ALi}
\end{equation}

Lithium abundances were only determined from synthetic fits based on 1D model atmospheres under LTE. The same subset of the MARCS model atmospheres available for ParSES is used for the synthetic spectra for the lithium region. A total of 200 spectra are synthesized per model atmosphere covering 20 Li abundances and five isotope ratios (each for two values for the microturbulence) and thus cover a 6D parameter space including $T_{\rm eff}$, $\log g$, $[$Fe/H$]$, and microturbulence. The synthesized wavelength range is 6706--6709\,\AA . As line list we adopted the list from Strassmeier \& Steffen (\cite{xiBoo}), which is based on Mel\'endez et al. (\cite{mel12}), except for the Li doublet with isotopic hyperfine components and now also includes the broadening constants (for a comparison with other Li line lists see Mott et al. \cite{mott}). 3D NLTE corrections are available for the dwarf stars (Mott et al. \cite{mott20}, Harutyunyan et al. \cite{gohar}) and are listed in our tables but are not applied. The fitting procedure with TurboMPfit involves the continuum level and a global wavelength shift as free parameters if needed. One example fit for one of our pilot stars, HD\,175225 (G9), is shown in Fig.~\ref{Fli}. A reference fit was also obtained for a spectrum of the Sun, which verified the method applied (see Strassmeier et al. \cite{Sun}).

Table~\ref{table-A7} summarizes the Li results for the pilot stars. The only dwarf stars for which the best fit required $^6$Li were HD\,142006 and HD\,155859. Their isotopic ratios are 6.7$\pm$0.8\,\%, and 1.8$\pm$0.7\,\%, respectively, both with similarly high Li abundance of A(Li)$\approx$2.2. Because these measurements are difficult and the fits involve many free parameters, we consider only the 8-$\sigma$ detection for HD\,142006 a significant result and discard the detection for HD\,155859. Most stars in the giant sample show no or very little Li, but there are exceptions with strong $^7$Li lines such as those in HD\,144903 and HD\,149843. No formal detection of $^6$Li in the giant-star sample is seen. We note that if we allow TurboMPfit to alter the continuum level, as well as to shift the spectrum in wavelength, we obtain sometimes larger best-fit isotope ratios with practically the same error but with a significantly larger ratio. This also indicates that the fitting error is not necessarily indicative for the isotope ratio error.

\subsection{Magnetic activity proxies}

We measured the absolute line-core flux in Balmer \Halpha\ (at 6563\,\AA\ in PEPSI CD\,V), and the Ca\,{\sc ii} infrared triplet (at 8498, 8542 and 8662\,\AA\ in CD\,VI, dubbed IRT-1, IRT-2, and IRT-3, respectively). For \Hbeta\ (4861\,\AA\ in CD\,III), we give only an equivalent width and its ratio to \Halpha\ because we lack a proper absolute flux calibration for this line region. The SES data also cover Ca\,{\sc ii} H\&K at 3950\,\AA\ but only with very low S/N. We do not use it in the analysis but provide the full wavelength coverage of the reduced data. Relative fluxes are converted to absolute fluxes according to the procedures below. Figure~\ref{Fact} shows the measured absolute emission-line fluxes for the pilot sample as a function of $B-V$ color as well as an example of a typical flux-flux relation (\Halpha\ vs. \ion{Ca}{ii} IRT-3).

Single-lined (SB1) binary spectra are treated just like single stars, which have continua presumably unaffected by the companion. The few double-lined (SB2) spectra were first multiplied by the respective photospheric light ratio. It is obtained from the depths of photospheric absorption-line pairs near above respective wavelengths. The resulting fluxes for SB2's are generally less precise because the available S/N in the continuum is distributed among the two components.

\subsubsection{Balmer \Halpha\ line-core flux}

The stellar surface \Halpha -core flux in erg\,cm$^{-2}$s$^{-1}$ is computed from the measured 1-\AA\ equivalent width, $W_{\rm core}$, under the spectrum and zero intensity,
\begin{equation}\label{eq1}
{\cal F}_{{\rm H}\alpha}  = W_{\rm core} \ {\cal F}_{\rm c} \ ,
\end{equation}
and is related to the expected absolute continuum flux, ${\cal F}_{\rm c}$, at these wavelengths. The continuum flux is obtained from the relations provided by Hall (\cite{hall96}) for various stellar luminosity classes and a color ranges.  The relations used in this paper are as follows:
\begin{equation}\label{eq1a}
\log {\cal F}_{\rm c} = 7.538 - 1.081 \ (B-V)
,\end{equation}
for MK I-V and  $0<(B-V)<1.4$;
\begin{equation}\label{eq1b}
\log {\cal F}_{\rm c} = 7.518 - 1.236 \ (V-R),
\end{equation}
for MK V and $0<(V-R)<1.4$; and
\begin{equation}\label{eq1c}
\log {\cal F}_{\rm c} = 7.576 - 1.447 \ (V-R),
\end{equation}
for MK I-IV  and $0<(V-R)<1.8$.

Rotational line broadening is accounted for by applying a similar correction to $W_{\rm core}$ as done by Strassmeier et al. (\cite{orbits}). However, the line-core fluxes are not corrected for photospheric light. Based on the discussion in Strassmeier et al. (\cite{orbits}), we refrained from a photospheric correction and give combined photospheric plus chromospheric fluxes as observed. The largest contribution to the flux error is due to the continuum-flux calibration and comes from systematic differences in the $T_{\rm eff}$-color relations and the error in $T_{\rm eff}$. Our prime choice for ${\cal F}_{\rm c}$ is based on a measured $B-V$ color from the Tycho-2 catalog (H\o g et al.~\cite{tycho}) rather than $V-R$. It gives the lowest residuals in our pilot sample with a rms of 0.45 in units of $10^5$~erg\,cm$^{-2}$s$^{-1}$.

\subsubsection{Ca\,{\sc ii} IRT line-core fluxes}

As above, we integrate the central 1-\AA\ portion of each \ion{Ca}{ii} IRT line profile (Figs.~\ref{F3}c and \ref{F4}c). This relative flux, or equivalent width, is again related to the absolute continuum flux, ${\cal F}_{\mathrm c}$, obtained from relations provided by Hall (\cite{hall96}). The relations used here are:
\begin{equation}\label{eq3}
  \log {\cal F}_{\mathrm c} = 7.223 - 1.330\, (B - V),
\end{equation}
for luminosity class I-V and $-0.1 < (B-V) < 0.22$, and
\begin{equation}\label{eq4}
  \log {\cal F}_{\mathrm c} = 7.083 - 0.685\, (B - V),
\end{equation}
for luminosity class I-V and $0.22 < (B-V) < 1.4$. The original relations were obtained from calibrations for a temperature range of 4000--7800~K, basically covering all our survey targets.

\begin{table*}[ht!]
\caption{Observing log for the survey. } \label{Tobslog}
\begin{scriptsize}
\begin{flushleft}
\begin{tabular}{lllll lllll ll}
\hline\hline
\noalign{\smallskip}
FITS file ID & Target ID & RA \ \ \ \ \ \ \ DEC   & $V$ & Sample & Date UT & Time UT & Exp. & S/N & Time BJD & Instrument & CD \\
             &           & (2000.0) & (mag) &      & (d/m/y) &(h:m:s)  &(h:m:s) &    & (245+)   &   &    \\
(1) &(2) &(3) &(4) &(5) &(6) &(7) &(8) &(9) &(10) &(11) & (12) \\
\noalign{\smallskip}\hline\noalign{\smallskip}
science20180222B-0013 &HD 134585 & 15:05:45 +71:52:58 &7.46 & giant & 22/02/2018 &01:02:38 &0:40:00 &202 &8171.60059 &STELLA+SES & \\
science20180414B-0365 &HD 134585 & 15:05:45 +71:52:58 &7.46 & giant & 14/04/2018 &19:33:36 &0:40:00 &200 &8223.41294 &STELLA+SES & \\
science20181206B-0040 &HD 134585 & 15:05:45 +71:52:58 &7.46 & giant & 06/12/2018 &04:59:03 &0:40:00 &104 &8458.76383 &STELLA+SES & \\
pepsir.20170522.004   &HD 135143 & 15:08:30 +72:22:10 &7.84 & dwarf & 22/05/2017 &02:58:52 &1:25:00 &185 &7895.65445 &VATT+PEPSI &  CD35 \\
pepsir.20170527.024   &HD 135143 & 15:08:30 +72:22:10 &7.84 & dwarf & 27/05/2017 &03:03:56 &1:25:00 &184 &7900.65783 &VATT+PEPSI &  CD36\\
science20181214B-0025 &HD 135143 & 15:08:30 +72:22:10 &7.84 & dwarf & 14/12/2018 &04:29:18 &1:00:00 &195 &8466.75035 &STELLA+SES & \\
science20190115B-0020 &HD 135143 & 15:08:30 +72:22:10 &7.84 & dwarf & 15/01/2019 &02:29:32 &1:00:00 &203 &8498.66775 &STELLA+SES & \\
science20191214B-0026 &HD 135143 & 15:08:30 +72:22:10 &7.84 & dwarf & 14/12/2019 &04:30:24 &1:00:00 &194 &8831.75111 &STELLA+SES & \\
pepsir.20180528.008   &HD 135045 & 15:08:35 +70:08:49 &8.03 & dwarf & 28/05/2018 &03:36:29 &1:25:00 &137 &8266.68065 &VATT+PEPSI &  CD35 \\
pepsir.20180602.000   &HD 135045 & 15:08:35 +70:08:49 &8.03 & dwarf & 02/06/2018 &03:08:06 &1:25:00 &146 &8271.66081 &VATT+PEPSI &  CD36\\
science20181219B-0022 &HD 135045 & 15:08:35 +70:08:49 &8.03 & dwarf & 19/12/2018 &04:20:35 &1:30:00 &184 &8471.75464 &STELLA+SES & \\
science20190425B-0364 &HD 135045 & 15:08:35 +70:08:49 &8.03 & dwarf & 25/04/2019 &19:54:14 &1:30:00 &285 &8599.44454 &STELLA+SES & \\
science20200116B-0029 &HD 135045 & 15:08:35 +70:08:49 &8.03 & dwarf & 16/01/2020 &02:22:02 &1:30:00 &233 &8864.67287 &STELLA+SES & \\
\noalign{\smallskip}
\hline
\end{tabular}
\tablefoot{The first 3 of the 1067 targets are listed here as an example, full version only electronically. File names appear truncated in the printed table.}
\end{flushleft}
\end{scriptsize}
\end{table*}

As for \Halpha , the effect of rotational line broadening on the emission-line fluxes is usually negligible for stars with $v\sin i < 10$~\kms\ from spectra with a spectral resolution of up to, say, 50\,000. However, as discussed in Strassmeier et al. (\cite{orbits}), the IRT-flux correction amounts to 30\%\ for a rapidly rotating (cool) star with $v\sin i = 60$~\kms. Because of our high spectral resolution, we employ the $v\sin i$ correction suggested by Strassmeier et al. (\cite{orbits}) consistently to all three IRT lines (for the $R$=200\,000 PEPSI spectra as well as for the $R$=55\,000 SES spectra). We again refrained from a photospheric line-core correction though.

The total radiative loss in the IRT lines, $R_{\rm IRT}$, is determined from the sum of the corrected fluxes from all three IRT lines in units of the stellar bolometric luminosity. It is a good indicator of the star's overall chromospheric activity:
\begin{equation}
R_{\rm IRT} = \frac{{\cal F}_{\rm IRT-1} + {\cal F}_{\rm IRT-2} + {\cal F}_{\rm IRT-3}}{\sigma \ T_{\rm eff}^4} .
\end{equation}

Table~\ref{table-A8} in the appendix summarizes the results for the pilot stars. Errors for $R_{\rm IRT}$ are driven by the error for $T_{\rm eff}$ and are not explicitly listed in the table but are on the order of 20\%.

\begin{table}[ht!]
\caption{Average uncertainties based on the pilot-star sample.} \label{T5}
\begin{flushleft}
\begin{tabular}{llll}
\hline\hline
\noalign{\smallskip}
Parameter  & \multicolumn{2}{c}{Uncertainty} & Unit \\
           & VATT & STELLA  &  \\
\noalign{\smallskip}\hline\noalign{\smallskip}
RV           & 15  & 50 & \ms \\
$T_{\rm eff}$& 70 & 70 & K \\
$\log g$     & 0.20  & 0.15 & cm\,s$^{-2}$ \\
$[$M/H$]$        & 0.1   & 0.1  & solar \\
$\xi_{\rm t}$& 0.10 & 0.08 & \kms \\
$v\sin i$$^1$ ParSES & 0.20 & 0.40 & \kms \\
$v\sin i$$^2$ ParSES & 5 & \dots & \kms \\
$v\sin i$$^1$ $i$SVD   & 0.79 & 0.48 & \kms \\
$v\sin i$$^2$ $i$SVD   & 1.5 & \dots & \kms \\
$\zeta_{\rm RT}$ & 0.71 & 0.47 & \kms \\
BIS          & 25 & 30 & \ms \\
$\log t$$^1$ & 0.05 & 0.10 & yr \\
$\log t$$^2$ & 0.1  & \dots & yr \\
$M$          & 0.04 & 0.15 & M$_\odot$ \\
A(Li)        & 0.03 & 0.05 & H=12 \\
$^{6}$Li/$^{7}$Li & 0.7 & (0.1) & \dots \\
$^{12}$C/$^{13}$C & 10 & 0.8--10 & \dots \\
A(X)         & \multicolumn{2}{l}{\ \ $<0.1$\,dex} & H=12\\
${\cal F}_{{\rm H}\alpha}$ & 3 & 0.7 & $10^5$\,erg\,cm$^{-2}$s$^{-1}$\\
${\cal F}_{\rm IRT}$ & 3 & 0.5 & $10^5$\,erg\,cm$^{-2}$s$^{-1}$\\
$\log R_{\rm IRT}$   & 0.02 & 0.02 & $\sigma T^4_{\rm eff}$ \\
\noalign{\smallskip}
\hline
\end{tabular}
\tablefoot{$^1$representative for stars with $v\sin i <50$\,\kms . $^2$representative for stars with $v\sin i >50$\,\kms .}
\end{flushleft}
\end{table}

\section{Initial data release}\label{S5}

Table~\ref{Tobslog} is the observing log for the full sample. It is available in its entirety
only in electronic form via CDS~Strasbourg. It lists all individual observations for all 1067 targets, including the pilot stars analyzed in this paper.
We note that the dwarf sample included targets on both telescope+spectrograph combinations, while the giants were predominantly only on STELLA+SES. The table lists 12 columns. Column (1) is the FITS file name; (2) the main target identifier; (3) the target coordinates right-ascension (RA) and declination (DEC) in equinox 2000.0 with increasing target RA; (4) the visual $V$ magnitude; (5) the sample pre-selection (dwarf or giant) as indicated above; (6) the UT date of observation; (7) the UT time of observation of mid exposure; (8) the exposure time; (9) the S/N per pixel. For SES it is the averaged value around \Halpha , for PEPSI it is the quantile 95\%\ over the entire wavelength range per cross disperser; (10) the barycentric Julian Date of mid exposure; (11) the telescope-spectrograph combination; and (12) the identification of the two cross dispersers for a PEPSI entry.

The final, reduced, and 1D, spectra will be made available through CDS~Strasbourg. For the current (initial) data release, we employed the reduction packages SDS4PEPSI in its version 2.0 for PEPSI spectra, and SESDR in its version 4.0 for SES spectra. Future data releases (if forthcoming) will employ higher versions of the DR packages. The pilot-star spectra from this paper are included in the full sample.

\section{Summary and conclusions}\label{S6}

In this work, we introduce a new high-resolution spectroscopic survey of bright stars surrounding the northern ecliptic pole, which we named the VPNEP survey. The VATT and STELLA telescopes in Arizona and Tenerife were employed. VATT was used with a 450-m fiber link to the LBT spectrograph PEPSI and obtained two $R$=200\,000 spectra of the (bona-fide) dwarf stars in the NEP field. The giant stars in this field were observed, usually three times, with the robotic STELLA facility in Tenerife at $R$=55\,000. Seventy stars were observed with both telescope-spectrograph combinations. The present paper summarizes the entire survey target sample and our analysis tools but presents only an initial analysis from a pilot survey of 54 stars observed in 2017 (27 dwarfs and 27 giants). Among the deliverables are the stellar astrophysical parameters $T_{\rm eff}$, $\log g$, $[$M/H$]$, $v\sin i$, micro- and macroturbulence, bisector velocity span, convective blue shift (when possible), chemical abundances, mass, and age. Chemical abundances are currently derived for 27 elements, among them lithium, many $\alpha$ elements, and CNO. Isotope ratios for lithium and carbon are derived in favorable cases. Proxy magnetic activity parameters are given in form of absolute emission-line fluxes and radiative losses for the Ca\,{\sc ii} IRT, \Halpha , and \Hbeta\ lines. Our current dwarf-star pilot sample of 27 targets contains 2 SB2 and 3 SB1 spectroscopic binaries, while the giant-star pilot sample (of also 27 targets) had no SB2s but 4 SB1s. The total number of targets of the full survey is 1067 stars. Forthcoming papers in this series will focus on the analysis of particular astrophysical parameters but for the full sample.

Table~\ref{T5} offers a summary of the expected uncertainties of the prime astrophysical parameters. These values are derived from the 54-star pilot sample and are representative for the respective instrumental telescope-spectrograph combination but are biased because the two sub-samples contained mostly either dwarfs (VATT+PEPSI) or giants (STELLA+SES). These data serve as a preliminary quality estimate and will be revised once the full sample of spectra has been analyzed.

The results from the pilot survey can be summarized as follows:
\begin{itemize}
\item Our spectra have an average S/N per pixel of $\approx$200 but range between 40 and 870 depending on target brightness.
\item A RV zero-point offset between STELLA+SES and VATT+PEPSI of 0.395$\pm$0.21~\kms\ was established.
\item For the STELLA+SES spectra external errors per measurement for $T_{\rm eff}$ are typically 70~K, for $\log g$ typically 0.2~dex, and for $[$M/H$]$ typically 0.1\,dex. For the VATT+PEPSI spectra the external errors are comparable, that is 70~K for $T_{\rm eff}$, 0.15\,dex for $\log g$, and 0.1\,dex for $[$M/H$]$.
\item Ensemble bias (rms) for $T_{\rm eff}$, $\log g$, [M/H], and $v\sin i$ are 70\,K, 0.14\,dex, 0.05\,dex, and $\approx$1\,\kms.
\item Systematic errors exist but are small and remain on the order of the ensemble 1$\sigma$; deviations between VATT+PEPSI and STELLA+SES for $T_{\rm eff}$, $\log g$, [M/H], and $v\sin i$ are 70\,K, 0.10\,dex, 0.08\,dex, and $\approx$1\,\kms\ in the sense SES minus PEPSI.
\item Rotational line broadening $v\sin i$ ranges between 1.1$\pm$0.8\,\kms\ for HD\,142006 and HD\,135143, and 185$\pm$16\,\kms\ for VX\,UMi.
\item Radial-tangential macro turbulence ranges between 1.8$\pm$0.7\,\kms\ for HD\,175225 and 5.5$\pm$0.5 for HD\,148374.
\item Bisector spans were measured between $-$559$\pm$189\,\ms\ for HD\,139797 and +185$\pm$89\,\ms\ for HD\,160052. In one case the bisector severely affects the $v\sin i$ measurement (HD\,194298).
\item Stellar masses range from 0.86 up to 3.0~M$_\odot$, and ages range between 0.4 and 12~Gyr. Errors basically reflect the errors of $T_{\rm eff}$ and $M_V$ and do not take into account model uncertainties.
\item Solar chemical abundances from SES and PEPSI spectra agree very well with Asplund et al. (\cite{asp21}). The median standard deviation is $0.04$\,dex for all elements but O and Cu (median rms 0.10) in [X/H]. The solar iron abundance from the deep PEPSI spectrum of A(Fe)=7.46$\pm$0.01 is even four times more precise than in Asplund et al. (\cite{asp21}). No Li abundance was obtained for the solar spectra.
\item Stellar chemical abundances are given for up to 27 elements; Li, C, N, O, Na, Mg, Al, Si, S, K, Ca, Sc, Ti, V, Cr, Mn, Fe, Co, Ni, Cu, Zn, Y, Zr, Ba, La, Ce, and Eu, with individual-element errors usually better than 0.10~dex.
\item $\alpha$-element abundances relative to iron range from $-0.03$ to $+0.31$~dex for the giant-star sub-sample, and from $+0.02$ to $+0.21$~dex for the dwarf-star sub-sample with typical errors of $\pm$0.05\,dex.
\item Absolute lithium abundances range from $-$0.87$\pm$0.13\,dex for HD\,151698 to +2.74$\pm$0.01 for $\psi^{01}$ Dra\,B.
\item Carbon $^{12}$C/$^{13}$C isotope ratios of the bona-fide giant stars appear to be significantly lower than solar ($\approx$90) with an average of $\approx$18. Besides the Sun, only two other stars in the pilot dwarf sample had a detectable $^{12}$C/$^{13}$C ratio; HD\,175225 (99$\pm$39) and HD\,160605 (40$\pm$20).
\item A significant (and believable) non-zero $^6$Li/$^7$Li isotope ratio was detected only in one star; HD\,142006 (6.7$\pm$0.8\,\%).
\item Relations of absolute-emission line fluxes with $B-V$ and their respective flux-flux distribution verify the spectral lines used as magnetic surface-activity proxies.
\end{itemize}

\begin{acknowledgements}

We thank all engineers and technicians involved in LBT/PEPSI, STELLA, and the VATT.  Thanks also go to an anonymous referee for the many useful comments. STELLA and PEPSI were made possible by funding through the State of Brandenburg (MWFK) and the German Federal Ministry of Education and Research (BMBF) through their Verbundforschung grants 05AL2BA1/3 and 05A08BAC. The STELLA facility is a collaboration of the AIP in Brandenburg with the IAC in Tenerife (see https://stella.aip.de/). PEPSI is an AIP instrument on the LBT and operated jointly by LBTO and AIP (see https://pepsi.aip.de/). The VATT part of the survey has been co-funded by the Vatican Observatory (see https://www.vaticanobservatory.va/) and the Vatican Observatory Foundation. Finally, the many helpful discussions with colleagues abroad are very much appreciated. In particular we thank Daniel Apai of Steward Observatory and David Latham of CfA for earlier discussions. Thanks also to students Y. J. Hew and C. Tulban from the University of Arizona. We also thank the ESA/Gaia team for providing such a splendid data set. This research also has made use of the SIMBAD database, operated at CDS, Strasbourg, France. \emph{Note.} Due to the outbreak of the Covid pandemic the VATT was shut down during most of the NEP visibility in 2020. Missing observations had to be deferred to 2022.
\end{acknowledgements}

\appendix

\section{Detailed tables}

\begin{table*}[!tbh]
\caption{Results from the ParSES analysis:\ RV radial velocity; $T_{\rm eff}$ effective temperature; $\log g$ logarithmic gravity; $[$M/H$]$ metallicity; $\xi_{\rm t}$ micro turbulence;  $v\sin i$ projected rotational broadening; and CD cross disperser for PEPSI observations.}\label{table-A1}
\begin{flushleft}
\begin{tabular}{llllllll}
\hline\hline\noalign{\smallskip}
Star      &         &  $T_{\rm eff}$ & $\log g$\,$^1$& $[$M/H$]$  & $\xi_{\rm t}$ & $v\sin i$ & \\
BJD +245  & RV      &                &               &            &               &           & CD \\
          & (\kms)  & (K)            & (cm\,s$^{-2}$)& (solar)    & (\kms )       & (\kms)    & \\
\noalign{\smallskip}\hline\noalign{\smallskip}
\multicolumn{2}{l}{BY Dra$^1$} & 4435$\pm$12 & 3.68$\pm$0.05 & -0.64$\pm$0.16 & 0.53$\pm$0.04 & 8.33$\pm$0.14 & \\
7894.8719 & -57.09$\pm$0.05 & 4331$\pm$36 & 3.14$\pm$0.12 & -0.78$\pm$0.74 & 0.50$\pm$0.10 & 7.87$\pm$0.41  & 3,5 \\
7899.9343 & -17.76$\pm$0.05 & 4432$\pm$32 & 3.75$\pm$0.14 & -0.64$\pm$0.48 & 0.53$\pm$0.10 & 8.53$\pm$0.37  & 3,6 \\
\hline
\multicolumn{2}{l}{HD 135119} & 6477$\pm$54 & 3.05$\pm$0.14 & -0.69$\pm$0.07 & 2.82$\pm$0.28 & 90.5$\pm$4.9 & \\
7893.7181 & -1.73$\pm$0.02 & 6529$\pm$178 & 2.55$\pm$0.69 & -0.86$\pm$0.20 & 3.41$\pm$0.93 & 91$\pm$19  & 3,5 \\
\hline
\end{tabular}
\tablefoot{The first block of entries is from VATT+PEPSI, the second block from STELLA+SES. $^1$Target is SB2 with unaccounted secondary during the fit.}
\end{flushleft}
\end{table*}

\setcounter{table}{0}
\begin{table*}[!tbh]
\caption{Continued.}
\begin{flushleft}
\begin{tabular}{llllllll}
\hline\hline\noalign{\smallskip}
Star      &         &  $T_{\rm eff}$ & $\log g$      & $[$M/H$]$  & $\xi_{\rm t}$ & $v\sin i$ & \\
BJD +245  & RV      &                &               &            &               &           & CD \\
          & (\kms)  & (K)            & (cm\,s$^{-2}$)& (solar)    & (\kms )       & (\kms)    & \\
\noalign{\smallskip}\hline\noalign{\smallskip}
\multicolumn{2}{l}{HD 180005} & 6055$\pm$77 & 2.85$\pm$0.21 & -0.88$\pm$0.12 & 1.43$\pm$0.41 & 117$\pm$10 & \\
7895.9277 & -6.54$\pm$0.03 & 5795$\pm$184 & 1.20$\pm$0.82 & -1.13$\pm$0.35 & 1.17$\pm$0.79 & 114$\pm$44  & 3,5 \\
7900.9337 & -0.10$\pm$0.03 & 5800$\pm$181 & 2.40$\pm$0.65 & -1.14$\pm$0.26 & 1.46$\pm$0.95 & 116$\pm$27  & 3,6 \\
\hline
\multicolumn{2}{l}{HD 180712} & 5765$\pm$40 & 4.26$\pm$0.07 & -0.30$\pm$0.03 & 1.16$\pm$0.07 & 2.92$\pm$0.23 & \\
7894.9320 & -13.20$\pm$0.01 & 5754$\pm$107 & 4.24$\pm$0.19 & -0.32$\pm$0.07 & 1.23$\pm$0.19 & 2.87$\pm$0.64  & 3,5 \\
\hline
\end{tabular}
\end{flushleft}
\end{table*}

\setcounter{table}{0}
\begin{table*}[!tbh]
\caption{Continued.}
\begin{flushleft}
\begin{tabular}{llllllll}
\hline\hline\noalign{\smallskip}
Star      &         &  $T_{\rm eff}$ & $\log g$      & $[$M/H$]$  & $\xi_{\rm t}$ & $v\sin i$ & \\
BJD +245  & RV      &                &               &            &               &           & CD \\
          & (\kms)  & (K)            & (cm\,s$^{-2}$)& (solar)    & (\kms )       & (\kms)    & \\
\noalign{\smallskip}\hline\noalign{\smallskip}
\multicolumn{2}{l}{HR 6180} & 4330$\pm$25 & 1.99$\pm$0.12 & -0.03$\pm$0.04 & 1.32$\pm$0.06 & 3.58$\pm$0.31 \\
8491.7781 & -35.57$\pm$0.03 & 4313$\pm$24 & 1.98$\pm$0.11 & -0.06$\pm$0.03 & 1.33$\pm$0.06 & 3.75$\pm$0.30 \\
8225.4540 & -35.42$\pm$0.03 & 4321$\pm$24 & 1.99$\pm$0.11 & -0.04$\pm$0.03 & 1.32$\pm$0.05 & 3.32$\pm$0.32 \\
\hline
\multicolumn{2}{l}{HD 145310} & 4480$\pm$27 & 2.20$\pm$0.12 & 0.10$\pm$0.04 & 1.29$\pm$0.06 & 2.74$\pm$0.36 \\
8225.5510 & -35.48$\pm$0.03 & 4476$\pm$27 & 2.20$\pm$0.11 & 0.09$\pm$0.04 & 1.30$\pm$0.05 & 2.75$\pm$0.36 \\
\hline
\end{tabular}
\end{flushleft}
\end{table*}

\begin{table}[!tbh]
\caption{Surface velocities from $i$SVD line profiles: $n_{\rm lines}$ number of spectral lines used; $v\sin i$ projected rotational velocity; $\zeta_{\rm RT}$ radial-tangential macro turbulence; $BIS=RV_{\rm top}-RV_{\rm bottom}$ bisector velocity span.}\label{table-A2}
\begin{flushleft}
\begin{tabular}{lllll}
\hline\hline\noalign{\smallskip}
Star  & $n_{\rm lines}$ & $v\sin i$ & $\zeta_{\rm RT}$ & $BIS$  \\
      &  (--)   & \multicolumn{2}{c}{(\kms)} & (\ms) \\
\noalign{\smallskip}\hline\noalign{\smallskip}
BY Dra a   & 193 & 3.39$\pm$0.57 & 3.82$\pm$0.59 & +71$\pm$31 \\
HD 135119  & 843 & 92.8$\pm$1.6 & 3.7$\pm$1.6 & \dots  \\
HD 135143  & 193 & 1.10$\pm$0.78 & 2.90$\pm$0.81 &  +43$\pm$27 \\
HD 138916  & 843 & 101$\pm$2.3 & 2.5$\pm$0.9 & \dots  \\
HD 139797  & 193 & 1.49$\pm$1.01 & 3.03$\pm$1.1 & --559$\pm$189  \\
HD 140341  & 843 & 86.4$\pm$1.4  & 5.1$\pm$0.6 & \dots  \\
\hline
\end{tabular}
\end{flushleft}
\end{table}

\begin{table}[ht!]
\caption{Masses and ages:$M_\star$ stellar mass; $\log t$ age; \emph{} "label"\ evolutionary status).} \label{table-A3}
\begin{flushleft}
\begin{tabular}{lllll}
\hline\hline
\noalign{\smallskip}
Star       &   $M_\star$   &    $\log t$  &  label & flag \\
           &   (M$_\odot$) &    (yr)      &        & \\
\noalign{\smallskip}\hline\noalign{\smallskip}
BY Dra &                       \dots &                       \dots &        \dots &        SB2 \\
\\ [-0.8em]
HD 135119 &                       \dots &                       \dots &        \dots &        v50 \\ \\ [-0.8em]
HD 135143 &    $0.90^{+0.06}_{-0.09}$ &   $10.07^{+0.24}_{-0.18}$ &         MS &        \dots \\
\\
\hline
\end{tabular}
\tablefoot{Errors are from mitigated observational errors. The notations in the label and flag columns are explained in the main text in Table\,\ref{isofit_results_flag_notation}.}
\end{flushleft}
\end{table}

\setcounter{table}{2}
\begin{table}[ht!]
\caption{Continued.}
\begin{flushleft}
\begin{tabular}{lllll}
\hline\hline
\noalign{\smallskip}
Star       &   $M_\star$   &    $\log t$  &  \emph{label} & flag \\
           &   (M$_\odot$) &    (yr)      &        & \\
\noalign{\smallskip}\hline\noalign{\smallskip}
HD 148978 &    $1.15^{+0.54}_{-0.60}$ &    $9.79^{+0.72}_{-0.78}$ &       CHeB &        \dots \\
\\ [-0.8em]
HD 149843 &    $2.26^{+1.29}_{-0.57}$ &    $8.94^{+0.30}_{-0.81}$ &   RGB/CHeB &        \dots \\
\\ [-0.8em]
HD 137292 &    $1.68^{+0.63}_{-1.32}$ &    $9.39^{+0.96}_{-0.48}$ &       CHeB &        \dots \\
\\ [-0.8em]
HD 138852 &    $1.25^{+0.57}_{-1.20}$ &    $9.65^{+1.20}_{-0.69}$ &       CHeB &        \dots \\
\\ [-0.8em]
HD 145742 &    $1.88^{+1.71}_{-1.80}$ &    $9.17^{+1.20}_{-1.41}$ &   RGB/CHeB &        \dots \\
\\ [-0.8em]
HR 5844 &    $2.46^{+0.57}_{-0.27}$ &    $8.91^{+0.15}_{-0.27}$ &        AGB &        \dots \\
\\ [-0.8em]
HD 138301 &    $1.51^{+0.45}_{-0.54}$ &    $9.39^{+0.45}_{-0.39}$ &        RGB &        \dots \\
\\ [-0.8em]
$\varepsilon$ Dra  &    $2.24^{+0.81}_{-0.60}$ &    $8.86^{+0.30}_{-0.45}$ &   RGB/CHeB &        \dots \\
\\ [-0.8em]
HD 194298 &    $1.86^{+0.54}_{-0.39}$ &    $9.17^{+0.24}_{-0.27}$ &    RGB/AGB &        \dots \\
\\ [-0.8em]
\noalign{\smallskip}
\hline
\end{tabular}
\end{flushleft}
\end{table}

\begin{table}[!tbh]
\caption{Spectral line list:\ ion, $\lambda$ central wavelength, $\chi$ excitation potential, $\log gf$ transition probability, EW equivalent widths measured from the PEPSI solar spectrum ($D$), and the SES solar spectrum ($G$).} \label{table-A4}
\begin{flushleft}
\begin{tabular}{lllrll}
\hline\hline
\noalign{\smallskip}
Ion    & $\lambda$    &  $\chi$   & $\log gf $ & EW$_\odot, D$ &  EW$_\odot, G$ \\
       & (\AA )       &  (eV) &   (--)         & (m\AA)     &  (m\AA)    \\
\noalign{\smallskip}\hline\noalign{\smallskip}
  \ion{C}{i}  &      5052.149     &      7.685    &      -1.304   &       37.2       &       \dots  \\
  \ion{C}{i}  &      5380.325     &      7.685    &      -1.615   &       20.8       &       21.4     \\
  \ion{C}{i}  &      7113.180     &      8.647    &      -0.773   &       22.6       &       \dots  \\
  \ion{C}{i}  &      7116.990     &      8.647    &      -0.907   &       18.2       &       \dots  \\
  \ion{O}{i}  &      7771.944     &      9.146    &       0.369   &       68.7       &       68.2     \\
  \ion{O}{i}  &      7774.166     &      9.146    &       0.223   &       58.7       &       58.3      \\
\hline
\end{tabular}
\end{flushleft}
\end{table}

\setcounter{table}{3}
\begin{table}[!tbh]
\caption{Continued.}
\begin{flushleft}
\begin{tabular}{lllrll}
\hline\hline\noalign{\smallskip}
Ion    & $\lambda$    &  $\chi$   & $\log gf $ & EW$_\odot, D$ &  EW$_\odot, G$ \\
       & (\AA )       &  (eV) &   (--)         & (m\AA)     &  (m\AA)    \\
\noalign{\smallskip}\hline\noalign{\smallskip}
  \ion{Ca}{i} &      5260.387     &      2.521    &      -1.719   &       32.5       &       33.6      \\
  \ion{Ca}{i} &      5261.704     &      2.521    &      -0.579   &       99.2       &       99.5      \\
  \ion{Ca}{i} &      5349.465     &      2.709    &      -0.310   &      101.8      &       100.0     \\
  \ion{Ca}{i} &      5512.980     &      2.933    &      -0.464   &      \dots   &       87.4      \\
  \ion{Ca}{i} &      5581.965     &      2.523    &      -0.555   &      \dots   &       98.3      \\
  \ion{Ca}{i} &      5588.749     &      2.526    &       0.358   &      \dots   &       \dots       \\
\hline
\end{tabular}
\end{flushleft}
\end{table}

\setcounter{table}{3}
\begin{table}[!tbh]
\caption{Continued.}
\begin{flushleft}
\begin{tabular}{lllrll}
\hline\hline\noalign{\smallskip}
Ion    & $\lambda$    &  $\chi$   & $\log gf $ & EW$_\odot, D$ &  EW$_\odot, G$ \\
       & (\AA )       &  (eV) &   (--)         & (m\AA)     &  (m\AA)    \\
\noalign{\smallskip}\hline\noalign{\smallskip}
  \ion{Ti}{ii} &     4764.525     &      1.237    &      -2.690   &       32.3       &       \dots  \\
  \ion{Ti}{ii} &     4798.531     &      1.079    &      -2.660   &       44.2       &       \dots  \\
  \ion{Ti}{ii} &     4806.323     &      1.084    &      -3.380   &       14.5       &       \dots  \\
  \ion{Ti}{ii} &     4865.612     &      1.116    &      -2.700   &       40.9       &       41.2     \\
  \ion{Ti}{ii} &     4874.009     &      3.095    &      -0.860   &       36.0       &       \dots  \\
  \ion{Ti}{ii} &     4911.195     &      3.124    &      -0.640   &       50.9       &       \dots  \\
\hline
\end{tabular}
\end{flushleft}
\end{table}

\setcounter{table}{3}
\begin{table}[!tbh]
\caption{Continued.}
\begin{flushleft}
\begin{tabular}{lllrll}
\hline\hline\noalign{\smallskip}
Ion    & $\lambda$    &  $\chi$   & $\log gf $ & EW$_\odot, D$ &  EW$_\odot, G$ \\
       & (\AA )       &  (eV) &   (--)         & (m\AA)     &  (m\AA)    \\
\noalign{\smallskip}\hline\noalign{\smallskip}
  \ion{Mn}{i} &      5420.351     &      2.143    &      -1.462   &      \dots   &       86.0      \\
  \ion{Mn}{i} &      5432.539     &      0.000    &      -3.795   &      \dots   &       50.4      \\
  \ion{Mn}{i} &      5516.766     &      2.178    &      -1.847   &      \dots   &       42.4      \\
  \ion{Mn}{i} &      6013.510     &      3.072    &      -0.354   &      \dots   &       83.9      \\
  \ion{Mn}{i} &      6016.670     &      3.073    &      -0.183   &      \dots   &       93.1      \\
  \ion{Mn}{i} &      6021.820     &      3.075    &      -0.054   &      \dots   &       93.1      \\
\hline
\end{tabular}
\end{flushleft}
\end{table}

\setcounter{table}{3}
\begin{table}[!tbh]
\caption{Continued.}
\begin{flushleft}
\begin{tabular}{lllrll}
\hline\hline\noalign{\smallskip}
Ion    & $\lambda$    &  $\chi$   & $\log gf $ & EW$_\odot, D$ &  EW$_\odot, G$ \\
       & (\AA )       &  (eV) &   (--)         & (m\AA)     &  (m\AA)    \\
\noalign{\smallskip}\hline\noalign{\smallskip}
  \ion{Fe}{i} &      6027.051     &      4.076    &      -1.089   &      \dots   &       62.5      \\
  \ion{Fe}{i} &      6065.482     &      2.609    &      -1.529   &      \dots   &       \dots       \\
  \ion{Fe}{i} &      6082.710     &      2.223    &      -3.576   &      \dots   &       \dots       \\
  \ion{Fe}{i} &      6120.246     &      0.915    &      -5.970   &      \dots   &       \dots       \\
  \ion{Fe}{i} &      6127.906     &      4.143    &      -1.399   &      \dots   &       49.2      \\
  \ion{Fe}{i} &      6151.617     &      2.176    &      -3.295   &      \dots   &       48.3      \\
\hline
\end{tabular}
\end{flushleft}
\end{table}

\setcounter{table}{3}
\begin{table}[!tbh]
\caption{Continued.}
\begin{flushleft}
\begin{tabular}{lllrll}
\hline\hline\noalign{\smallskip}
Ion    & $\lambda$    &  $\chi$   & $\log gf $ & EW$_\odot, D$ &  EW$_\odot, G$ \\
       & (\AA )       &  (eV) &   (--)         & (m\AA)     &  (m\AA)    \\
\noalign{\smallskip}\hline\noalign{\smallskip}
  \ion{Co}{i} &      4813.476     &      3.216    &       0.120   &       44.2       &       \dots  \\
  \ion{Co}{i} &      5230.209     &      1.740    &      -1.840   &       28.1       &       \dots  \\
  \ion{Co}{i} &      5247.920     &      1.785    &      -2.070   &       18.0       &       \dots  \\
  \ion{Co}{i} &      5331.452     &      1.785    &      -1.997   &      \dots   &       18.3      \\
  \ion{Co}{i} &      5352.051     &      3.576    &      -0.503   &       23.9       &       24.1     \\
  \ion{Co}{i} &      5530.775     &      1.710    &      -2.060   &      \dots   &       \dots       \\
\hline
\end{tabular}
\end{flushleft}
\end{table}

\setcounter{table}{3}
\begin{table}[!tbh]
\caption{Continued.}
\begin{flushleft}
\begin{tabular}{lllrll}
\hline\hline\noalign{\smallskip}
Ion    & $\lambda$    &  $\chi$   & $\log gf $ & EW$_\odot, D$ &  EW$_\odot, G$ \\
       & (\AA )       &  (eV) &   (--)         & (m\AA)     &  (m\AA)    \\
\noalign{\smallskip}\hline\noalign{\smallskip}
  \ion{Y}{ii} &      4854.861     &      0.992    &      -0.270   &       55.4       &       \dots  \\
  \ion{Y}{ii} &      4883.682     &      1.084    &       0.190   &       57.9       &       57.9     \\
  \ion{Y}{ii} &      5087.416     &      1.084    &      -0.160   &       45.5       &       44.3      \\
  \ion{Y}{ii} &      5119.112     &      0.992    &      -1.330   &       14.6       &       \dots  \\
  \ion{Y}{ii} &      5123.211     &      0.992    &      -0.790   &       43.5       &       \dots  \\
  \ion{Y}{ii} &      5200.406     &      0.992    &      -0.470   &       36.2       &       39.0     \\
\hline
\end{tabular}
\end{flushleft}
\end{table}

\begin{table*}[!tbh]
\caption{LTE chemical abundances A(X) of the pilot stars.} \label{table-A5}
\begin{flushleft}
\begin{tabular}{lllllllll}
\hline\hline
\noalign{\smallskip}
Star   & C  &  N   &  O  &  Na  &  Mg  &  Al  & Si  &  S  \\
\noalign{\smallskip}\hline\noalign{\smallskip}
BY Dra          &       \dots                   &       \dots                   &       \dots                   &       \dots                   &       \dots                    &       \dots                   &       \dots                   &       \dots    \\
HD 135119       &       \dots                   &       \dots                   &       \dots                   &       \dots                   &       \dots                    &       \dots                   &       \dots                   &       \dots    \\
HD 135143       &       \dots                   &       \dots                   &       \dots                   &       6.03    $\pm$   0.06    &        7.54    $\pm$   0.05    &       6.35    $\pm$   0.03    &        7.62    $\pm$   0.03    &       7.50    $\pm$   0.08\\
HD 138916       &       \dots                   &       \dots                   &       \dots                   &       \dots                   &       \dots                    &       \dots                   &       \dots                   &       \dots            \\
HD 139797       &       8.21    $\pm$   0.15    &       \dots                   &       \dots                   &       6.17    $\pm$   0.04     &       7.57    $\pm$   0.06    &       6.36    $\pm$    0.04    &      7.52    $\pm$   0.03    &       7.31    $\pm$    0.09\\
HD 140341       &       \dots                   &       \dots                   &       \dots                   &       \dots                   &       \dots                    &       \dots                   &       \dots                   &       \dots            \\
\hline
\end{tabular}
\end{flushleft}
\end{table*}

\setcounter{table}{4}
\begin{table*}[!tbh]
\caption{Continued.}
\begin{flushleft}
\begin{tabular}{lllllllll}
\hline\hline
\noalign{\smallskip}
Star  & K       &       Ca      &       \ion{Sc}{i}     &       \ion{Sc}{ii}    &       \ion{Ti}{i}     &       \ion{Ti}{ii}    &       V       &       Cr       \\
\noalign{\smallskip}\hline\noalign{\smallskip}
BY Dra          &       \dots                   &       \dots                   &       \dots                   &        \dots                   &       \dots                    &      \dots                   &       \dots                   &       \dots    \\
HD 135119       &       \dots                   &       \dots                   &       \dots                   &       \dots                   &       \dots                    &       \dots                   &       \dots                   &       \dots    \\
HD 135143       &       5.16    $\pm$   0.05    &       6.24    $\pm$   0.07    &       \dots                   &       3.15    $\pm$    0.05    &        4.83   $\pm$   0.05    &       4.90    $\pm$    0.04   &       4.03    $\pm$   0.09    &       5.59     $\pm$   0.04\\
HD 138916       &       \dots                   &       \dots                   &       \dots                   &        \dots                   &       \dots                    &      \dots                   &       \dots                   &       \dots    \\
HD 139797       &       5.29    $\pm$   0.06    &       6.22    $\pm$   0.06    &       \dots                   &       2.97     $\pm$    0.06   &       4.72    $\pm$   0.06    &       4.82    $\pm$    0.04    &       3.63    $\pm$   0.06    &       5.51     $\pm$  0.05\\
HD 140341       &       \dots                   &       \dots                   &       \dots                   &        \dots                   &       \dots                    &      \dots                   &       \dots                   &       \dots    \\
\hline
\end{tabular}
\end{flushleft}
\end{table*}

\setcounter{table}{4}
\begin{table*}[!tbh]
\caption{Continued.}
\begin{flushleft}
\begin{tabular}{lllllllll}
\hline\hline
\noalign{\smallskip}
Star  & Mn      &       \ion{Fe}{i}     &       \ion{Fe}{ii}    &       Co      &       Ni      &       Cu      &       Zn      &       Y       \\
\noalign{\smallskip}\hline\noalign{\smallskip}
BY Dra          &       \dots                   &       \dots                   &       \dots                   &       \dots                   &       \dots                    &       \dots                   &       \dots                   &        \dots            \\
HD 135119       &       \dots                   &       \dots                   &       \dots                   &       \dots                   &       \dots                    &       \dots                   &       \dots                   &       \dots            \\
HD 135143       &       5.66    $\pm$   0.05    &        7.40   $\pm$   0.04    &       7.48    $\pm$   0.04    &       4.89     $\pm$   0.06    &       6.20    $\pm$   0.04    &       4.18     $\pm$  0.07    &        4.52    $\pm$   0.02    &        2.27   $\pm$   0.06\\
HD 138916       &       \dots                   &       \dots                   &       \dots                   &       \dots                   &       \dots                    &       \dots                   &       \dots                   &        \dots            \\
HD 139797       &       5.53     $\pm$  0.06    &       7.30    $\pm$   0.05    &       7.27    $\pm$   0.04    &       4.84     $\pm$   0.07    &       6.07    $\pm$   0.05    &       4.05     $\pm$  0.07    &       4.58    $\pm$   0.05    &        2.05    $\pm$   0.08\\
HD 140341       &       \dots                   &       \dots                   &       \dots                   &       \dots                   &       \dots                    &       \dots                   &       \dots                   &        \dots            \\
\hline
\end{tabular}
\end{flushleft}
\end{table*}

\setcounter{table}{4}
\begin{table*}[!tbh]
\caption{Continued.}
\begin{flushleft}
\begin{tabular}{llllllll}
\hline\hline
\noalign{\smallskip}
Star  & \ion{Zr}{i}     &       \ion{Zr}{ii}    &       Ba      &       La      &       Ce      &       Eu      &       [$\alpha$/Fe] \\
\noalign{\smallskip}\hline\noalign{\smallskip}
BY Dra          &       \dots   &       \dots   &       \dots   &       \dots   &       \dots   &       \dots   &       \dots   \\
HD 135119       &       \dots   &       \dots   &       \dots   &       \dots   &       \dots   &       \dots   &       \dots   \\
HD 135143       &       \dots   &       2.59    $\pm$   0.03    &       2.00    $\pm$   0.04    &       \dots   &       1.73     $\pm$   0.03    &       \dots   &       0.03    $\pm$   0.04     \\
HD 138916       &       \dots   &       \dots   &       \dots   &       \dots   &       \dots   &       \dots   &       \dots   \\
HD 139797       &       \dots   &       2.72    $\pm$   0.03    &       2.09    $\pm$   0.05    &        \dots   &       \dots   &        0.41   $\pm$   0.03    &       0.09    $\pm$   0.04     \\
HD 140341       &       \dots   &       \dots   &       \dots   &       \dots   &       \dots   &       \dots   &       \dots   \\
HD 142006       &       \dots   &       2.53    $\pm$   0.15    &       1.85     $\pm$   0.05    &       \dots   &       \dots   &        \dots  &       0.09    $\pm$   0.04    \\
\hline
\end{tabular}
\end{flushleft}
\end{table*}

\begin{table}[!tbh]
\caption{Oxygen and potassium NLTE abundances (A(O) logarithmic O abundance from the NIR triplet; error for A(O); A(K) logarithmic K abundance from the 7699\,\AA\ line; error for A(K)). } \label{table-A6}
\begin{flushleft}
\begin{tabular}{lllll}
\hline\hline
\noalign{\smallskip}
Star       & A(\ion{O}{i} NIR)   &  error &  A(\ion{K}{i} 7699) &  error  \\
\noalign{\smallskip}\hline\noalign{\smallskip}
BY Dra     & \dots  & \dots & \dots  & \dots \\
HD 135119  & \dots  & \dots & \dots  & \dots \\
HD 135143  & 8.76 & 0.05 & 4.76 & 0.05 \\
HD 138916  & \dots  & \dots & \dots  & \dots \\
HD 139797  & 8.67 & 0.05 & 5.03 & 0.06 \\
HD 140341  & \dots  & \dots & \dots  & \dots \\
HD 142006  & 8.42 & 0.05 & 4.84 & 0.16 \\
\hline
\end{tabular}
\tablefoot{$^1$Effective temperature and/or micro turbulence is outside the available NLTE correction grid.}
\end{flushleft}
\end{table}

\begin{table}[!t]
\caption{Li abundance and Li and C isotope ratios:\ Li logarithmic 1D LTE Li abundance; $^6$Li/$^7$Li lithium isotope ratio;  $\Delta_{\rm Li}$ Li 3D NLTE correction; and $^{12}$C/$^{13}$C carbon isotope ratio.}\label{table-A7}
\begin{flushleft}
\begin{tabular}{lllll}
\hline\hline
\noalign{\smallskip}
Star     & A(Li)         & $^6$Li/$^7$Li & $\Delta_{\rm Li}$ & $^{12}$C/$^{13}$C \\
\noalign{\smallskip}\hline\noalign{\smallskip}
Sun      & +1.03$\pm$0.01 & 0.00  &0.095 & 110$\pm$30 \\
BY Dra$^1$   & --0.16$\pm$0.16& \dots  & \dots & \dots  \\
HD 135119$^2$& --0.03$\pm$0.05& \dots & \dots & \dots \\
HD 135143& +2.26$\pm$0.01 & 0.00  & 0.083 & \dots \\
HD 138916$^2$ & --0.19$\pm$0.43  & \dots & \dots  & \dots \\
HD 139797& +1.72$\pm$0.02 & 0.00  & 0.065 & \dots \\
HD 140341$^2$ & +2.25$\pm$0.18  & \dots & \dots & \dots \\
\hline
\end{tabular}
\tablefoot{$\Delta_{\rm Li}$ is the 3D NLTE correction according to Mott et al. (\cite{mott20}) to be added to A(Li) for a 3D NLTE Li abundance. $^1$Target is SB2, value is an estimate for the primary. $^2$Spectral lines are very broad.}
\end{flushleft}
\end{table}

\begin{table*}[!tbh]
\caption{Absolute emission line fluxes ($W$(\Hbeta) \Hbeta\ equivalent width; $W$(\Hbeta)/$W$(\Halpha) ratio of equivalent widths; ${\cal F}_{{\rm H}\alpha}$ absolute emission-line flux in \Halpha ; ${\cal F}_{\rm IRT}$ absolute emission-line fluxes for the three IRT lines IRT-1 (8498\,\AA ), IRT-2 (8542\,\AA ), and IRT-3 (8662\,\AA ), respectively; $\log R_{\rm IRT}$ logarithmic total radiative loss in the \ion{Ca}{ii} infrared triplet lines).} \label{table-A8}
\begin{flushleft}
\begin{tabular}{lllllllll}
\hline\hline
\noalign{\smallskip}
Star   & $(B-V)_0$   &  $W$(\Hbeta)   &  $W$(\Hbeta)/$W$(\Halpha)   & ${\cal F}_{{\rm H}\alpha}$  & ${\cal F}_{\rm IRT-1}$   &  ${\cal F}_{\rm IRT-2}$ &  ${\cal F}_{\rm IRT-3}$  & $\log R_{\rm IRT}$ \\
       & (mag)       &  (m\AA ) &   (--)   & \multicolumn{4}{c}{($10^5$\,erg\,cm$^{-2}$s$^{-1}$)} & ($\sigma T^4_{\rm eff}$) \\
\noalign{\smallskip}\hline\noalign{\smallskip}
BY\,Dra         & 1.40        & 769$\pm$32        & 0.64  &    \dots        &     \dots      &    \dots       & \dots       & \dots     \\
HD\,135119      & 0.33        & 431$\pm$4         & 0.91  &    35.5$\pm$8.9   &   40.8$\pm$7.9   &    27.9$\pm$6.4  & 30.1$\pm$6.5  & $-$4.16        \\
HD\,135143      & 0.66        & 324$\pm$6         & 1.30  &    16.5$\pm$2.7   &   21.5$\pm$3.0   &    15.1$\pm$2.2  & 15.8$\pm$2.2  & $-$4.06        \\
HD\,138916      & 0.36        & 554$\pm$2         & 1.07  &    42.3$\pm$8.8   &   42.8$\pm$7.9   &    30.4$\pm$7.0  & 34.7$\pm$6.9  & $-$4.07        \\
HD\,139797      & 0.65        & 318$\pm$10        & 1.33  &    16.1$\pm$2.5   &   20.7$\pm$3.2   &    13.7$\pm$2.0  & 14.6$\pm$2.1  & $-$4.09        \\
HD\,140341      & 0.43        & 485$\pm$16        & 1.06  &    27.7$\pm$6.6   &   33.2$\pm$6.5   &    21.1$\pm$5.0  & 24.4$\pm$5.4  & $-$4.13        \\
HD\,142006      & 0.53        & 289$\pm$6         & 1.21  &    21.8$\pm$3.5   &   27.4$\pm$3.8   &    19.1$\pm$2.8  & 19.9$\pm$2.8  & $-$4.09        \\
HD\,142089      & 0.54        & 294$\pm$10        & 1.21  &    21.0$\pm$3.1   &   25.9$\pm$3.8   &    17.5$\pm$2.6  & 18.5$\pm$2.8  & $-$4.11        \\
HD\,144061      & 0.69        & 324$\pm$8         & 1.27  &    15.9$\pm$2.8   &   21.5$\pm$3.2   &    15.6$\pm$2.4  & 16.0$\pm$2.4  & $-$4.02        \\
HD\,145710      & 0.43        & 443$\pm$20        & 1.14  &    15.1$\pm$5.9   &   27.4$\pm$6.4   &    16.2$\pm$4.9  & 16.7$\pm$5.3  & $-$4.25        \\
HD\,150826      & 0.51        & 312$\pm$13        & 1.07  &    21.0$\pm$3.4   &   26.7$\pm$4.3   &    17.1$\pm$3.3  & 18.8$\pm$3.2  & $-$4.13        \\
HD\,155859      & 0.61        & 282$\pm$7         & 1.25  &    16.9$\pm$3.1   &   22.5$\pm$3.2   &    15.0$\pm$2.1  & 15.9$\pm$2.3  & $-$4.09        \\
HD\,160052      & 0.42        & 318$\pm$4         & 1.24  &    30.7$\pm$5.3   &   39.6$\pm$5.5   &    32.0$\pm$5.1  & 32.6$\pm$5.2  & $-$4.02        \\
HD\,160076      & 0.35        & 382$\pm$6         & 1.11  &    29.3$\pm$6.6   &   35.3$\pm$6.3   &    20.6$\pm$5.1  & 23.0$\pm$5.1  & $-$4.23        \\
HD\,160605      & 1.19        & 344$\pm$26        & 1.40  &    4.4$\pm$0.9    &   7.0$\pm$1.0    &    4.1$\pm$0.6   & 4.4$\pm$0.7   & $-$4.18       \\
HD\,161897      & 0.81        & 343$\pm$6         & 1.29  &    12.3$\pm$2.2   &   17.5$\pm$2.4   &    12.8$\pm$1.8  & 12.9$\pm$1.8  & $-$4.01        \\
HD\,162524      & 0.55        & 358$\pm$14        & 1.39  &    19.8$\pm$3.2   &   25.8$\pm$4.0   &    17.1$\pm$2.7  & 18.6$\pm$2.9  & $-$4.10        \\
HD\,165700      & 0.51        & 215$\pm$2         & 0.71  &    \dots        &   \dots        &    \dots       & \dots       & \dots      \\
HD\,175225      & 0.95        & 329$\pm$11        & 1.21  &    8.8$\pm$1.5    &   11.5$\pm$1.6   &    7.4$\pm$1.1   & 7.6$\pm$1.1   & $-$4.11       \\
HD\,176841      & 0.75        & 307$\pm$9         & 1.28  &    12.8$\pm$2.3   &   17.1$\pm$2.4   &    11.6$\pm$1.7  & 12.1$\pm$1.8  & $-$4.08        \\
HD\,180005      & 0.36        & 564$\pm$9         & 1.02  &    47.4$\pm$9.3   &   44.3$\pm$8.1   &    33.8$\pm$7.0  & 35.4$\pm$6.9  & $-$4.06        \\
HD\,180712      & 0.68        & 325$\pm$9         & 1.27  &    16.2$\pm$2.6   &   21.9$\pm$3.2   &    15.8$\pm$2.3  & 15.8$\pm$2.3  & $-$4.03        \\
HD\,192438      & 0.40        & 358$\pm$7         & 0.94  &    19.6$\pm$6.0   &   30.9$\pm$6.3   &    19.1$\pm$5.2  & 21.3$\pm$5.0  & $-$4.21        \\
HD\,199019      & 0.87        & 392$\pm$2         & 1.03  &    14.9$\pm$2.1   &   19.4$\pm$2.8   &    16.5$\pm$2.5  & 15.8$\pm$2.6  & $-$3.88        \\
VX\,UMi         & 0.31        & 641$\pm$16        & 1.06  &    62.4$\pm$11.9  &   50.1$\pm$9.2   &    41.2$\pm$8.1  & 42.5$\pm$8.1  & $-$4.05         \\
$\psi^{01}$\,Dra\,A     & 0.46& 244$\pm$7         & 0.96  &    26.2$\pm$3.7   &   29.7$\pm$4.2   &    19.5$\pm$2.9  & 21.0$\pm$3.1  & $-$4.15        \\
$\psi^{01}$\,Dra\,B     & 0.56& 270$\pm$5         & 1.16  &    19.7$\pm$3.3   &   24.8$\pm$3.4   &    17.0$\pm$2.4  & 18.0$\pm$ 2.5 & $-$4.12        \\
                        &                 &                   &       &                   &                  &                  &                    &    \\
11\,UMi         & 1.63        & 252$\pm$27        & 1.16  &    1.3$\pm$0.3    &   2.2$\pm$0.4    &     1.1$\pm$0.2  & 1.2$\pm$0.3   & $-$4.36       \\
HD\,136919      & 1.20        & 308$\pm$18        & 1.31  &    4.1$\pm$0.8    &   6.6$\pm$0.9    &     3.8$\pm$0.6  & 4.0$\pm$0.6   & $-$4.21       \\
HD\,138020      & 1.62        & 199$\pm$34        & 1.02  &    1.2$\pm$0.2    &   2.5$\pm$0.4    &     1.3$\pm$0.2  & 1.4$\pm$0.3   & $-$4.33       \\
20\,UMi         & 1.55        & 255$\pm$26        & 1.17  &    1.6$\pm$0.3    &   2.7$\pm$0.5    &     1.4$\pm$0.3  & 1.5$\pm$0.3   & $-$4.36       \\
HD\,138116      & 1.23        & 301$\pm$19        & 1.28  &    3.7$\pm$0.7    &   6.0$\pm$0.0    &     3.5$\pm$0.5  & 3.6$\pm$0.6   & $-$4.22       \\
HD\,143641      & 1.22        & 310$\pm$17        & 1.31  &    3.9$\pm$0.8    &   6.4$\pm$0.0    &     3.8$\pm$0.6  & 4.0$\pm$0.6   & $-$4.20       \\
HD\,142961      & 1.08        & 269$\pm$19        & 1.23  &    5.1$\pm$0.9    &   7.3$\pm$1.1    &     4.4$\pm$0.7  & 4.7$\pm$0.8   & $-$4.23       \\
HD\,150275      & 1.14        & 296$\pm$22        & 1.29  &    4.6$\pm$0.8    &   7.5$\pm$1.1    &     4.3$\pm$0.7  & 4.7$\pm$0.7   & $-$4.18       \\
HD\,151698      & 1.22        & 290$\pm$24        & 1.27  &    3.7$\pm$0.7    &   6.1$\pm$0.9    &     3.4$\pm$0.5  & 3.7$\pm$0.6   & $-$4.23       \\
HD\,155153      & 1.15        & 272$\pm$18        & 1.23  &    4.3$\pm$0.8    &   5.9$\pm$0.9    &     3.4$\pm$0.6  & 3.6$\pm$0.6   & $-$4.28       \\
HR\,6069        & 1.16        & 269$\pm$22        & 1.25  &    4.2$\pm$0.8    &   6.6$\pm$1.0    &     3.8$\pm$0.6  & 4.0$\pm$0.6   & $-$4.23       \\
HR\,6180        & 1.60        & 297$\pm$23        & 1.25  &    1.5$\pm$0.3    &   2.7$\pm$0.4    &     1.4$\pm$0.3  & 1.5$\pm$0.3   & $-$4.31       \\
HD\,145310      & 1.47        & 291$\pm$24        & 1.23  &    2.1$\pm$0.4    &   3.4$\pm$0.5    &     1.8$\pm$0.3  & 1.9$\pm$0.4   & $-$4.33       \\
HD\,144903      & 1.38        & 272$\pm$24        & 1.21  &    2.5$\pm$0.5    &   3.9$\pm$0.6    &     2.1$\pm$0.4  & 2.3$\pm$0.4   & $-$4.32       \\
HD\,148374      & 1.11        & 276$\pm$19        & 1.19  &    5.0$\pm$0.9    &   6.9$\pm$1.0    &     4.0$\pm$0.6  & 4.3$\pm$0.7   & $-$4.24       \\
HD\,150142      & 1.64        & 242$\pm$29        & 1.12  &    1.2$\pm$0.2    &   2.2$\pm$0.4    &     1.1$\pm$0.2  & 1.2$\pm$0.3   & $-$4.36       \\
HD\,147764      & 1.20        & 289$\pm$20        & 1.24  &    4.0$\pm$0.8    &   5.8$\pm$0.9    &     3.2$\pm$0.5  & 3.5$\pm$0.6   & $-$4.26       \\
$\upsilon$\,Dra & 1.35    & 272$\pm$23        & 1.22  &    2.7$\pm$0.5    &   4.1$\pm$0.6    &     2.2$\pm$0.4  & 2.3$\pm$0.4   & $-$4.33         \\
HD\,148978      & 1.20        & 281$\pm$22        & 1.25  &    3.9$\pm$0.7    &   6.1$\pm$0.9    &     3.4$\pm$0.5  & 3.6$\pm$0.6   & $-$4.24       \\
HD\,149843      & 1.21        & 290$\pm$17        & 1.23  &    4.0$\pm$0.8    &   5.7$\pm$0.8    &     3.2$\pm$0.5  & 3.4$\pm$0.5   & $-$4.27       \\
HD\,137292      & 1.37        & 307$\pm$18        & 1.23  &    2.8$\pm$0.5    &   4.0$\pm$0.6    &     2.2$\pm$0.4  & 2.3$\pm$0.4   & $-$4.31       \\
HD\,138852      & 1.13        & 274$\pm$21        & 1.24  &    4.5$\pm$0.8    &   7.0$\pm$1.0    &     4.0$\pm$0.6  & 4.3$\pm$0.7   & $-$4.22       \\
HD\,145742      & 1.23        & 285$\pm$20        & 1.24  &    3.7$\pm$0.7    &   5.4$\pm$0.8    &     3.1$\pm$0.5  & 3.2$\pm$0.5   & $-$4.27       \\
HR\,5844        & 1.65        & 245$\pm$27        & 1.14  &    1.2$\pm$0.3    &   2.1$\pm$0.4    &     1.0$\pm$0.2  & 1.1$\pm$0.2   & $-$4.37       \\
HD\,138301      & 1.19        & 276$\pm$22        & 1.23  &    3.9$\pm$0.7    &   6.3$\pm$0.9    &     3.5$\pm$0.6  & 3.7$\pm$0.7   & $-$4.24       \\
$\varepsilon$\,Dra      & 1.06& 271$\pm$21        & 1.24  &    5.4$\pm$1.0    &   8.4$\pm$1.2    &     4.9$\pm$0.8  & 5.3$\pm$0.8   & $-$4.18       \\
HD\,194298$^{1}$        & 1.85& 197$\pm$32        & 1.05  &    \dots        &   \dots        &     \dots      & \dots       & \dots      \\
\noalign{\smallskip}
\hline
\end{tabular}
\tablefoot{$^1$Target $(B-V)_0$ is out of available calibration range.}
\end{flushleft}
\end{table*}

\section{Individual notes}

\subsection{VATT+PEPSI pilot subsample}

There are two double-lined spectroscopic binaries (SB2) in this subsample.

\noindent {\sl BY Dra (HD\,234677).} Prototype of BY-Dra spotted variables. It is SB2; \Halpha\ and Ca\,{\sc ii} IRT core emission from primary and secondary, Li absorption from primary. Strong potassium line \ion{K}{i} 7699-\AA\ from both components. Weak \ion{O}{i} triplet. \\
{\sl HD 135119.} Very broad lines. No \ion{K}{i} line detected. \\
{\sl HD 135143.} Li detected. Weak [O\,{\sc i}] 6300\,\AA\ line. \\
{\sl HD 138916.} Very broad lines. No \ion{K}{i} line detected. Large telluric contamination. \\
{\sl HD 139797.} Li detected.\\
{\sl HD 140341.} Very broad lines. No \ion{K}{i} line detected.\\
{\sl HD 142006.} Li with $^6$Li line detected. \\
{\sl HD 142089.} Li detected.\\
{\sl HD 144061.} Li detected.\\
{\sl HD 145710.} Very broad lines. \\
{\sl HD 150826.} Broad lines.\\
{\sl HD 155859.} Li with possible $^6$Li line detected.  \\
{\sl HD 160052.} Li detected.\\
{\sl HD 160076.} Broad lines. \\
{\sl HD 160605.} Weak \ion{O}{i} 7774-\AA\ triplet. $^{12}$C/$^{13}$C ratio detected from CN. The star is actually a giant.\\
{\sl HD 161897.} APOFIS finds it as a PMS star, but no Li is detected. \\
{\sl HD 162524.} Strong Li. Ca\,{\sc ii} IRT core emission. \\
{\sl HD 165700.} SB2. Li and K-line absorption from both components. \\
{\sl HD 175225.} Li detected. $^{12}$C/$^{13}$C ratio detected from CN. The star is actually a giant.\\
{\sl HD 176841.} Li detected.\\
{\sl HD 180005.} Very broad lines. No \ion{K}{i} line detected.\\
{\sl HD 180712.} Li detected.\\
{\sl HD 192438.} Very broad lines.\\
{\sl HD 199019.} Li detected. Red-shifted night-sky \ion{K}{i} line present. \\
{\sl VX UMi (HD\,155154).} Very broad lines. No \ion{K}{i} line detected.   \\
{\sl $\psi^{01}$ Dra A (HD\,162003).} Li detected. $^{12}$C/$^{13}$C ratio undetected from CN.\\
{\sl $\psi^{01}$ Dra B (HD\,162004).} Li detected. $^{12}$C/$^{13}$C ratio undetected. A Jupiter-sized planet candidate was announced by Endl et al. (\cite{endl}).\\

\subsection{STELLA+SES pilot subsample}

No SB2 in this subsample. All (giant) stars show a strong \ion{K}{i} 7699-\AA\ line in our spectra.

\noindent {\sl 11 UMi (HD\,136726).}  No significant Li line detected, small rotational broadening. \\
{\sl HD 136919.} \\
{\sl HD 138020.} \ion{K}{i} 7699-\AA\ blended with terrestrial O$_2$ doublet. Its large distance and low gravity suggest a supergiant. \\
{\sl 20 UMi (HD\,147142).} \\
{\sl HD 138116.} Weak Li. \\
{\sl HD 143641.} Both Na\,D lines have a blue-shifted, sharp, ISM line of comparable strength, while \ion{K}{i} 7699-\AA\ appears predominantly single with only a weak blue-shifted component.\\
{\sl HD 142961.} Weak Li.\\
{\sl HD 150275.} \\
{\sl HD 151698.} Both Na\,D lines appear with a red-shifted, sharp, ISM line of approximately half of its strength. No sign of ISM-related duplicity in \ion{K}{i} 7699-\AA .\\
{\sl HD 155153.} Li detected.\\
{\sl HR 6069 (HD\,146603).} \\
{\sl HR 6180 (HD\,150010).} \\
{\sl HD 145310.} \\
{\sl HD 144903.} Moderately strong Li detected.\\
{\sl HD 148374.} Li detected. \\
{\sl HD 150142.} Very weak Li abundance.\\
{\sl HD 147764.} Li detected. \\
{\sl $\upsilon$ Dra (HD\,176524).} \\
{\sl HD 148978.} Li detected.\\
{\sl HD 149843.} Strong Li line.\\
{\sl HD 137292.} Li detected. The spectrum synthesis suggests a super-solar metallicity of +0.19\,dex.\\
{\sl HD 138852.} Smallest $^{12}$C/$^{13}$C ratio in our giant sample. \\
{\sl HD 145742.} \\
{\sl HR 5844 (HD\,140227).} Most rapidly rotating giant in our pilot sample ($v\sin i\approx 5$\,\kms ).\\
{\sl HD 138301.} Both Na\,D lines appear with a red-shifted, sharp, ISM blend of less than half of its strength. A very weak red-shifted blend is may be seen in \ion{K}{i} 7699\,\AA , not in \ion{K}{i} 7664\,\AA\ though.\\
{\sl $\varepsilon$ Dra A (HD\,188119).} The spectrum is single lined in agreement with previous high-resolution spectra (e.g., McWilliam  \cite{will}). The star was included in the visual double star classification catalog of Lutz \& Lutz (\cite{landl}) as G8\,III for component A, and K5\,III for component B, but no photometry was derived. Simbad lists F7\,III for the B component. AB separation is 3.1\arcsec . Our spectrum is for the brighter A component. \\
{\sl HD 194298.} Li detected. Only very weak \ion{O}{i} triplet detected but a strong line within the \ion{O}{i} triplet is seen at 7774\,\AA . The cool temperature and low gravity suggest a bright giant classification rather than a normal class III. It also exhibits the largest bisector span in our sample which explains the discrepant $v\sin i$ measurements. The spectrum appears comparably metal deficient ($-0.44\pm0.16$\,dex).

\section{Extra figures}

\begin{figure*}[th]
 \centering
\includegraphics[angle=0,width=15cm,clip]{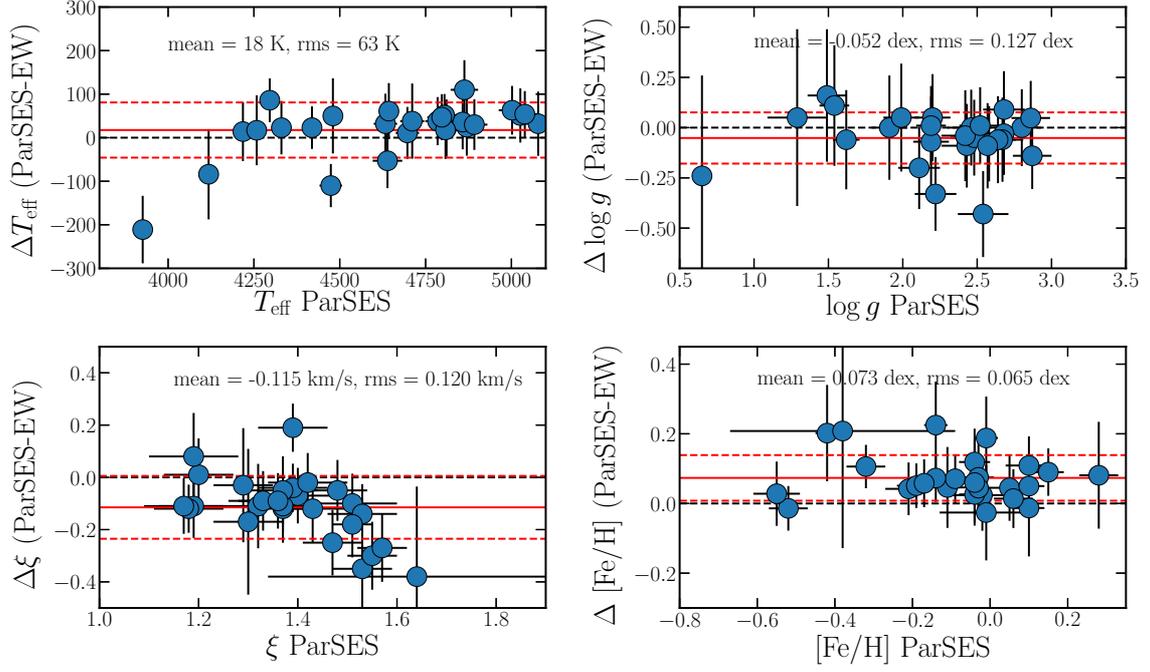}
 \caption{Comparison of results from the ParSES spectrum synthesis and the EW method (EW) for the STELLA+SES giant-star pilot sample. The four panels show, from top left to bottom right, effective temperature ($T_{\rm eff}$), logarithmic gravity ($\log g$), microturbulence ($\xi$), and logarithmic metallicity ([Fe/H]). The respective thick red line in each panel represents the mean difference and its $\pm$1$\sigma$ values (dashed red lines) also given numerically in the top of each panel. The black dashed line is the zero difference line. The dots are the individual target results with their $\pm$1$\sigma$ error bars. See text for details.}
 \label{F_C1}
\end{figure*}

\begin{figure*}[bh]
 \centering
\includegraphics[angle=0,width=15cm,clip]{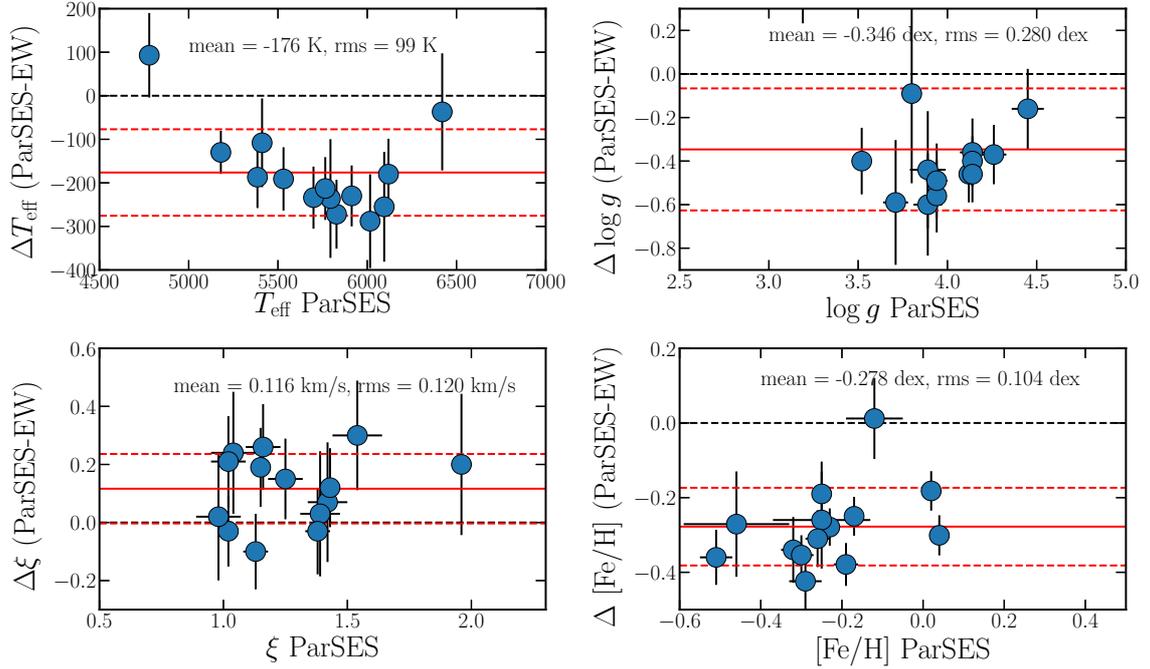}
 \caption{Comparison between spectral synthesis and EW-method results for the VATT+PEPSI dwarf-star pilot sample. Only stars with $v\sin i<20$\,\kms\ were used for this comparison. Also note that there are five giants within this sample. Otherwise, the details are the same as in Fig.~\ref{F_C1}. }
 \label{F_C2}
\end{figure*}

\begin{figure*}[ht]
{\bf a.} \\
\includegraphics[angle=0,width=14cm,clip]{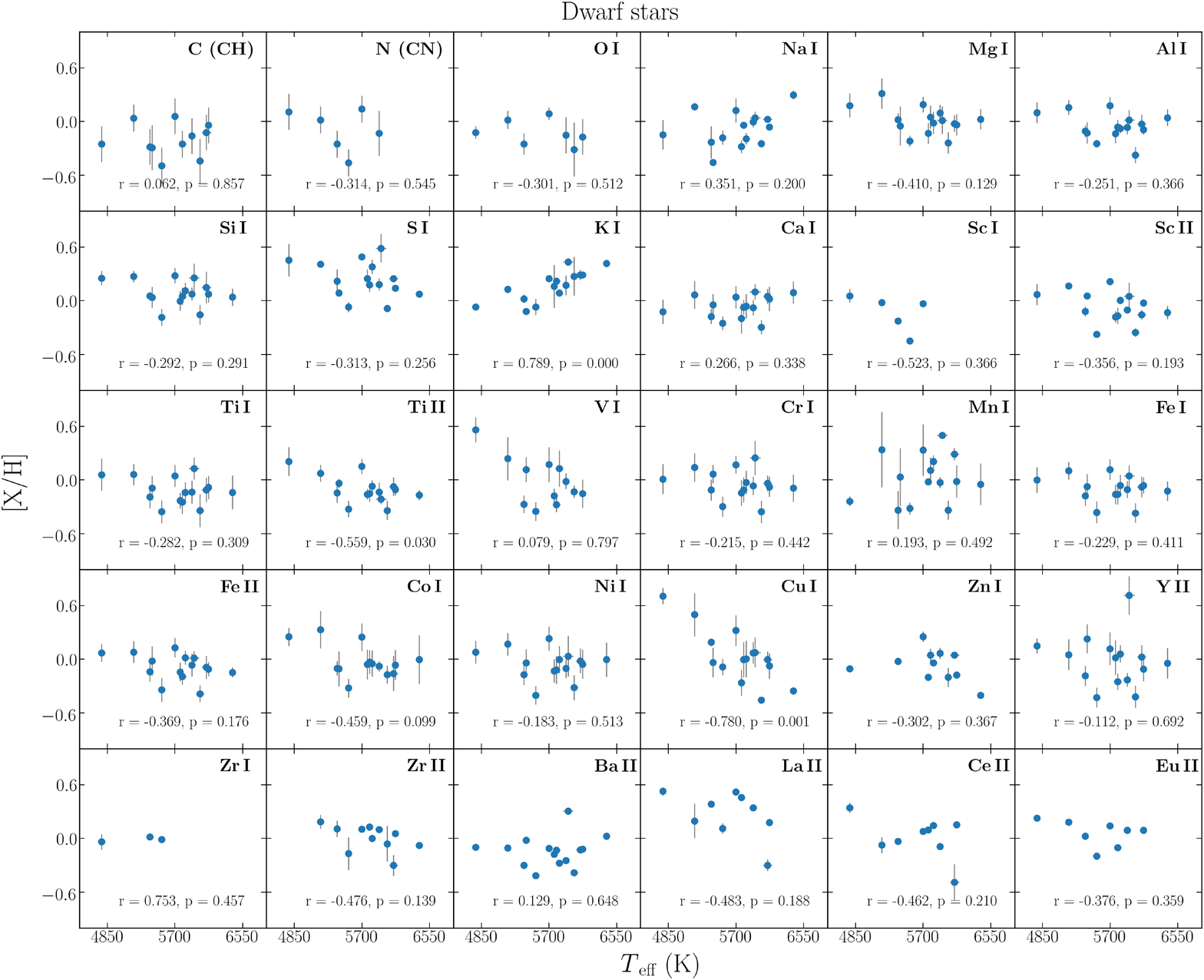}

{\bf b.} \\
\includegraphics[angle=0,width=14cm,clip]{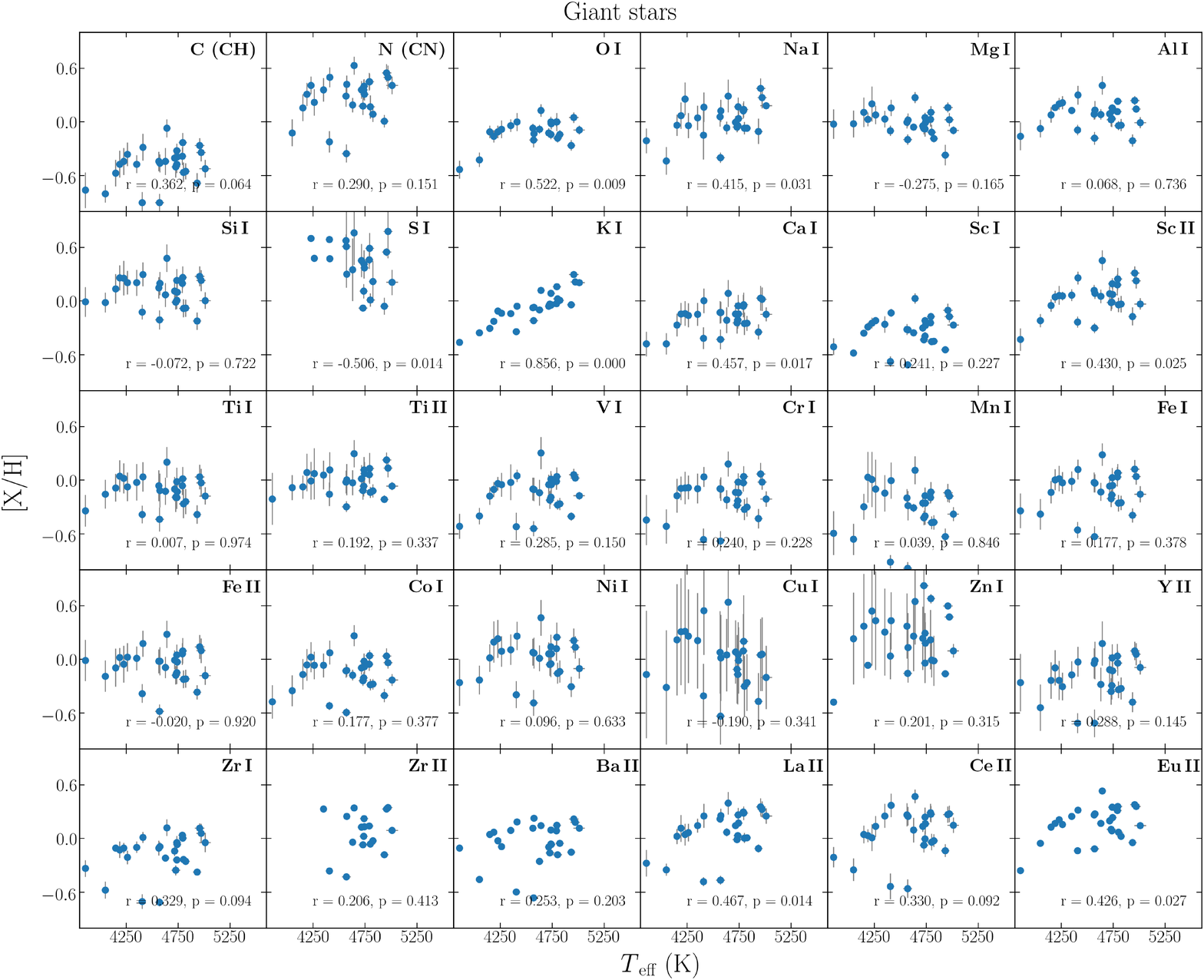}
 \caption{Logarithmic LTE elemental abundances, [X/H], as a function of effective temperature. Only pilot stars with $v\sin i<20$\,\kms\ were used for this check. Panel \emph{a}: Dwarf stars observed with VATT+PEPSI. Panel \emph{b}: Giant stars observed with STELLA+SES. Each sub-panel lists the $r,p$-values from a linear regression fit. }
 \label{F_C3}
\end{figure*}

\end{document}